\documentclass[11pt,a4paper]{article}

\usepackage{amsmath,mathtools,amssymb,amsfonts,bbm}
\usepackage[top=3.11cm,bottom=4.11cm,right=3.11cm,left=3.11cm]{geometry}
\usepackage{physics}
\usepackage{slashed}
\usepackage{xcolor,graphicx}
\usepackage{subfigure}
\usepackage{etoolbox}
\usepackage{bbold}
\usepackage{xspace}
\usepackage{appendix}
\usepackage{amsthm}
\usepackage{makecell}

\usepackage[utf8]{inputenc}
\usepackage[T1]{fontenc}
\usepackage[UKenglish]{babel}

\numberwithin{equation}{section}
\usepackage[hidelinks]{hyperref} % Must usually be the last package to be loaded, but before cleveref.
\usepackage[capitalise]{cleveref}

\usepackage{tabularx}
\newcolumntype{C}{>{\centering\arraybackslash}X}
\newcolumntype{R}{>{\raggedleft\arraybackslash}X}
\newcolumntype{L}{>{\raggedright\arraybackslash}X}

\usepackage{multirow, bigdelim}

\usepackage[style=ieee, citestyle=numeric-comp, backend=biber]{biblatex}
\allowdisplaybreaks

\newcommand{\intx}{\int d^4x}
\newcommand{\Dintx}{\int d^Dx}
\newcommand{\hypL}{\mathcal{Y}_L}
\newcommand{\hypR}{\mathcal{Y}_R}
\newcommand{\hypLR}{\mathcal{Y}_{LR}}
\newcommand{\hypRL}{\mathcal{Y}_{RL}}
\newcommand{\hypS}{\mathcal{Y}_S}
\newcommand{\projR}{\mathbb{P}_{\mathrm{R}}}
\newcommand{\projL}{\mathbb{P}_{\mathrm{L}}}
\newcommand{\DeltaHat}{\widehat{\Delta}}

\newcommand{\STIopD}{\mathcal{S}_{D}}

\theoremstyle{definition}
\newtheorem{definition}{Option}
\newcounter{tempOption} % Create a temporary counter for manual control
\renewcommand{\thedefinition}{\arabic{definition}} % Standard numbering for 'Option 1', 'Option 2'
\newtheorem{newdefinition}{Option}

\title{Shedding Light on Evanescent Shadows --- Exploration of non-anticommuting $\gamma_5$ in Dimensional Regularisation}

\newcommand{\email}{}
\author{
Paul L.\ Ebert,$^a$\thanks{\email{paul.ebert@mailbox.tu-dresden.de}}
\quad
Paul Kühler,$^a$\thanks{\email{paul.kuehler@tu-dresden.de}}\quad
Dominik  Stöckinger,$^a$\thanks{\email{dominik.stoeckinger@tu-dresden.de}}\quad
Matthias Weißwange,$^a$\thanks{\email{matthias.weisswange@tu-dresden.de}}\\[2em]
$^a$Institut für Kern- und Teilchenphysik, TU Dresden,\\ Zellescher Weg 19, DE-01069 Dresden, Germany}

\begin{document}
\thispagestyle{empty}

\maketitle

\setcounter{footnote}{0}
\vspace{2ex}
\begin{abstract}
\noindent
The mathematical consistency of the BMHV scheme
of dimensional regularisation (DReg) comes at the cost of requiring
symmetry-restoring counterterms
to cancel the regularisation-induced 
breaking of gauge and BRST invariance. 
There is no unique way to extend a 4-dimensional theory to $D$ dimensions, and different choices can be made for the dimensionally regularised fermions, evanescent parts of their kinetic terms and evanescent gauge interactions. Here we present a detailed study of the impact of changing such evanescent details. We leverage this freedom to identify
a particularly convenient formulation that simplifies practical calculations.
In order to thoroughly study the available options,
we focus on a general abelian chiral gauge theory including 
scalar fields and adopt a general 
approach to the BMHV implementation.
This allows for specialisation to various models
and different approaches, including those from the literature.
Importantly, our model can be specialised to the abelian sector 
of the Standard Model (SM).
Consequently, this article also serves as a roadmap for upcoming
applications to the full SM.
\end{abstract}

\newpage
\setcounter{page}{1}

\tableofcontents\newpage

%%%%% Main %%%%%

% +++++++++++++++++++++++++++++++++++++++++++++++++++++++++++++++++++++++++++++
\section{Introduction}

The Standard Model of particle physics (SM) has been tested by a wide range of both high- and low-energy experiments and is in fantastic overall agreement with measurements. Persistent efforts of increasing experimental precision as well as prospective future colliders call for more and more precise theory calculations \cite{Blondel:2018mad}.

The electroweak Standard Model (EWSM) is a chiral gauge theory where Dirac fields $\psi$ must be decomposed into their left- and right-handed parts, ${\psi}_{L,R}=\frac{1}{2}(1\mp\gamma_5)\psi$ with different gauge interactions. Dimensional Regularisation (DReg) of the SM is therefore  affected by the $\gamma_5$-problem, i.e.\ the absence of a consistent definition of $\gamma_5$ in $D$ dimensions which respects the familiar $4$-dimensional relations. Indeed requiring simultaneously full anticommutativity with all $\gamma^{\mu}$-matrices, cyclic traces and the validity of $4$-dimensional limits of traces, immediately leads to contradictions and one has to abandon some of the desired properties, see Ref.~\cite{Jegerlehner:2000dz,Belusca-Maito:2023wah} for reviews and further references. 
Early on, the BMHV scheme with non-anticommuting $\gamma_5$ has been proposed as a rigorous, mathematically fully consistent scheme \cite{tHooft:1972tcz, Breitenlohner:1976te, Breitenlohner:1977hr}. 
Alternative proposals include schemes with anticommuting $\gamma_5$, Ref.\ \cite{Chanowitz:1979zu}, or non-cyclic traces, 
Refs.\ \cite{Kreimer:1989ke, Korner:1991sx}. 
In recent years interest in solutions of the $\gamma_5$-problem has been increasing, and a number of results have been obtained in the well-established BMHV scheme as well as in alternative approaches \cite{Chen:2023lus, Chen:2024zju, Bruque:2018bmy, Rosado:2024pyy, Fuentes-Martin:2022vvu, Carmona:2021xtq, DiNoi:2023ygk, Naterop:2023dek}.
Here we continue the study of the BMHV scheme where chiral gauge invariance is spuriously broken on the regularised level by a non-anticommuting $\gamma_5$, and needs to be restored by appropriate counterterms. 
For early discussions of the scheme see e.g.\ Refs.\ \cite{Jones:1982zf,Freitas:2002ja}, while results for gauge-invariance restoring counterterms can be found first in Refs.\ \cite{Martin:1999cc, Sanchez-Ruiz:2002pcf} and more recently in Refs.\ \cite{Belusca-Maito:2020ala, Cornella:2022hkc, OlgosoRuiz:2024dzq} at 1-loop and \cite{Belusca-Maito:2021lnk, Stockinger:2023ndm, Kuhler:2024fak} at 2- and 3-loop order.
We also mention that in the field of flavour physics the BMHV scheme is regularly applied in cases where weak interactions appear at tree-level and QCD/QED corrections are considered \cite{Ciuchini:1993fk,Buchalla:1995vs}.

A major focus of our present work is that there is no unique $D$-dimensional extension of a 4-dimensional action, and in particular there is no unique way to implement the BMHV scheme. Instead, the regularised action can be chosen in different ways. If treated consistently, each option will lead to a correct, renormalised theory, though with different intermediate results and different symmetry-restoring counterterms. To appreciate this point, we contrast the usual $D$-dimensional QED interaction with the BMHV Z-boson interaction as treated e.g.\ in Refs.\ \cite{Ciuchini:1993fk,Buchalla:1995vs} and advocated in Ref.~\cite{Jegerlehner:2000dz},
\begin{equation}\label{Eq:QED_vs_BMHV-Z-Boson}
    eQ\overline{{\psi}}\gamma^{\mu}\psi A_{\mu}
    \text{\quad vs.\quad}\overline{\psi}(g_{L}\overline{\gamma}^{\mu}\projL+g_{R}\overline{\gamma}^{\mu}\projR)\psi Z_{\mu},
\end{equation}
with $\mathbb{P}_\mathrm{L/R}=\frac{1}{2}(1\mp\gamma_5)$ and where $\gamma^{\mu}$ denotes the fully $D$-dimensional $\gamma^{\mu}$-matrices, whereas $\overline{\gamma}^{\mu}$ denotes the purely $4$-dimensional parts. Thus, the vectorial photon interaction is treated fully in $D$ dimensions, while the Z-boson interaction is purely $4$-dimensional. Alternatively one could choose a purely $4$-dimensional interaction $\overline{\gamma}^{\mu}$ for the photon or include $(D-4)$-dimensional terms in the Z-boson interaction. While such different options have been mentioned in Refs.~\cite{Belusca-Maito:2020ala, OlgosoRuiz:2024dzq}, all actual computations of symmetry-restoring counterterms have opted for the purely $\overline{\gamma}^{\mu}$-interaction \cite{Martin:1999cc, Sanchez-Ruiz:2002pcf,Belusca-Maito:2020ala, Belusca-Maito:2021lnk, Stockinger:2023ndm, Cornella:2022hkc, OlgosoRuiz:2024dzq}.

In the present paper we analyse the impact of a general class of modifications of the BMHV implementation. These include all approaches discussed in the literature as special cases but also further generalisations. We consider an abelian chiral gauge theory modelled after the hypercharge sector in the EWSM, with suitable generalisations.
We omit non-abelian complications since this is sufficient to focus the study on different realisations of dimensionally regularised fermions. In Sec.\ \ref{Sec:The-Model} we motivate and characterise different options for the dimensional continuation of fermions including the different choices for fermion interactions and subsequently define our general model. Sec.\ \ref{Sec:SpecialCases} describes the BMHV treatments found in the literature and explains how they can be recovered by specialising our general setup. In Sec.\ \ref{Sec:STI+QAP} we provide an exhaustive list of Slavnov-Taylor identities which function as consistency conditions on the successful renormalisation of our model and our technical implementation. As the main general result Sec.\ \ref{Sec:1-Loop-Ren-Results} contains the full and explicit $1$-loop results both for the UV-renormalisation of Green functions as well as the required finite symmetry-restoring counterterms for the general model. In Sec.\ \ref{Sec:AnalysisOfResults} we discuss a number of special cases. After  recovering results of the literature we specialise to the abelian sector of the SM and make explicit how the counterterm structure depends on details of the $D$-dimensional continuation of the fermion action. We conclude with a summary of the main insights and their implications for studies of the full SM.

% -----------------------------------------------------------------------------

% +++++++++++++++++++++++++++++++++++++++++++++++++++++++++++++++++++++++++++++
\section{Abelian Chiral Gauge Theory with Scalar Fields}\label{Sec:The-Model}

In this section, we define the model considered in the present paper,
focusing on a massless abelian chiral gauge theory 
with both left- and right-handed fermions, as well as complex scalars.
This allows to study various aspects of
implementing the BMHV scheme of DReg for chiral gauge theories. 
We begin with a preliminary discussion of fermion fields and issues related to their 
kinetic and gauge interaction terms in 4 and in $D$ dimensions.
This will motivate the choice of our model which is defined and discussed subsequently.

We follow the notation established in
Refs.\ 
\cite{Belusca-Maito:2020ala,Belusca-Maito:2021lnk,Belusca-Maito:2023wah,Stockinger:2023ndm,Kuhler:2024fak} 
for the BMHV algebra, where e.g.\ $\overline{\gamma}^{\mu}$, $\widehat{\gamma}^{\mu}$, and $\gamma^{\mu}$ denote purely 4-dimensional, $(-2\epsilon)$-dimensional, and fully $D=(4-2\epsilon)$-dimensional objects, respectively. 
Every object carrying a Lorentz index decomposes into physical and evanescent contributions such as
$\gamma^{\mu}=\overline{{\gamma}}^{\mu}+\widehat{{\gamma}}^{\mu}$, where
\begin{equation}
    \begin{tabular}{c c c c c}
        $\overline{g}^{\mu\nu}\gamma_{\nu}=\overline{\gamma}^{\mu},$ &  $\widehat{g}^{\mu\nu}\gamma_{\nu}=\widehat{\gamma}^{\mu},$&
        $\overline{\gamma}^{\mu}\overline{\gamma}_{\mu}=4,$&
        $\widehat{\gamma}^{\mu}\widehat{\gamma}_{\mu}=-2\epsilon,$&
        $\overline{\gamma}^{\mu}\widehat{\gamma}_{\mu}=0.$\\
    \end{tabular}
\end{equation}
A central feature of the BMHV scheme is that the $\gamma_5$-matrix retains the usual anticommutator with the 4-dimensional part of every other gamma matrix, but commutes with their evanescent contribution
\begin{equation}
    \begin{tabular}{c c c}
        $\{\gamma_5,\overline{\gamma}^{\mu}\}=0,$ &  $\{\gamma_5,\gamma^{\mu}\}=2\gamma_5\widehat{\gamma}^{\mu},$&
        $[\gamma_5,\widehat{\gamma}^{\mu}]=0.$\\
    \end{tabular}
\end{equation}

\subsection{Fermions in $4$ and in $D$ Dimensions}\label{Sec:FermionsInGeneral}

We begin by highlighting basic issues related to the $D$-dimensional
treatment of fermion fields such as the kinetic and gauge interaction terms.
These issues are in principle well known, but they motivate our general setup.
For concreteness, we use the electron and a neutrino as explicit examples.

\paragraph{Fermion kinetic term:} 
In the pure electroweak Standard Model (EWSM) neutrinos are massless, and there exist only left-handed neutrinos and right-handed antineutrinos. 
Each (anti)neutrino can be described by a
2-component Weyl spinor field $\chi_{\nu}{}_{,\alpha}(x)$, or
equivalently by a 4-component Dirac spinor field $\nu(x)$ which
is an eigenstate of $\gamma_5$, such that
\begin{align}
  \nu &= \mathbb{P}_{\mathrm{L}}\nu\equiv \nu{}_{L},
\end{align}
where $\mathbb{P}_{\mathrm{L/R}}=\frac12(1\mp\gamma_5)$.
We use the latter 4-component description, since it is more amenable to
DReg.
In 4 dimensions, an associated kinetic term can
be written as
\begin{align}
    \mathcal{L}^{4D}_{\mathrm{kin},\nu} =
    {\overline{\nu{}_L}} i
    \overline{\slashed{\partial}} {\nu{}_L} .
\end{align}

As a matter of principle, DReg requires that propagator denominators contain $D$-dimensional momenta. On the Lagrangian level this implies that a fully $D$-dimensional fermion kinetic term is required to ensure
properly regularised propagators in loop diagrams.
Hence the
neutrino kinetic term must be extended to
\begin{align}
    \mathcal{L}_{\mathrm{kin},\nu} =
    {\overline{\nu}} i
    {\slashed{\partial}}{\nu} 
\end{align}
without $\mathbb{P}_{\mathrm{L/R}}$ projection operators, since only in
this way a fully $D$-dimensional ${\slashed{\partial}}$ appears and a
propagator Feynman rule 
$i\slashed{p}/(p^2+i\varepsilon)$ with fully $D$-dimensional
propagator is obtained.
Thus, DReg implicitly necessitates 
the introduction of a fictitious right-handed neutrino spinor field $\nu^{\text{st}}_R$ such that
\begin{align}
  \nu=\nu_L+\nu^{\text{st}}_R
  =\mathbb{P}_{\mathrm{L}}\nu+\mathbb{P}_{\mathrm{R}}\nu,
\end{align}
where the superscript ``st'' indicates that the respective field is
sterile, i.e., it is non-interacting. 
The kinetic term effectively decomposes as
\begin{align}
          \mathcal{L}_{\mathrm{kin},\nu} =
                  {\overline{\nu_L}} i
                  \overline{\slashed{\partial}} {\nu_L}
                  +
                  {\overline{\nu^{\text{st}}_R}} i
                  \overline{\slashed{\partial}} {\nu^{\text{st}}_R}
                  +
                  {\overline{\nu_L}} i
                  \widehat{\slashed{\partial}} {\nu^{\text{st}}_R}
                  +
                  {\overline{\nu^{\text{st}}_R}} i
                  \widehat{\slashed{\partial}} {\nu_L}.
\end{align}
In the BMHV scheme, it is inevitable that the kinetic term mixes chiralities and thus the physical and the fictitious sterile field.\footnote{%
In an anticommuting $\gamma_5$-scheme the terms mixing ${\nu_L}$--$\nu^{\text{st}}_R$ would be absent.
}
It is possible to assign to $\nu^{\text{st}}_R$ the global symmetry
transformation laws of  $\nu_L$, such that the kinetic term is globally
symmetric. However, since $\nu^{\text{st}}_R$ has no gauge
interactions, its local gauge transformation and the corresponding
BRST transformation must differ from the one of
$\nu_L$. Hence, the kinetic term inevitably breaks
BRST invariance --- the main drawback of the BHMV scheme.

Next we focus on the electron and positron, both of which can be
described by one 4-component Dirac spinor field $e(x)$. In the
EWSM, this field is split into the two chiralities
\begin{align}
  e=
  \mathbb{P}_{\mathrm{L}}e + \mathbb{P}_{\mathrm{R}}e
  \equiv e_L + e_R ,
\end{align}
where the gauge quantum numbers of $e_{L/R}$ are different.
The 4-dimensional kinetic term is as usual,
\begin{align}
          \mathcal{L}^{4D}_{\mathrm{kin},e} & =
                  {\overline{e}} i
                  \overline{\slashed{\partial}} {e}
                  =
                  {\overline{e_L}} i
                  \overline{\slashed{\partial}} {e_L}
                  +
                  {\overline{e_R}} i
                  \overline{\slashed{\partial}} {e_R}.
\end{align}
Note that for some purposes, e.g.\ in the context of
supersymmetric or grand unified theories, it can be useful to describe
the two chiralities by two distinct 2-component Weyl spinor fields,
and/or to replace the right-handed spinor field $e_R$ by an
equivalent left-handed spinor field $e^C_L$ with opposite charges. In such a case the
second term could be rewritten as 
${\overline{e^C_L}} i \overline{\slashed{\partial}} {e^C_L}$.

For the extension to $D$ dimensions there are now several options:
\begin{definition}[$\psi_L + \psi_R \longrightarrow \psi$]\label{Opt:Option1}\
    The most obvious approach is to work with
    the natural Dirac spinor combination
    \begin{align}
        e= e_L + e_R
    \end{align}
    without introducing additional fictitious fields. 
    In this case the $D$-dimensional kinetic term is written as
    \begin{align}
        \mathcal{L}_{\mathrm{kin},e} =
                  {\overline{e}} i
                  {\slashed{\partial}} {e},
    \end{align}
    and decomposes into
    \begin{align}\label{Eq:Lkin-Electron-D-dim-Decomposed}
          \mathcal{L}_{\mathrm{kin},e} =
                  {\overline{e_L}} i
                  \overline{\slashed{\partial}} {e_L}
                  +
                  {\overline{e_R}} i
                  \overline{\slashed{\partial}} {e_R}
                  +
                  {\overline{e_L}} i
                  \widehat{\slashed{\partial}} {e_R}
                  +
                  {\overline{e_R}} i
                  \widehat{\slashed{\partial}} {e_L}.
    \end{align}
    Although natural, this approach has the drawback that the
    evanescent $\widehat{\slashed{\partial}}$-terms mix physical fields 
    with different gauge quantum numbers (e.g.\ different hypercharges) 
    and thus not only local
    gauge and BRST invariance, but even 
    global gauge invariance is broken.
\end{definition}
%%%%%%%%%%%%%%%%%%%%%%%%%%%%%%%%%%%%%%%%%%%%%%%%%%%%%%%%%%%%
    %Manually increment and control numbering for 'Option 2a'
    \stepcounter{definition} % Increment to Option 2
    \setcounter{tempOption}{\value{definition}} % Store the current Option number in temp counter
    \setcounter{definition}{0} % Reset the definition counter for 'a', 'b' formatting
    \renewcommand{\thedefinition}{\arabic{tempOption}\alph{definition}} % Now number as 2a, 2b, etc.
    %Manual label for group reference, to reference both Options 2a and 2b
    \phantomsection
    \label{Opt:Option2a-2b-GroupReference}
%%%%%%%%%%%%%%%%%%%%%%%%%%%%%%%%%%%%%%%%%%%%%%%%%%%%%%%%%%%%
\begin{definition}[$\psi_L + \psi_R^{\text{st}} \longrightarrow \psi_1$ and $\psi_L^\text{st} + \psi_R \longrightarrow \psi_2$]\label{Opt:Option2}\
    Another option seeks to separate the different gauge multiplets
    $e_L$ and $e_R$ and to retain at least global gauge
    invariance. 
    This replicates the procedure for the neutrino, introduces two fictitious sterile fields $e^{\text{st}}_L$ and $e^{\text{st}}_R$ and defines the two Dirac spinors
    \begin{align}\label{Eq:Electron-2-Spinors-Option2}
        e_1 &= e_L + e^{\text{st}}_R,
        &
        e_2 &= e^{\text{st}}_L + e_R,
    \end{align}
    as well as the kinetic term
    \begin{align}
          \mathcal{L}_{\mathrm{kin},e} =
                  {\overline{e_1}} i
                  {\slashed{\partial}} {e_1}
                  +
                  {\overline{e_2}} i
                  {\slashed{\partial}} {e_2} .
    \end{align}
    The explicit
    decomposition of the kinetic terms reads
    \begin{equation}\label{Eq:Lkin-Electron-Option2-Decomposition}
        \begin{aligned}
          \mathcal{L}_{\mathrm{kin},e} & =
                  {\overline{e_L}} i
                  \overline{\slashed{\partial}} {e_L}
                  +
                  {\overline{e^{\text{st}}_R}} i
                  \overline{\slashed{\partial}} {e^{\text{st}}_R}
                  +
                  {\overline{e_L}} i
                  \widehat{\slashed{\partial}} {e^{\text{st}}_R}
                  +
                  {\overline{e^{\text{st}}_R}} i
                  \widehat{\slashed{\partial}} {e_L}\\
                  %&+(L\leftrightarrow R).
                  &+ {\overline{e_R}} i
                  \overline{\slashed{\partial}} {e_R}
                  +
                  {\overline{e^{\text{st}}_L}} i
                  \overline{\slashed{\partial}} {e^{\text{st}}_L}
                  +
                  {\overline{e_R}} i
                  \widehat{\slashed{\partial}} {e^{\text{st}}_L}
                  +
                  {\overline{e^{\text{st}}_L}} i
                  \widehat{\slashed{\partial}} {e_R}.
        \end{aligned}
    \end{equation}
    Hence, the global symmetry can be preserved at the cost of a proliferation of terms.
    Still, local gauge and BRST invariance is lost, as in all other cases.
\end{definition}
\begin{definition}[$\psi_L + \psi_R^{\text{st}} \longrightarrow \psi_1$ and $\psi_L^C + {\psi^C_R}^{\text{st}} \longrightarrow \psi_2$]\label{Opt:Option2b}\
    This approach is similar to Option \ref{Opt:Option2}, but uses the rewritten
    spinor $e^C_L$ instead of $e_R$ as a basis. 
    In this way $e_2$ becomes
    $e_2=e^C_L+{e^C_R}^{\text{st}}$. As a potential advantage, 
    all physical fields are left-handed and all
    fictitious sterile fields are right-handed. 
    The other properties are
    the same as for Option \ref{Opt:Option2}.
\end{definition}

The preceding discussion used a neutrino and the electron as examples
but can be transferred to other fermions $\psi$. 
It illustrates that the general treatment of fermions in the
context of DReg and the BMHV scheme uses Dirac spinors $\psi$ which
can be decomposed into $\psi_L+\psi_R$, where either both $\psi_L$ and
$\psi_R$ may be physical fields, with possibly different gauge quantum
numbers, or where one of them may be fictitious and sterile.

\paragraph{Propagators and drawback of  Option \hyperref[Opt:Option2a-2b-GroupReference]{2}:} 
    A peculiar consequence of introducing sterile partner fields in Options \ref{Opt:Option2} and \ref{Opt:Option2b} arises from the mixed kinetic terms in Eq.\ (\ref{Eq:Lkin-Electron-Option2-Decomposition}), necessitating a diagonalisation.
    Although our focus thus far has been exclusively on massless fermions, let us temporarily consider
    physical fermion mass terms, 
    which, in chiral gauge theories, can only emerge from spontaneous symmetry breaking (SSB).
    These introduce a noteworthy complication for Option \hyperref[Opt:Option2a-2b-GroupReference]{2}
    and are essential in models such as the EWSM.
    We continue to work with the electron as an example such that the decomposition of the kinetic terms in the massive case reads
    \begin{equation}\label{Eq:Lkin-Electron-MassiveCase}
        \begin{aligned}
          \mathcal{L}_{\mathrm{kin},e} =
                  \Big[
                  {\overline{e_L}} i
                  \overline{\slashed{\partial}} {e_L}
                  +
                  {\overline{e^{\text{st}}_R}} i
                  \overline{\slashed{\partial}} {e^{\text{st}}_R}
                  +
                  {\overline{e_L}} i
                  \widehat{\slashed{\partial}} {e^{\text{st}}_R}
                  +
                  {\overline{e^{\text{st}}_R}} i
                  \widehat{\slashed{\partial}} {e_L}
                  +(L\leftrightarrow R)
                  \Big]
                  - m(\overline{e_L}e_R+\overline{e_R}e_L),
        \end{aligned}
    \end{equation}
    cf.\ Eq.\ (\ref{Eq:Lkin-Electron-Option2-Decomposition}).
    In order to perform diagrammatic computations in DReg,
    we again combine the chiral fermion fields into full Dirac spinors using Eq.\ (\ref{Eq:Electron-2-Spinors-Option2}).
    With these definitions, the Lagrangian can be rewritten as follows:
    \begin{equation}\label{Eq:Lkin-Electron-MassiveCase-MatrixForm}
        \begin{aligned}
            \mathcal{L}_{\mathrm{kin},e} & =
                \overline{E} \mathcal{D} E =
                \begin{pmatrix}
                    \overline{e_1} & \overline{e_2}\\
                \end{pmatrix}
                \begin{pmatrix}
                    i \slashed{\partial} & -m\projR\\
                    -m\projL & i \slashed{\partial} \\
                \end{pmatrix}
                \begin{pmatrix}
                    e_1\\
                    e_2
                \end{pmatrix}.
        \end{aligned}
    \end{equation}
The propagator matrix $\mathcal{P}$ is obtained by inversion of the 
above introduced matrix $\mathcal{D}$, i.e.\ 
$\mathcal{P}\mathcal{D}=\mathcal{D}\mathcal{P}=\mathbb{1}$. 
In momentum space, with incoming momentum $p$, the propagator matrix reads,
\begin{equation}\label{Eq:MassivePropagatorMatrix}
    \widetilde{\mathcal{P}} = 
    \begin{pmatrix}
            \frac{p^2\slashed{p}-m^2\overline{\slashed{p}}\projL}{p^4-m^2\overline{p}^2}& \frac{(\overline{p}^2\projL+\widehat{p}^2\projR)m+\widehat{\slashed{p}}\overline{\slashed{p}}(\projL-\projR)m}{p^4-m^2\overline{p}^2}\\
            \frac{(\widehat{p}^2\projL+\overline{p}^2\projR)m+\widehat{\slashed{p}}\overline{\slashed{p}}(\projR-\projL)m}{p^4-m^2\overline{p}^2} & \frac{p^2\slashed{p}-m^2\overline{\slashed{p}}\projR}{p^4-m^2\overline{p}^2}\\
        \end{pmatrix}.
\end{equation}
In the massless case, this result reduces
to the usual $D$-dimensional denominators.
Hence, we obtain the same propagators as in the case of Option \ref{Opt:Option1},
with no further complications.
However, in the presence of masses the crucial complication arises from the
appearance of explicitly $4$-dimensional terms in the denominators, in addition to the
usual $D$-dimensional ones.
Such denominators are not the standard subject of integration techniques in DReg, 
rendering the computation rather involved.
This and the general proliferation of terms constitutes the major drawback of  Option \hyperref[Opt:Option2a-2b-GroupReference]{2},
in particular for practical calculations in massive theories such as the EWSM
in its broken phase.

\paragraph{Fermion-gauge boson interaction current:} 
Next, we highlight issues related to the $D$-dimensional
treatment of fermion fields appearing once we introduce gauge
interactions. 
We continue to use the electron and a neutrino as examples,
but the discussion can again be transferred to other fermions.

In pure QED, the interaction of the electron with the photon
$A_\mu$ can naturally be written as $  \overline{e}\gamma^\mu e A_\mu$ 
in $D$ dimensions, using the above Option \ref{Opt:Option1} for the
electron spinor,
in order to maintain its vector-like nature.
As mentioned in the Introduction this is the usual way to treat the QED interaction
in DReg, but using the decomposition into chiralities it amounts to
\begin{align}
  \label{Eq:LeegammaQED}
\mathcal{L}_{ee\gamma} =
  \overline{e}\projR\overline{\gamma}^\mu \projL e\overline{A}_\mu
  +
  \overline{e}\projL\overline{\gamma}^\mu \projR e\overline{A}_\mu
  +
  \overline{e}\projR\widehat{\gamma}^\mu \projR e\widehat{A}_\mu
  +
  \overline{e}\projL\widehat{\gamma}^\mu \projL e\widehat{A}_\mu.
\end{align}
This demonstrates that in conventional DReg, QED contains purely
4-dimensional interactions with $\overline{A}_\mu$, which preserve
chirality, as well as evanescent interactions with the
$(D-4)$-dimensional photon $\widehat{A}_\mu$, which do not preserve chirality.
Applied to the EWSM, the $\widehat{A}_\mu$ terms would, for example, violate
global hypercharge conservation since they couple fields of
different hypercharges.  In this context it might be motivated to change the Lagrangian and drop
the evanescent interaction terms.

In contrast, the EWSM electron--neutrino interaction
with the $W^-$-boson can be written as
$\overline{e}\,\overline{\gamma}^\mu\projL\nu W^-_\mu$ in 4 dimensions. In $D$ dimensions, it is in principle possible to modify the structure $\overline{\gamma}^\mu\projL$ by any evanescent linear combination of the form $a\widehat{\gamma}^\mu\projL+b\widehat{\gamma}^\mu\projR$, as long as the $D$-dimensional Lagrangian is hermitian and formally Lorentz invariant.
Two concrete, motivated $D$-dimensional continuations 
(using the above Option \ref{Opt:Option1} for the electron spinor) 
would be
\begin{newdefinition}[$\overline{\gamma}^\mu\projL \longrightarrow \gamma^\mu\projL=\projR\overline{\gamma}^\mu\projL+\projL\widehat{\gamma}^\mu\projL$]\label{Opt:Option-(i)}\
    \begin{equation}\label{Eq:W-Treatment-Option-(i)}
        \begin{aligned}
            \mathcal{L}_{e\nu W} =
            \overline{e}\projR\overline{\gamma}^\mu\projL\nu \overline{W}^-_\mu+
            \overline{e}\projL\widehat{\gamma}^\mu\projL\nu  \widehat{W}^-_\mu + \mathrm{h.c.},
        \end{aligned}
    \end{equation}
\end{newdefinition}
\begin{newdefinition}[$\overline{\gamma}^\mu\projL \longrightarrow \overline{\gamma}^\mu\projL =\projR\overline{\gamma}^\mu\projL$]\label{Opt:Option-(ii)}\
    \begin{equation}\label{Eq:W-Treatment-Option-(ii)}
        \begin{aligned}
            \mathcal{L}_{e\nu W} =
            \overline{e}\projR\overline{\gamma}^\mu\projL\nu \overline{W}^-_\mu + \mathrm{h.c.}.
        \end{aligned}
    \end{equation}
\end{newdefinition}
The first version corresponds to a naive $D$-dimensional continuation, while the second version, 
i.e.\ Option \ref{Opt:Option-(ii)},
uses a strictly
4-dimensional $\gamma$-matrix in the regularised theory. As a result, the two options
differ by an additional evanescent interaction, which 
couples the wrong-chirality electron $e_R$ to
$\nu_L$ and the $(D-4)$-dimensional gauge boson
$\widehat{W}^-_\mu$.
Here it appears obvious that the second version, which
contains only the 4-dimensional gauge field and no such evanescent
interaction, is advantageous. 

Note that the treatment of the fermion-gauge boson interactions is independent of
the method used to construct Dirac spinors discussed
above. For example, the ordinary DReg treatment of QED can be
transferred to Option \ref{Opt:Option2}, where 
$e_1=e_L+e_R^{\text{st}}$ 
and $e_2=e_R+e_L^{\text{st}}$.
In this case, we could define the above $D$-dimensional electron--photon
interaction as
\begin{equation*}
    \begin{aligned}
\mathcal{L}_{ee\gamma} =
        \overline{e}_1 \projR \overline{\slashed{A}} \projL e_1 
        +
        \overline{e}_2 \projL \overline{\slashed{A}} \projR e_2
        +
        \overline{e}_1 \projR \widehat{\slashed{A}} \projR e_2 
        +
        \overline{e}_2 \projL \widehat{\slashed{A}} \projL e_1,
    \end{aligned}
\end{equation*}
which is equal to Eq.\ (\ref{Eq:LeegammaQED}) 
but purely expressed in terms of the Dirac spinors $e_{1,2}$.

The discussion shows that it generally depends on
the context whether it is preferable to include interactions with
evanescent gauge fields and which form they should take. 
In general, e.g.\ for the interactions with the $Z$ boson in the EWSM, 
there is no obvious best choice. 
For instance, as mentioned in the Introduction, in the context of the construction of effective
Hamiltonians for $B$-meson decays \cite{Ciuchini:1993fk,Buchalla:1995vs}, weak vertices were treated in the BMHV scheme in purely 4 dimensions as in
Eq.\ (\ref{Eq:W-Treatment-Option-(ii)}).

In the present paper we will therefore allow all such choices and
study their consequences.

\subsection{Definition of the Theory in $D$-Dimensions}\label{Sec:D-DimLagrangian}

Having discussed issues related to the $D$-dimensional treatment
of fermions, we now lay out the $D$-dimensional theory to be investigated.
The key feature of our model is that it allows to study all the options of handling fermions illustrated in the previous subsection. In this way it generalises the literature, but it also contains all implementations of gauge--fermion interactions used in the literature as special cases.
To focus the attention on the treatment of fermions,
we restrict ourselves to abelian gauge theories and
impose several other physically motivated restrictions.

As a starting point, the $D$-dimensional tree-level Lagrangian is given by
\begin{equation}\label{Eq:model-Lagrangian}
    \begin{aligned}
        \mathcal{L} = 
        \mathcal{L}_{\mathrm{fermion}} + \mathcal{L}_{\mathrm{gauge}}
        + \mathcal{L}_{\mathrm{scalar}} + \mathcal{L}_{\mathrm{Yukawa}}
        + \mathcal{L}_{\mathrm{ghost+fix}} + \mathcal{L}_{\mathrm{ext}},
    \end{aligned}
\end{equation}
and the corresponding ordinary $4$-dimensional Lagrangian is obtained
in the limit $D \to 4$ and after setting all evanescent contributions to zero.
We denote this procedure as
\begin{equation}\label{Eq:4DlimitL}
    \begin{aligned}
        \mathop{\text{LIM}}_{D \, \to \, 4} \mathcal{L} = \mathcal{L}^{4D}.
    \end{aligned}
\end{equation}
The following paragraphs give a detailed description of every contribution to the Lagrangian \eqref{Eq:model-Lagrangian} of our model:

\paragraph{Physical field content and basic symmetries:}
As alluded to earlier, we consider a massless
abelian chiral gauge theory with $U(1)$ (hypercharge) gauge group and corresponding gauge field $B^\mu$. 
The model contains a set of Dirac fermions $\psi_i$ which can be decomposed into left- and right-handed parts,
$\psi_i=\psi_{L,i}+\psi_{R,i}$.
For every field $\psi_i$, both parts may be physical, or one of them may be fictitious and sterile.
The left- and right-handed fermions couple to the gauge boson via real and diagonal hypercharge matrices $\hypL{}_{ij}$ and $\hypR{}_{ij}$, respectively.
Further, we also allow for complex scalars $\phi_a$ with real hypercharge $\hypS$, which is assumed to be universal for all scalar fields.
Besides the gauged $U(1)$ symmetry of the hypercharges, we also impose a
global $U(1)_{Q}$ and a global $SU(3)_c$ symmetry. 
The electromagnetic charge $Q$ and the colour charge
are assumed to be diagonal in the fields, and in contrast to
the hypercharge they are both attributed to vector-like symmetries, i.e.\ for each index
$i$, the electromagnetic and colour charges of $\psi_{L,i}$ and
$\psi_{R,i}$ are equal.
As a further simplifying restriction, we assume that the
Yukawa interactions are fermion-number conserving. 
This constitutes a physical scenario similar to the
abelian sector of the SM or SM-like models such as the two-Higgs doublet model (2HDM), however,
only with \emph{global} $U(1)_Q$ and $SU(3)_c$ symmetries.

\paragraph{Fermion kinetic and gauge boson interaction terms:} 
The $D$-dimensional Lagrangian for the fermions is defined as
\begin{equation}
  \begin{aligned}
  \label{Eq:Lfermion}
        \mathcal{L}_{\mathrm{fermion}} = \overline{\psi}_j i \slashed{\partial} \psi_j 
        &- g {\hypR}_{ij} \overline{\psi}_i \projL \overline{\slashed{B}} \projR \psi_j
        - g {\hypL}_{ij} \overline{\psi}_i \projR \overline{\slashed{B}} \projL \psi_j\\
        &- g {\hypRL}_{ij} \overline{\psi}_i \projL \widehat{\slashed{B}} \projL \psi_j
        - g {\hypLR}_{ij} \overline{\psi}_i \projR \widehat{\slashed{B}} \projR \psi_j.
\end{aligned}
\end{equation}
The first line contains the usual $D$-dimensional fermion kinetic terms
and gauge interactions corresponding to the 4-dimensional covariant derivative
\begin{equation}
    \begin{aligned}
        &\overline{D}_{\mu} \psi = \big(\overline{\partial}_{\mu} + i g \hypL
      \overline{B}_{\mu} \projL + i g \hypR \overline{B}_{\mu} \projR\big) \psi
    \end{aligned}
\end{equation}
with the real and diagonal hypercharge matrices $\hypL$ and $\hypR$. 
These hypercharges $\hypL$ and $\hypR$
are the same as in 4 dimensions, such that the 4-dimensional part
of the $D$-dimensional action preserves BRST-invariance.
The last line of Eq.\ (\ref{Eq:Lfermion}) constitutes the evanescent 
interactions of the fermions with the $(D-4)$-dimensional part 
of the gauge field $\widehat{B}_\mu$.
Following the discussion of the previous subsection we opt for an extension which is as
general as possible by allowing such evanescent couplings via evanescent hypercharges $\hypLR$ and $\hypRL$.
These evanescent hypercharge matrices can generally be independent of the
$4$-dimensional gauge interaction matrices $\hypL$ and $\hypR$,
in particular they may be off-diagonal.
The only requirement that they must satisfy is
\begin{align}\label{Eq:Requirement-YLR-YRL}
  \hypRL = \hypLR^{\dagger}
\end{align}
due to hermiticity.
In contrast to the usual gauge interactions,
they couple left-handed to right-handed fermions and thus do not
preserve chirality.  
As a result it may happen that fields with different hypercharges are
coupled together, such that the evanescent interaction terms break
local and even global hypercharge conservation, similar to the
evanescent parts of the kinetic terms.
These terms present a key novelty of our model and have so far not been used in the literature (cf.\ Sec.\ \ref{Sec:ModelsFromLiterature}).

In order to preserve the global symmetries $U(1)_Q$ and $SU(3)_c$,
we choose the evanescent hypercharges $\hypLR$ and $\hypRL$
to commute with $Q$ and respect colour charge conservation.
Any further specifications or restrictions on $\hypLR$ and $\hypRL$,
where applicable, are stated in the following sections presenting
results, notably in Sec.\ \ref{Sec:1-Loop-Ren-Results} and \ref{Sec:AnalysisOfResults}.

A restriction on the 4-dimensional hypercharges is given by the
anomaly cancellation condition  
\begin{equation}\label{Eq:AnomalyCancellationCondition}
    \begin{aligned}
        \mathrm{Tr}\big(\hypR^3\big) - \mathrm{Tr}\big(\hypL^3\big) = 0,
    \end{aligned}
\end{equation}
which is assumed to hold as part of the definition of the theory. 
Finally, by definition the sterile part of a
fermion field $\psi^{\text{st}}_{X,k}$ ($X=L,R$) is characterised by the absence
of interactions and thus the absence of higher-order
corrections. Technically this implies the constraint
\begin{equation}\label{Eq:NonRenOfSterileFields}
    \begin{aligned}
        \frac{\delta \Gamma_{\text{cl}}}{\delta
          \psi_{X,k}^{\text{st}}} 
        =
        \frac{\delta \Gamma}{\delta \psi_{X,k}^{\text{st}}} = \text{linear in quantum fields}
    \end{aligned}
\end{equation}
at all orders of perturbation theory. In particular,
all matrix elements of (evanescent and non-evanescent) hypercharge
matrices that correspond to sterile fields must vanish.

\paragraph{Gauge boson, scalar and Yukawa terms:}
For the remaining parts of the Lagrangian, no complications arise. The
gauge boson kinetic term in the Lagrangian is provided as usual by
\begin{equation}
    \begin{aligned}
        \mathcal{L}_{\mathrm{gauge}} = - \frac{1}{4} F^{\mu\nu}F_{\mu\nu},
    \end{aligned}
\end{equation}
with field strength tensor $F_{\mu\nu} = \partial_{\mu}B_{\nu} -
\partial_{\nu}B_{\mu}$, continued to $D$ dimensions in the obvious way.

As mentioned earlier, in the scalar sector we use complex scalar
fields $\phi_a$, which are eigenstates of electric charge $Q$,\footnote{
Real scalars are not eigenstates of $Q$ (and generally also not of the hypercharge).} 
and which all have the same real hypercharge $\mathcal{Y}_S$.
The associated
4-dimensional covariant derivative is
\begin{align}
  \overline{D}_{\mu}\phi_{a} = \big(\overline{\partial}_{\mu} 
  + i g \hypS \overline{B}_{\mu}\big)\phi_a, 
\end{align}
and the obvious $D$-dimensional Lagrangian for the scalar kinetic,
gauge interaction and scalar potential terms reads
\begin{equation}
    \begin{aligned}
        \mathcal{L}_{\mathrm{scalar}} &=        
        (\partial_{\mu}\phi_a^{\dagger})(\partial^{\mu}\phi_a)
        - ig\hypS B^{\mu} \big( \phi_a^{\dagger} \partial_{\mu}\phi_a -
        \phi_a \partial_{\mu} \phi_a^{\dagger} \big)
        + g^2 \hypS^2 B^{\mu}B_{\mu}\phi_a^{\dagger}\phi_a\\
        &- \frac{\lambda_{klmn}}{6} \phi^{\dagger}_k \phi_l \phi^{\dagger}_m \phi_n,
    \end{aligned}
\end{equation}
where the potential parameters satisfy         $\lambda_{klmn} =
\lambda_{nmlk}^{*}$ because of hermiticity and
\begin{equation}
    \begin{aligned}
        \lambda_{klmn} \equiv 0, \quad \text{unless} \quad -Q_{\phi_k}+Q_{\phi_l}-Q_{\phi_m}+Q_{\phi_n}=0,
    \end{aligned}
\end{equation}
with index permutation symmetry $\lambda_{klmn}=\lambda_{mlkn}=\lambda_{knml}=\lambda_{mnkl}$.

Further, we impose perturbative fermion number conservation such that a fermion flux can be defined in all Feynman diagrams.
As a consequence, only Yukawa interactions involving the fundamental
fermions $\psi_i$ may appear, while Yukawa interactions involving
charge-conjugated fields $\psi^C_j$ (which were allowed in the purely
right-handed theory in  Ref.\ \cite{Belusca-Maito:2020ala}) are
forbidden. This leaves us with the Yukawa interaction Lagrangian
\begin{equation}
    \begin{aligned}
        \mathcal{L}_{\mathrm{Yukawa}} = 
        - G^{a}_{ij} {\overline{\psi_L}}_i \phi_{a} {\psi_R}_j
        - K^{a}_{ij} {\overline{\psi_L}}_i \phi_{a}^{\dagger} {\psi_R}_j
        + \mathrm{h.c.},
    \end{aligned}
\end{equation}
where the Yukawa couplings are restricted by global hypercharge conservation
and BRST invariance via
\begin{equation}\label{Eq:YukawaHyperchargeBRSTCondition}
    \begin{aligned}
        {\hypL}_{ik} G^{a}_{kj} - G^{a}_{ik} {\hypR}_{kj} - \hypS G^{a}_{ij} &= 0,\\
        {\hypL}_{ik} K^{a}_{kj} - K^{a}_{ik} {\hypR}_{kj} + \hypS K^{a}_{ij} &= 0,
    \end{aligned}
\end{equation}
by electric charge conservation via
\begin{equation}
    \begin{aligned}
        G^{a}_{ij} &\equiv 0, \quad \text{unless} \quad -Q_{{\psi_L}_i}+Q_{\phi_a}+Q_{{\psi_R}_j} = 0,\\
        K^{a}_{ij} &\equiv 0, \quad \text{unless} \quad -Q_{{\psi_L}_i}-Q_{\phi_a}+Q_{{\psi_R}_j} = 0,
    \end{aligned}
\end{equation}
and by colour charge conservation.

\paragraph{Gauge fixing and external field Lagrangian:}
The familiar $R_\xi$-gauge fixing and ghost parts of the Lagrangian are
defined straightforwardly in $D$ dimensions as
\begin{equation}\label{Eq:Ddim-LGaugeFixing+Ghost}
    \begin{aligned}
        \mathcal{L}_{\mathrm{ghost+fix}} = 
        - \frac{1}{2\xi} (\partial^{\mu}B_{\mu})^2
        - \overline{c} \Box c.
    \end{aligned}
\end{equation}
Working with the Slavnov-Taylor identity, along with the quantum action
principle of DReg, requires us to couple the BRST transformations of
the fields to external sources $\{\rho^{\mu},R_i,\Upsilon^a,\zeta\}$ 
and taking them into account via
\begin{equation}\label{Eq:Ddim-Lext}
    \begin{aligned}
        \mathcal{L}_{\mathrm{ext}} = 
        \rho^{\mu}sB_{\mu} 
        + {\overline{R}}{}^{i}s{\psi}_i + R^{i}s\overline{\psi}_i
        + {\Upsilon^{a}}^{\dagger}s\phi_{a} + \Upsilon^{a}s\phi_{a}^{\dagger}
        + \zeta sc,
    \end{aligned}
\end{equation}
where $s$ denotes the $D$-dimensional BRST operator.

The BRST transformations, also extended straightforwardly to $D$ dimensions,
are explicitly given by 
\begin{equation}\label{Eq:Ddim-BRSTTrafos}
    \begin{aligned}
        sB_{\mu} &= \partial_{\mu}c,\\
        s\psi_i &= s{\psi_R}_i + s{\psi_L}_i = -igc\Big({\hypR}_{ij}{\psi_{R}}_j+{\hypL}_{ij}{\psi_{L}}_j\Big),\\
        s\overline{\psi}_i &= s{\overline{\psi_R}}_i + s{\overline{\psi_L}}_i = igc\Big({\overline{\psi_{R}}}_j{\hypR}_{ji}+{\overline{\psi_{L}}}_j{\hypL}_{ji}\Big),\\
        s\phi_a &= -igc \hypS \phi_a,\\
        sc&=0,\\
        s\overline{c}&=\mathcal{B}= - \frac{1}{\xi} \partial^{\mu}B_{\mu},\\
        s\mathcal{B}&=0,
    \end{aligned}
\end{equation}
where the Nakanishi-Lautrup field $\mathcal{B}$ is not to be confused with 
the gauge field $B_{\mu}$.

\subsection{Breaking of Gauge and BRST Invariance}

In the BMHV scheme, the fermionic parts of the $D$-dimensional
Lagrangian break local gauge invariance. Equivalently, the total
$D$-dimensional tree-level action including gauge fixing,
ghost, and external field terms, breaks BRST invariance. 
This regularisation-induced BRST breaking  
at tree-level is  explicitly given by
\begin{equation}\label{Eq:TreeLevelBreaking}
    \begin{aligned}
        \DeltaHat &= \STIopD\Big(\Dintx\,\mathcal{L}\Big)\\
        &= - g \Dintx \, c \,
        \Bigg\{
        \overline{\psi}_i 
        \bigg[
        \projR 
        \bigg(
        {\hypR}_{ij} \overset{\leftarrow}{\widehat{\slashed{\partial}}}
        + {\hypL}_{ij} \overset{\rightarrow}{\widehat{\slashed{\partial}}}
        - {\hypLR}_{ij} \Big( \overset{\leftarrow}{\widehat{\slashed{\partial}}} 
                            + \overset{\rightarrow}{\widehat{\slashed{\partial}}} \Big)
        \bigg)
        \projR\\
        &\hspace{3cm} +
        \projL 
        \bigg(
        {\hypR}_{ij} \overset{\rightarrow}{\widehat{\slashed{\partial}}}
        + {\hypL}_{ij} \overset{\leftarrow}{\widehat{\slashed{\partial}}}
        - {\hypRL}_{ij} \Big( \overset{\leftarrow}{\widehat{\slashed{\partial}}} 
                            + \overset{\rightarrow}{\widehat{\slashed{\partial}}} \Big)
        \bigg)
        \projL
        \bigg]
        \psi_j
        \Bigg\}\\
        &\phantom{= } + i g^2 \Dintx \, c \,
        \Bigg\{
        \overline{\psi}_i 
        \bigg[
        \Big(
        {\hypRL}_{ik}{\hypL}_{kj} - {\hypR}_{ik}{\hypRL}_{kj}
        \Big)
        \projL \widehat{\slashed{B}} \projL\\
        &\hspace{3.2cm} +
        \Big(
        {\hypLR}_{ik}{\hypR}_{kj} - {\hypL}_{ik}{\hypLR}_{kj}
        \Big)
        \projR \widehat{\slashed{B}} \projR
        \bigg]
        \psi_j
        \Bigg\}\\
        &\eqqcolon \widehat{\Delta}_1\big[c,\overline{\psi},\psi\big] + \widehat{\Delta}_2\big[c,B,\overline{\psi},\psi\big],
    \end{aligned}
\end{equation}
with $\STIopD$ being the $D$-dimensional Slavnov-Taylor operator.
As usual in the BMHV framework, the evanescent part of
the fermion kinetic term gives rise to the first contribution of the spurious 
BRST breaking, $\widehat{\Delta}_1$, in Eq.\ (\ref{Eq:TreeLevelBreaking}),
cf.\ \cite{Belusca-Maito:2020ala,Belusca-Maito:2021lnk,Belusca-Maito:2023wah,Stockinger:2023ndm}.
However, compared to the literature, there are two additional kinds
of breaking terms resulting from the evanescent gauge boson
interactions in the last line of Eq.\ (\ref{Eq:Lfermion}). First,
$\widehat{\Delta}_1$ is modified by terms involving the
chirality-violating hypercharge matrices $\hypLR$ and $\hypRL$. 
Second, the entire quantity $\widehat{\Delta}_2$,
which involves the gauge boson, is new and represents an
additional tree-level breaking.

Although BRST invariance is broken in $D$ dimensions due to the regularisation, 
the Lagrangian is still required to be hermitian,
i.e.\ $\mathcal{L}^{\dagger}=\mathcal{L}$,
even in $D$ dimensions.

% -----------------------------------------------------------------------------

% +++++++++++++++++++++++++++++++++++++++++++++++++++++++++++++++++++++++++++++
\section{Special Cases}\label{Sec:SpecialCases}

Before discussing the renormalisation of our general model from 
Sec.\ \ref{Sec:D-DimLagrangian}, we discuss how it can be related to the 
existing literature, how it generalises other approaches, and how it can be specialised to concrete models of interest.

Essentially, the fermion multiplets $\psi_L$ and $\psi_R$ 
of the general model can be 
populated with physical and sterile fields.
Further, a number of $N_S$ scalar fields $\phi_a$ may be included, 
though their presence is not mandatory.
The specific models that can then be obtained 
depend on the particular field content,
the detailed arrangement of these fields along
with any associated sterile fields (if necessary)
within the fermion multiplets, and the choice of symmetries and 
conservation laws that affect the coupling constants.
An overview of considered models is provided in Tab.\ \ref{Tab:SpecialCasesOverview}
together with details regarding the coupling constants.
\begin{table}[h!]
    \centering
    \begin{tabular}{|c||c|c|c|c|c|c|c|c|} \hline 
        Theory & $\hypR$ & $\hypL$ & $\hypLR$ & $\hypRL$ & $\hypS$ & $\lambda$ & $G$ & $K$\\ \hline \hline
        QED & $Q$ & $Q$ & $Q$ & $Q$ & $0$ & $0$ & $0$ & $0$\\ \hline
        $\chi$QED \cite{Belusca-Maito:2021lnk,Belusca-Maito:2023wah,Stockinger:2023ndm} & $\hypR$ & $0$ & $0$ & $0$ & $0$ & $0$ & $0$ & $0$\\ \hline
        Martin et al. \cite{Martin:1999cc} & (\ref{Eq:CPMartin-YLYR}) & (\ref{Eq:CPMartin-YLYR}) & $0$ & $0$ & $0$ & $0$ & $0$ & $0$\\ \hline
        Cornella et al. \cite{Cornella:2022hkc} & $T_R^a$ & $T_L^a$ & $0$ & $0$ & $0$ & $0$ & $0$ & $0$\\ \hline
        ASM & (\ref{Eq:ASSM-YLYR}) & (\ref{Eq:ASSM-YLYR}) & (\ref{Eq:ASSM-YLR}) & (\ref{Eq:ASSM-YLR}) & (\ref{Eq:ASSM-YS}) & (\ref{Eq:ASSM-lambda}) & (\ref{Eq:ASSM-Yuk-G}) & (\ref{Eq:ASSM-Yuk-K})\\ \hline
        A2HDM & (\ref{Eq:ASTHDM-YLYR}) & (\ref{Eq:ASTHDM-YLYR}) & (\ref{Eq:ASTHDM-YLR}) & (\ref{Eq:ASTHDM-YLR}) & (\ref{Eq:ASTHDM-YS}) & (\ref{Eq:ASTHDM-lambda}) & (\ref{Eq:ASTHDM-Yuk-G}) & (\ref{Eq:ASTHDM-Yuk-K})\\ \hline
    \end{tabular}
    \caption{Special cases of the abelian chiral gauge theory defined in Sec.\ \ref{Sec:D-DimLagrangian}.}
    \label{Tab:SpecialCasesOverview}
\end{table}

In Sec.\ \ref{Sec:ModelsFromLiterature}
we comment on models which have already been analysed in the literature,
while in sections \ref{Sec:ASSM} and \ref{Sec:ASTHDM} 
we explore two models 
which are of particular interest due to their explicit phenomenological applications.
For this reason each of the latter two models is given its own subsection.
The first is the abelian sector of the Standard Model (ASM), described in Sec.\ \ref{Sec:ASSM} 
(cf.\ row five of Tab.\ \ref{Tab:SpecialCasesOverview}), and the second is the 
abelian sector of the two-Higgs doublet model (A2HDM), covered in 
Sec.\ \ref{Sec:ASTHDM} (cf.\ row six of Tab.\ \ref{Tab:SpecialCasesOverview}).

\subsection{Selected Models from the Literature}\label{Sec:ModelsFromLiterature}

Starting with the first row of Tab.\ \ref{Tab:SpecialCasesOverview}, it is possible
to recover an abelian vector-like gauge theory such as QED from the considered theory described 
in Sec.\ \ref{Sec:D-DimLagrangian}.
Due to the introduction of the evanescent interaction currents in
the last line of Eq.\ (\ref{Eq:Lfermion}),
the ordinary DReg treatment of QED given in Eq.\ (\ref{Eq:LeegammaQED})
can be obtained by setting
$\hypL=\hypR=\hypLR=\hypRL=Q$.
The possibility to reproduce standard QED is an advantage 
of allowing the evanescent gauge interactions 
governed by $\hypLR$, $\hypRL$.
In addition, to reproduce QED with our setup,
all Yukawa and scalar 
couplings are set to zero, as illustrated in Tab.\ \ref{Tab:SpecialCasesOverview}.

We can also recover the purely right-handed models discussed 
in our previous publications~\cite{Belusca-Maito:2021lnk,Belusca-Maito:2023wah,Stockinger:2023ndm}, which we
refer to as ``chiral QED'' (cf.\ row two of Tab.\ \ref{Tab:SpecialCasesOverview}).
This is achieved if only the right-handed multiplet $\psi_R$ contains physical fermions,
while the left-handed multiplet $\psi_L$ is filled solely with sterile ones.
In the language of Sec.\ \ref{Sec:FermionsInGeneral} all fermions are
treated according to Option \ref{Opt:Option2b}.
The purely right-handed models of the literature do not allow evanescent gauge 
interactions, hence in this case $\hypRL=\hypLR=0$.
Additionally, all couplings involving scalars must be set to zero
for this configuration.

Furthermore, we can also reproduce two models discussed in the 
literature in Refs.\ \cite{Martin:1999cc} and \cite{Cornella:2022hkc}.
The authors in both publications performed a 1-loop
renormalisation of a chiral gauge theory with a compact and simple
(not necessarily abelian) gauge group, but without scalar fields, 
within the BMHV scheme.
Our statement that the abelian theory defined in Sec.\ \ref{Sec:D-DimLagrangian}
reproduces these models means that it replicates the abelian
special case of these models in which the considered gauge group is simply $U(1)$.

First, let us consider Ref.\ \cite{Martin:1999cc}.
Essentially, they have chosen to work in the spirit of Option \ref{Opt:Option2},
described above in Sec.\ \ref{Sec:FermionsInGeneral}. 
They introduce fermion multiplets $\psi$ and $\psi'$, where 
only the left-handed $\psi_L$ and the right-handed $\psi'_R$
represent physical fermions and participate in interactions,
while the right-handed $\psi_R$ and the left-handed $\psi'_L$
are sterile. In our framework, this can be realised via a single 
left- and right-handed fermion multiplet that admits 
a block structure, such that, for example, the first half of $\psi_L$ is physical 
and the second half is sterile, and vice-versa for $\psi_R$.
Consequently, the coupling matrices also admit an analogous block structure, such that
\begin{equation}\label{Eq:CPMartin-YLYR}
    \begin{aligned}
            \hypR =
            \begin{pmatrix}
                0 & 0\\
                0 & T_R^a
            \end{pmatrix},
            \quad
            \hypL =
            \begin{pmatrix}
                T_L^a & 0\\
                0 & 0
            \end{pmatrix}.
    \end{aligned}
\end{equation}
Ref.\ \cite{Martin:1999cc} does not consider evanescent gauge interactions, 
and therefore we must choose $\hypRL=\hypLR=0$ in this case.
Similarly, all couplings involving scalars must be set 
to zero (cf.\ row three of Tab.\ \ref{Tab:SpecialCasesOverview}), as the authors of 
Ref.\ \cite{Martin:1999cc} do not consider scalar fields.

Second, the authors of Ref.\ \cite{Cornella:2022hkc}
work in the spirit of Option \ref{Opt:Option1}. 
Where possible, they combine physical left-handed and right-handed fermions
to one Dirac spinor. 
Hence, unlike the previous case, it is not necessary to employ a block structure to reproduce the abelian special
case of their model with our theory defined in Sec.\ \ref{Sec:D-DimLagrangian}.
Still, they do not allow evanescent gauge interactions either; 
hence, we again must choose $\hypRL=\hypLR=0$ in this case.
Also, they do not consider scalar fields, meaning that all couplings involving scalars must be set to zero.
An overview of all details regarding this model can be found in row four of Tab.\ \ref{Tab:SpecialCasesOverview}.

Recently, Ref.\ \cite{OlgosoRuiz:2024dzq} generalised the model 
considered in Ref.\ \cite{Cornella:2022hkc} to include scalar fields.
Further, this reference also considered the option of evanescent gauge
interactions. However, only a specific form has been considered, which corresponds, in our notation, to the matrix relations
\begin{equation}
    \begin{aligned}
        \hypLR = \frac{c \hypR + c^{*} \hypL}{2}, \qquad \hypRL = \frac{c \hypL + c^{*} \hypR}{2},
    \end{aligned}
\end{equation}
with a constant c, and in their explicit 1-loop calculations the authors set $c=0$.

In conclusion, to the best of our knowledge, the influence of such evanescent interaction terms 
on the counterterm structure in the BMHV scheme for chiral gauge theories
has not been discussed so far.
In the following, we discuss how our setup can be specialised to two concrete models of interest,
while allowing for the existence of evanescent interactions.

\subsection{Abelian Sector of the Standard Model}\label{Sec:ASSM}

As announced, the abelian sector of the SM can be derived as a special
case of the theory defined in Sec.\ \ref{Sec:D-DimLagrangian}. 
For simplicity, we focus on the SM with only one generation of fermions.
In the following, we discuss this model for both
Options \ref{Opt:Option1} and \ref{Opt:Option2}, introduced in
Sec.\ \ref{Sec:FermionsInGeneral}, regarding the treatment of fermions in $D$ dimensions.

\paragraph{ASM with fermion multiplets according to Option \ref{Opt:Option1}:}
The explicit representation of the fermion multiplets in this case is given by
\begin{equation}\label{Eq:ASSM-FermionMultiplets}
    \begin{aligned}
        \psi_L &= 
        \begin{pmatrix}
            \nu_L, & e_L, & u^r_L, & u^g_L, & u^b_L, & d^r_L, & d^g_L, & d^b_L
        \end{pmatrix}^{\top},\\
        \psi_R &= 
        \begin{pmatrix}
            \nu_R^{\text{st}}, & e_R, & u^r_R, & u^g_R, & u^b_R, & d^r_R, & d^g_R, & d^b_R
        \end{pmatrix}^{\top},
    \end{aligned}
\end{equation}
where we
introduce only a sterile right-handed neutrino and combine ${\psi_L}_i$ with ${\psi_R}_i$
into one Dirac spinor $\psi_i$ for each fermion.
This means, for example, that there is only a single Dirac spinor 
$e$ for the electron, which contains both $e_L$ and $e_R$, and similarly for the other fermions.
Given this choice of fermionic matter content, the hypercharge matrices are provided by
\begin{equation}\label{Eq:ASSM-YLYR}
    \begin{aligned}
        \hypR =
        \begin{pmatrix}
            0 & 0 & 0 & 0\\
            0 & -1 & 0 & 0\\
            0 & 0 & \frac{2}{3} \mathbb{1}_{3\times3} & 0\\
            0 & 0 & 0 & -\frac{1}{3} \mathbb{1}_{3\times3}
        \end{pmatrix},
        \quad
        \hypL =
        \begin{pmatrix}
            -\frac{1}{2} & 0 & 0 & 0\\
            0 & -\frac{1}{2} & 0 & 0\\
            0 & 0 & \frac{1}{6} \mathbb{1}_{3\times3} & 0\\
            0 & 0 & 0 & \frac{1}{6} \mathbb{1}_{3\times3}
        \end{pmatrix},
    \end{aligned}
\end{equation}
containing the well-known hypercharges of the SM fermions and
obviously satisfying the anomaly cancellation condition 
in Eq.\ (\ref{Eq:AnomalyCancellationCondition}). 

Considering only a single generation of fermions
with multiplets given in Eq.\ (\ref{Eq:ASSM-FermionMultiplets}),
the evanescent hypercharges $\hypLR$ and $\hypRL$
must be diagonal, as
required by electric and colour charge conservation.
Further, we choose the evanescent hypercharges to be real.
Together with the hermiticity relation, Eq.\ (\ref{Eq:Requirement-YLR-YRL}), it
follows that $\hypLR=\hypRL$, which can be written as
\begin{equation}\label{Eq:ASSM-YLR}
    \begin{aligned}
        \hypLR = \hypRL =
        \begin{pmatrix}
            0 & 0 & 0 & 0\\
            0 & \mathcal{Y}_{LR}^{e} & 0 & 0\\
            0 & 0 & \mathcal{Y}_{LR}^{u} \mathbb{1}_{3\times3} & 0\\
            0 & 0 & 0 & \mathcal{Y}_{LR}^{d} \mathbb{1}_{3\times3}
        \end{pmatrix}.
    \end{aligned}
\end{equation}
The explicit values of the diagonal elements have not yet been fixed. 
We have the freedom to assign their values as desired,
which amounts to fixing a specific choice
of $D$-dimensional extension of the theory.
For instance, they could simply be set to zero.
Note, however, that the first entry must be zero due to the fact that the right-handed
neutrino $\nu_{R}^{\text{st}}$ is sterile.
Further discussions on the implications of the choice of these
hypercharges are to be found in Sec.\ \ref{Sec:TheRoleOf-YLR-YRL}.

As mentioned in Sec.\ \ref{Sec:D-DimLagrangian}, we impose global
and vector-like $U(1)_{Q}$ and $SU(3)_c$ symmetries. 
To replicate the SM with the given matter content, 
provided in Eq.\ (\ref{Eq:ASSM-FermionMultiplets}), we find
\begin{equation}\label{Eq:ASSM-Q}
    \begin{aligned}
        Q =
        \begin{pmatrix}
            0 & 0 & 0 & 0\\
            0 & -1 & 0 & 0\\
            0 & 0 & \frac{2}{3} \mathbb{1}_{3\times3} & 0\\
            0 & 0 & 0 & -\frac{1}{3} \mathbb{1}_{3\times3}
        \end{pmatrix},
    \end{aligned}
\end{equation}
for the electric charge, and 
\begin{equation}\label{Eq:ASSM-Colour}
    \begin{aligned}
        T^{a}_c =
        \begin{pmatrix}
            0 & 0 & 0 & 0\\
            0 & 0 & 0 & 0\\
            0 & 0 & \frac{\lambda^{a}}{2} & 0\\
            0 & 0 & 0 & \frac{\lambda^{a}}{2}
        \end{pmatrix},
    \end{aligned}
\end{equation}
for the colour charge generators, with $\lambda^a$ being the Gell-Mann matrices.
Indeed, with these quantum numbers we can identify $\nu$, $e$, $u$ and $d$
with the physical neutrino, electron, up-quark and down-quark, respectively.

Furthermore, we introduce $N_S=2$ scalar fields $\{\phi_a\}_{a=1}^{2}$ that both have the same hypercharge
\begin{equation}\label{Eq:ASSM-YS}
    \begin{aligned}
        \hypS = \frac{1}{2}.
    \end{aligned}
\end{equation}
However, $\phi_1$ is electrically charged with $Q_{\phi_1}=1$, 
while $\phi_2$ is electrically neutral with $Q_{\phi_2}=0$. 
In order to reproduce the SM, we impose
\begin{equation}\label{Eq:ASSM-lambda}
    \begin{aligned}
        \lambda_{1111} &= \lambda_{2222} = 6 \lambda_{SM},\\
%        \quad
        \lambda_{1122} &= \lambda_{1221} = \lambda_{2112} = \lambda_{2211} = 3 \lambda_{SM},\\
%        \quad 
        \lambda_{klmn} &\equiv 0, \quad \text{else},
    \end{aligned}
\end{equation}
on the scalar self-interaction coupling,
such that we can write
\begin{equation}\label{Eq:ASSM-HiggsPotential}
    \begin{aligned}
        V(\Phi) = \lambda_{SM} \big(\Phi^{\dagger}\Phi\big)^2,
    \end{aligned}
\end{equation}
where we combined both scalars into one doublet
\begin{equation}
    \begin{aligned}
        \Phi =
        \begin{pmatrix}
            \phi_1\\
            \phi_2
        \end{pmatrix}.
    \end{aligned}
\end{equation}
Now, we can identify the SM scalar fields $\phi_1=G^{+}$ and 
$\phi_2=(\varphi+iG^{0})/\sqrt{2}$.
Note that we consider only completely massless theories in this work,
which is the reason why the scalar mass term $\mu^2|\Phi|^2$ is missing
in the Higgs potential in Eq.\ (\ref{Eq:ASSM-HiggsPotential}).
We note that neglecting this term affects neither the renormalisation of the dimension-$4$ operators nor the spurious breaking of BRST invariance induced by the regularisation scheme, nor the running of couplings.

Finally, the coupling matrices for the Yukawa interactions in the 
SM with one generation of fermions, as given in Eq.\ (\ref{Eq:ASSM-FermionMultiplets}), 
are provided by
\begin{equation}\label{Eq:ASSM-Yuk-G}
    \begin{aligned}
        G^1 =
        \begin{pmatrix}
            0 & y_e & 0 & 0\\
            0 & 0 & 0 & 0\\
            0 & 0 & 0 & y_d \mathbb{1}_{3\times3}\\
            0 & 0 & 0 & 0
        \end{pmatrix},
        \quad
        G^2 =
        \begin{pmatrix}
            0 & 0 & 0 & 0\\
            0 & y_e & 0 & 0\\
            0 & 0 & 0 & 0\\
            0 & 0 & 0 & y_d \mathbb{1}_{3\times3}
        \end{pmatrix},
    \end{aligned}
\end{equation}
and
\begin{equation}\label{Eq:ASSM-Yuk-K}
    \begin{aligned}
        K^1 =
        \begin{pmatrix}
            0 & 0 & 0 & 0\\
            0 & 0 & 0 & 0\\
            0 & 0 & 0 & 0\\
            0 & 0 & -y_u \mathbb{1}_{3\times3} & 0
        \end{pmatrix},
        \quad
        K^2 =
        \begin{pmatrix}
            0 & 0 & 0 & 0\\
            0 & 0 & 0 & 0\\
            0 & 0 & y_u \mathbb{1}_{3\times3} & 0\\
            0 & 0 & 0 & 0
        \end{pmatrix},
    \end{aligned}
\end{equation}
with $y_e, \, y_u, \, y_d \in \mathbb{R}$. 

With this choice for the matter content and the coupling constants,
our theory automatically exhibits a global $SU(2)_L$ symmetry, which is generally
absent in the theory defined by Eq.\ (\ref{Eq:model-Lagrangian}) in Sec.\ \ref{Sec:D-DimLagrangian}.
This symmetry is essential to replicate the abelian sector of the SM.
However, unlike the full SM and analogous to the $SU(3)_c$ symmetry
discussed above, this $SU(2)_L$ symmetry is solely a global symmetry of
our theory.
Since we do not admit non-abelian gauge groups in this work, 
the gauge bosons of the weak and the strong interactions are excluded.

\paragraph{ASM with fermion multiplets according to Option \ref{Opt:Option2}:}
Here, we introduce a sterile partner for each chiral fermion.
In this case, the multiplets are given by
\begin{equation}\label{Eq:ASSM-FermionMultiplets-Option2}
    \begin{aligned}
        \psi_L &= 
        \begin{pmatrix}
            \nu_L, & e_L, & u_L, & d_L,
          & \nu_L^{\text{st}}, & e_L^{\text{st}}, & u^{\text{st}}_L, & d^{\text{st}}_L
        \end{pmatrix}^{\top},\\
        \psi_R &= 
        \begin{pmatrix}
            \nu_R^{\text{st}}, & e_R^{\text{st}}, & u^{\text{st}}_R, & d^{\text{st}}_R,
          & \nu_R^{\text{st}}, & e_R, & u_R, & d_R
        \end{pmatrix}^{\top},
    \end{aligned}
\end{equation}
where up- and down-type quarks, $u$ and $d$, respectively, are included with all 
three colours $\{r,g,b\}$, analogous to Eq.\ (\ref{Eq:ASSM-FermionMultiplets}).
In the spirit of Eq.\ (\ref{Eq:Electron-2-Spinors-Option2}), we can define two Dirac spinor
multiplets
\begin{equation}\label{Eq:Fermions-psi1-psi2-Option2}
    \begin{aligned}
        \psi_1 &=
        \begin{pmatrix}
            \nu_L + \nu_R^{\text{st}}, & 
            e_L + e_R^{\text{st}}, & 
            u_L + u^{\text{st}}_R, & 
            d_L + d^{\text{st}}_R
        \end{pmatrix}^{\top},\\
        \psi_2 &=
        \begin{pmatrix}
            \nu_L^{\text{st}} + \nu_R^{\text{st}}, & 
            e_L^{\text{st}} + e_R, & 
            u^{\text{st}}_L + u_R, & 
            d^{\text{st}}_L + d_R
        \end{pmatrix}^{\top},
    \end{aligned}
\end{equation}
such that $\psi_1$ contains all SM left-handed physical fermion fields with suitable sterile partners, and $\psi_2$ contains analogously the SM right-handed physical fermion fields.\footnote{Note that $\psi_2$  contains an entirely sterile neutrino Dirac fermion, which has the advantage of producing a more systematic block structure of interaction coupling matrices.}
These can further be combined to a full multiplet of Dirac fermions
\begin{equation}\label{decomposition}
    \begin{aligned}
        \psi=\begin{pmatrix}\psi_1, \,\psi_2\end{pmatrix}^{\top},
    \end{aligned}
\end{equation}
which, upon applying $\mathbb{P}_{\mathrm{L/R}}$, 
reproduces Eq.\ (\ref{Eq:ASSM-FermionMultiplets-Option2}).

With this enlarged fermionic matter content,
the SM coupling matrices adopt a block structure associated with the decomposition (\ref{decomposition}).
Moreover, because sterile fields by definition 
do not participate in any interactions, cf.\ Eq.\ (\ref{Eq:NonRenOfSterileFields}),  blocks corresponding to sterile fields vanish. In contrast blocks corresponding to physical fields can be taken over from the matrices appearing in the previous case of Option \ref{Opt:Option1}.

For the $4$-dimensional $16 \times 16$ hypercharge matrices, we now find
\begin{equation}\label{Eq:ASSM-YLYR-Option2}
    \begin{aligned}
        \hypR =
        \begin{pmatrix}
            0 & 0\\
            0 & \hypR^{\text{ASM,opt\ref{Opt:Option1}}}
        \end{pmatrix},
        \quad
        \hypL =
        \begin{pmatrix}
            \hypL^{\text{ASM,opt\ref{Opt:Option1}}} & 0\\
            0 & 0 
        \end{pmatrix},
    \end{aligned}
\end{equation}
with $\hypL^{\text{ASM,opt\ref{Opt:Option1}}}$ and $\hypR^{\text{ASM,opt\ref{Opt:Option1}}}$
as defined in Eq.\ (\ref{Eq:ASSM-YLYR}). 
Although we again choose to assign identical
real, evanescent hypercharges coupling to both the $LR$- and 
the $RL$-interaction currents 
for each physical fermion, 
the resulting hypercharge matrices exhibit a block off-diagonal
structure
\begin{equation}\label{Eq:ASSM-YLR-YRL-Option2}
    \begin{aligned}
        \hypLR =
        \begin{pmatrix}
            0 & \hypLR^{\text{ASM,opt\ref{Opt:Option1}}}\\
            0 & 0
        \end{pmatrix},
        \quad
        \hypRL =
        \begin{pmatrix}
            0 & 0\\
            \hypRL^{\text{ASM,opt\ref{Opt:Option1}}} & 0 
        \end{pmatrix},
    \end{aligned}
\end{equation}
with $\hypLR^{\text{ASM,opt\ref{Opt:Option1}}}=\hypRL^{\text{ASM,opt\ref{Opt:Option1}}}$,
as specified in Eq.\ (\ref{Eq:ASSM-YLR}).
Hence, due to the block structure of the fermion multiplets in 
Eq.\ (\ref{Eq:ASSM-FermionMultiplets-Option2}), we find $\hypRL\neq\hypLR$
for the evanescent $16\times16$ hypercharge matrices in Option \ref{Opt:Option2},
while still satisfying the condition in Eq.\ (\ref{Eq:Requirement-YLR-YRL}).
However, working with non-vanishing evanescent hypercharges $\hypLR$ and $\hypRL$
conflicts with the actual goal of Option \ref{Opt:Option2} to preserve at 
least global hypercharge conservation.
The reason for this that these evanescent gauge
interactions do not preserve chirality and, as a result, may generally 
cause a violation of not only local but even global hypercharge conservation, 
as mentioned in Sec.\ \ref{Sec:D-DimLagrangian}.
This is easily illustrated by the example of the electron, where we obtain
\begin{equation}\label{Eq:ChiralityViolation-YLR-YRL-Electron}
    \begin{aligned}
        - g \hypRL^e \overline{e_2} \projL \widehat{\slashed{B}} \projL e_1
        - g \hypLR^e \overline{e_1} \projR \widehat{\slashed{B}} \projR e_2
        =
        - g \hypRL^e \overline{e_R} \widehat{\slashed{B}} e_L
        - g \hypLR^e \overline{e_L} \widehat{\slashed{B}} e_R
    \end{aligned}
\end{equation}
for the evanescent gauge interactions,\footnote{
As discussed, we work with real evanescent hypercharges; 
thus, $\hypRL^e=(\hypLR^e)^{*}=\hypLR^e\in\mathbb{R}$.} 
which clearly break global hypercharge
conservation due to the differing hypercharges of $-1/2$ for $e_L$ and $-1$ for $e_R$.
Nonetheless, for the sake of generality, we allow for non-vanishing $\hypLR$ and $\hypRL$,
as specified in Eq.\ (\ref{Eq:ASSM-YLR-YRL-Option2}), and discuss its implications in Sec.\ \ref{Sec:AnalysisOfASSMResults}.

Likewise, the Yukawa matrices display a similar block structure, 
and are given by
\begin{equation}\label{Eq:ASSM-Yuk-G-Option2}
    \begin{aligned}
        G^1 =
        \begin{pmatrix}
            0 & G^1_{\text{ASM,opt\ref{Opt:Option1}}}\\
            0 & 0
        \end{pmatrix},
        \quad
        G^2 =
        \begin{pmatrix}
            0 & G^2_{\text{ASM,opt\ref{Opt:Option1}}}\\
            0 & 0 
        \end{pmatrix},
    \end{aligned}
\end{equation}
and
\begin{equation}\label{Eq:ASSM-Yuk-K-Option2}
    \begin{aligned}
        K^1 =
        \begin{pmatrix}
            0 & K^1_{\text{ASM,opt\ref{Opt:Option1}}}\\
            0 & 0
        \end{pmatrix},
        \quad
        K^2 =
        \begin{pmatrix}
            0 & K^2_{\text{ASM,opt\ref{Opt:Option1}}}\\
            0 & 0 
        \end{pmatrix},
    \end{aligned}
\end{equation}
with $G^a_{\text{ASM,opt\ref{Opt:Option1}}}$ and $K^a_{\text{ASM,opt\ref{Opt:Option1}}}$
as defined in Eqs.\ (\ref{Eq:ASSM-Yuk-G}) and (\ref{Eq:ASSM-Yuk-K}), respectively.

The remainder of the setup, particularly the scalar sector, 
remains unaffected by the choice of the $D$-dimensional treatment
of the fermions, and is thus consistent with the previously discussed case under 
Option \ref{Opt:Option1}.

A discussion of the results for the renormalisation procedure at the
1-loop level in this model, considering both options for the treatment of
fermions in $D$ dimensions, is presented in Sec.\ \ref{Sec:AnalysisOfASSMResults}.

\subsection{Abelian Sector of the Two-Higgs Doublet Model}\label{Sec:ASTHDM}

The abelian sector of the 2HDM can also be reproduced 
as a special case of the theory defined in Sec.\ \ref{Sec:D-DimLagrangian}.
The approach to obtaining the A2HDM from Eq.\ (\ref{Eq:model-Lagrangian}) 
is analogous to the ASM case from Sec.\ \ref{Sec:ASSM}.
In contrast to the ASM, we consider the A2HDM with all three generations,
but restrict the discussion to Option \ref{Opt:Option1} for the treatment of fermions
in $D$ dimensions. 
The transition to Option \ref{Opt:Option2} is straightforward.
With an appropriate choice for the fermion multiplets within
the framework of Option \ref{Opt:Option1}, we obtain
the following physical hypercharges:
\begin{equation}\label{Eq:ASTHDM-YLYR}
    \begin{aligned}
        \hypR =
        \begin{pmatrix}
            0 & 0 & 0 & 0\\
            0 & - \mathbb{1}_{3\times3} & 0 & 0\\
            0 & 0 & \frac{2}{3} \mathbb{1}_{9\times9} & 0\\
            0 & 0 & 0 & -\frac{1}{3} \mathbb{1}_{9\times9}
        \end{pmatrix},
        \quad
        \hypL =
        \begin{pmatrix}
            -\frac{1}{2} \mathbb{1}_{3\times3} & 0 & 0 & 0\\
            0 & -\frac{1}{2} \mathbb{1}_{3\times3} & 0 & 0\\
            0 & 0 & \frac{1}{6} \mathbb{1}_{9\times9} & 0\\
            0 & 0 & 0 & \frac{1}{6} \mathbb{1}_{9\times9}
        \end{pmatrix}.
    \end{aligned}
\end{equation}
For the evanescent hypercharges, we choose
\begin{equation}\label{Eq:ASTHDM-YLR}
    \begin{aligned}
        \hypLR = \hypRL =
        \begin{pmatrix}
            0 & 0 & 0 & 0\\
            0 & \mathcal{Y}_{LR}^{e} \mathbb{1}_{3\times3} & 0 & 0\\
            0 & 0 & \mathcal{Y}_{LR}^{u} \mathbb{1}_{9\times9} & 0\\
            0 & 0 & 0 & \mathcal{Y}_{LR}^{d} \mathbb{1}_{9\times9}
        \end{pmatrix},
    \end{aligned}
\end{equation}
analogous to Eq.\ (\ref{Eq:ASSM-YLR}). 
Note that in the case of three generations, the evanescent 
hypercharges, in principle, may also admit off-diagonal elements,
which would mix the generations of leptons or generations of quarks, 
while still preserving electric and colour charge conservation.
The electric and colour charge are also given similarly
to the ASM case, consistent with global $U(1)_{Q}$ and $SU(3)_c$ symmetry,
respectively, and aligned with the matter content of the 2HDM. 

For the scalar sector, we introduce $N_S=4$ scalar fields 
$\{\phi_a\}_{a=1}^4$ having identical hypercharge
\begin{equation}\label{Eq:ASTHDM-YS}
    \begin{aligned}
        \hypS = \frac{1}{2}.
    \end{aligned}
\end{equation}
The scalars $\phi_1$ and $\phi_3$ are electrically charged,
while $\phi_2$ and $\phi_4$ are electrically neutral.
In order to replicate the scalar sector of the 2HDM,
we restrict the couplings 
$\lambda_{klmn}$ such that the scalar self-interaction
term reproduces the 2HDM Higgs potential
\begin{equation}\label{Eq:ASTHDM-lambda}
    \begin{aligned}
        V(\Phi_1,\Phi_2) 
        &= \frac{\lambda_1}{2} \big(\Phi_1^{\dagger}\Phi_1\big)^2
        + \frac{\lambda_2}{2} \big(\Phi_2^{\dagger}\Phi_2\big)^2
        + \lambda_3 \big(\Phi_1^{\dagger}\Phi_1\big)\big(\Phi_2^{\dagger}\Phi_2\big)
        + \lambda_4 \big(\Phi_1^{\dagger}\Phi_2\big)\big(\Phi_2^{\dagger}\Phi_1\big)\\
        &+ \bigg[
        \frac{\lambda_5}{2} \big(\Phi_1^{\dagger}\Phi_2\big)^2
        + \lambda_6 \big(\Phi_1^{\dagger}\Phi_1\big)\big(\Phi_1^{\dagger}\Phi_2\big)
        + \lambda_7 \big(\Phi_2^{\dagger}\Phi_2\big)\big(\Phi_1^{\dagger}\Phi_2\big)
        + \mathrm{h.c.}
        \bigg],
    \end{aligned}
\end{equation}
with the four scalars combined into two Higgs doublets 
\begin{equation}
    \begin{aligned}
        \Phi_1 =
        \begin{pmatrix}
            \phi_1\\
            \phi_2
        \end{pmatrix},
        \quad
        \Phi_2 =
        \begin{pmatrix}
            \phi_3\\
            \phi_4
        \end{pmatrix}.
    \end{aligned}
\end{equation}
Again we neglected the scalar mass terms, as
explained above in Sec.\ \ref{Sec:ASSM}.

In contrast to the ASM, the 2HDM features two types of Yukawa interactions,
each associated with one of the two scalar doublets, $\Phi_1$ and $\Phi_2$.
The Yukawa couplings to $\Phi_1$ are
represented by $G^1$, $G^2$, $K^1$ and $K^2$, while the couplings to 
$\Phi_2$ are denoted by $G^3$, $G^4$, $K^3$ and $K^4$.
More explicitly, these couplings are given by
\begin{equation}\label{Eq:ASTHDM-Yuk-G}
    \begin{aligned}
        G^1 &=
        \begin{pmatrix}
            0 & Y_e & 0 & 0\\
            0 & 0 & 0 & 0\\
            0 & 0 & 0 & Y_d\\
            0 & 0 & 0 & 0
        \end{pmatrix},
        \quad
        G^2 =
        \begin{pmatrix}
            0 & 0 & 0 & 0\\
            0 & Y_e & 0 & 0\\
            0 & 0 & 0 & 0\\
            0 & 0 & 0 & Y_d
        \end{pmatrix},\\
        G^3 &=
        \begin{pmatrix}
            0 & Z_e & 0 & 0\\
            0 & 0 & 0 & 0\\
            0 & 0 & 0 & Z_d\\
            0 & 0 & 0 & 0
        \end{pmatrix},
        \quad
        G^4 =
        \begin{pmatrix}
            0 & 0 & 0 & 0\\
            0 & Z_e & 0 & 0\\
            0 & 0 & 0 & 0\\
            0 & 0 & 0 & Z_d
        \end{pmatrix},
    \end{aligned}
\end{equation}
\begin{equation}\label{Eq:ASTHDM-Yuk-K}
    \begin{aligned}
        K^1 &=
        \begin{pmatrix}
            0 & 0 & 0 & 0\\
            0 & 0 & 0 & 0\\
            0 & 0 & 0 & 0\\
            0 & 0 & -Y_u & 0
        \end{pmatrix},
        \quad
        K^2 =
        \begin{pmatrix}
            0 & 0 & 0 & 0\\
            0 & 0 & 0 & 0\\
            0 & 0 & Y_u & 0\\
            0 & 0 & 0 & 0
        \end{pmatrix},\\
        K^3 &=
        \begin{pmatrix}
            0 & 0 & 0 & 0\\
            0 & 0 & 0 & 0\\
            0 & 0 & 0 & 0\\
            0 & 0 & -Z_u & 0
        \end{pmatrix},
        \quad
        K^4 =
        \begin{pmatrix}
            0 & 0 & 0 & 0\\
            0 & 0 & 0 & 0\\
            0 & 0 & Z_u & 0\\
            0 & 0 & 0 & 0
        \end{pmatrix},
    \end{aligned}
\end{equation}
where $Y_e,\,Y_u,\,Y_d,\,Z_e,\,Z_u,\,Z_d$ are in general complex $3\times3$ matrices.

Clearly, this model, with the specified matter content and
coupling constants, also features a global $SU(2)_L$ symmetry,
similar to the case of the ASM in Sec.\ \ref{Sec:ASSM}.

% -----------------------------------------------------------------------------

% +++++++++++++++++++++++++++++++++++++++++++++++++++++++++++++++++++++++++++++
\section{Slavnov-Taylor Identities}\label{Sec:STI+QAP}

Our methodology for computing the complete list of counterterms, necessary for a full renormalisation of the theory, is described in detail in previous publications~\cite{Belusca-Maito:2020ala,Belusca-Maito:2021lnk,Stockinger:2023ndm,Kuhler:2024fak},
with a comprehensive review in Ref.\ \cite{Belusca-Maito:2023wah}.
In a nutshell, the theory is defined at tree-level in $D$ dimensions by the Lagrangian given in Sec.~\ref{Sec:D-DimLagrangian} and the corresponding classical action $S_0$. At higher orders, counterterms $S_{\mathrm{ct}}$ are added to the action, resulting in the effective action $\Gamma_\mathrm{DRen}$, whose 4-dimensional limit defines the renormalised theory. The counterterms are obtained by requiring the renormalised theory to be finite and to fulfil the ultimate symmetry requirement
\begin{equation}\label{Eq:UltimateSymmetryRequirement}
    \begin{aligned}
        \mathop{\text{LIM}}_{D \, \to \, 4} \, (\mathcal{S}_D(\Gamma_\mathrm{DRen})) = 0.
    \end{aligned}
\end{equation}
This ensures that the BRST symmetry is intact at the quantum level and ensures the validity of indispensable Ward identities. 

In practice, all singular counterterms, including the BRST-breaking contributions,
are obtained from the UV divergences of standard, power-counting divergent,
1PI Green functions. 
The symmetry-restoring counterterms, including
also the finite contributions, are obtained from power-counting divergent, 
single $\Delta$-operator inserted 1PI Green functions, utilising the quantum action
principle of DReg
\cite{Breitenlohner:1975hg, Breitenlohner:1976te, Breitenlohner:1977hr},
\begin{equation}\label{Eq:QAPofDReg}
    \begin{aligned}
        \mathcal{S}_D(\Gamma_\mathrm{DRen})=(\widehat{\Delta}+\Delta_{\mathrm{ct}})\cdot\Gamma_\mathrm{DRen}.
    \end{aligned}
\end{equation}
The LHS expresses the action of the Slavnov-Taylor operator on the dimensionally regularised effective 1PI quantum action computed including counterterms.

The RHS of Eq.\ (\ref{Eq:QAPofDReg}) is defined by
\begin{equation}\label{Eq:DefDeltaBreaking}
    \begin{aligned}
        \Delta = \widehat{\Delta} + \Delta_{\mathrm{ct}} = \mathcal{S}_D(S_{0} + S_{\mathrm{ct}}),
    \end{aligned}
\end{equation}
i.e.\ by the BRST variation of the $D$-dimensional tree-level action as well as the counterterm action up to a given order.
The lowest-order contribution is the evanescent $\DeltaHat$-operator, i.e.\ the tree-level breaking shown in Eq.\ (\ref{Eq:TreeLevelBreaking}). The one-loop contribution $\Delta_{\mathrm{ct}}$ results from the one-loop counterterms $S_{\mathrm{ct}}$ which are to be determined by the procedure.

Although the RHS of Eq.\ (\ref{Eq:QAPofDReg}) presents an immense practical simplification over its LHS, 
for the sake of double-checks we provide a full list of
relations following from Eq.\ (\ref{Eq:QAPofDReg}) which are relevant to renormalisation. 
They express the correlation of products of standard Green functions with $\Delta$-inserted ones during renormalisation, and provide consistency checks for divergent and finite parts.
In the following we let 
\begin{equation}
    i(\widehat{\Delta}\cdot\widetilde{\Gamma})_{\Phi_n(p_n)\dots \Phi_1(p_1)}
\end{equation} 
denote the $1$-loop-, dimensionally regularised, momentum-space expression of Green functions of fields $\Phi_1\dots \Phi_n$ with all momenta incoming and computed without counterterms. With this notation we work out Eq.~(\ref{Eq:QAPofDReg}) in terms of all relations corresponding to explicit power-counting divergent Green functions. In the following relations the appearing Green functions corresponding to BRST sources do not renormalise. Hence their tree-level expression from Eq.\ (\ref{Eq:Ddim-Lext}) is substituted. 
With this in mind, this list of relations is explicitly given as follows:

\begin{align}\label{Eq:STI-QAP-Bc}
    p^{\nu} \, \widetilde{\Gamma}_{B^{\mu}(-p)B^{\nu}(p)}
    &= i \big( \widehat{\Delta} \cdot \widetilde{\Gamma} \big)_{B^{\mu}(-p) c(p)} 
\end{align}
\begin{align}
\begin{split}\label{Eq:STI-QAP-BBc}
    q^{\rho} \, \widetilde{\Gamma}_{B^{\rho}(q)B^{\nu}(p_2)B^{\mu}(p_1)}
    &= i \big( \widehat{\Delta} \cdot \widetilde{\Gamma} \big)_{B^{\nu}(p_2)B^{\mu}(p_1) c(q)}
\end{split}\\[1.5ex]
\begin{split}\label{Eq:STI-QAP-FFc}
    q^{\mu} \, \widetilde{\Gamma}_{\psi_{j,\beta}(p_2)\overline{\psi}_{i,\alpha}(p_1)B^{\mu}(q)}
    &+ g \big({\hypL}{}_{,lj}{\projL}{}_{,\delta\beta}+{\hypR}{}_{,lj}{\projR}{}_{,\delta\beta}\big)
    \widetilde{\Gamma}_{\psi_{l,\delta}(-p_1)\overline{\psi}_{i,\alpha}(p_1)}\\
    &- g \big({\hypL}{}_{,il}{\projR}{}_{,\alpha\delta}+{\hypR}{}_{,il}{\projL}{}_{,\alpha\delta}\big)
    \widetilde{\Gamma}_{\psi_{j,\beta}(p_2)\overline{\psi}_{l,\delta}(-p_2)}\\
    &= i \big( \widehat{\Delta} \cdot \widetilde{\Gamma} \big)_{\psi_{j,\beta}(p_2)\overline{\psi}_{i,\alpha}(p_1)c(q)}
\end{split}\\[1.5ex]
\begin{split}\label{Eq:STI-QAP-SSbarc}
    q^{\mu} \, \widetilde{\Gamma}_{\phi_{b}(p_2)\phi^{\dagger}_{a}(p_1)B^{\mu}(q)}
    &+ g \hypS
    \widetilde{\Gamma}_{\phi_{b}(-p_1)\phi^{\dagger}_{a}(p_1)}
    - g \hypS
    \widetilde{\Gamma}_{\phi_b(p_2)\phi^{\dagger}_{a}(-p_2)}\\
    &= i \big( \widehat{\Delta} \cdot \widetilde{\Gamma} \big)_{\phi_{b}(p_2)\phi^{\dagger}_{a}(p_1)c(q)}
\end{split}\\[1.5ex]
\begin{split}\label{Eq:STI-QAP-SSc}
    q^{\mu} \, \widetilde{\Gamma}_{\phi_{b}(p_2)\phi_{a}(p_1)B^{\mu}(q)}
    &+ g \hypS
    \widetilde{\Gamma}_{\phi_{a}(p_1)\phi_{b}(-p_1)}
    + g \hypS
    \widetilde{\Gamma}_{\phi_b(p_2)\phi_{a}(-p_2)}\\
    &= i \big( \widehat{\Delta} \cdot \widetilde{\Gamma} \big)_{\phi_{b}(p_2)\phi_{a}(p_1)c(q)}
\end{split}\\[1.5ex]
\begin{split}\label{Eq:STI-QAP-SbarSbarc}
    q^{\mu} \, \widetilde{\Gamma}_{\phi^{\dagger}_{b}(p_2)\phi^{\dagger}_{a}(p_1)B^{\mu}(q)}
    &- g \hypS
    \widetilde{\Gamma}_{\phi^{\dagger}_{a}(p_1)\phi^{\dagger}_{b}(-p_1)}
    - g \hypS
    \widetilde{\Gamma}_{\phi^{\dagger}_b(p_2)\phi^{\dagger}_{a}(-p_2)}\\
    &= i \big( \widehat{\Delta} \cdot \widetilde{\Gamma} \big)_{\phi^{\dagger}_{b}(p_2)\phi^{\dagger}_{a}(p_1)c(q)}
\end{split}
\end{align}
\begin{align}
\begin{split}\label{Eq:STI-QAP-BBBc}
    q^{\sigma} \, \widetilde{\Gamma}_{B^{\sigma}(q)B^{\rho}(p_3)B^{\nu}(p_2)B^{\mu}(p_1)}
    &= i \big( \widehat{\Delta} \cdot \widetilde{\Gamma} \big)_{B^{\rho}(p_3)B^{\nu}(p_2)B^{\mu}(p_1) c(q)}
\end{split}\\[1.5ex]
\begin{split}\label{Eq:STI-QAP-FFbarBc}
 g \big({\hypL}{}_{,lj}{\projL}{}_{,\delta\beta}+{\hypR}{}_{,lj}{\projR}{}_{,\delta\beta}\big)
    &\widetilde{\Gamma}_{\psi_{l,\delta}(-p_1-p_2)\overline{\psi}_{i,\alpha}(p_2)B^{\mu}(p_1)}\\
    - g \big({\hypL}{}_{,il}{\projR}{}_{,\alpha\delta}+{\hypR}{}_{,il}{\projL}{}_{,\alpha\delta}\big)
    &\widetilde{\Gamma}_{\psi_{j,\beta}(p_3)\overline{\psi}_{l,\delta}(-p_1-p_3)B^{\mu}(p_1)}\\
    &= i \big( \widehat{\Delta} \cdot \widetilde{\Gamma} \big)_{\psi_{j,\beta}(p_3)\overline{\psi}_{i,\alpha}(p_2)B^{\mu}(p_1)c(q)}
\end{split}\\[1.5ex]
\begin{split}\label{Eq:STI-QAP-FFbarSc}
g\hypS &\widetilde{\Gamma}_{\psi_{j,\beta}(p_3)\overline{\psi}_{i,\alpha}(p_2)\phi_a(-p_2-p_3)}\\
    + g \big({\hypL}{}_{,lj}{\projL}{}_{,\delta\beta}+{\hypR}{}_{,lj}{\projR}{}_{,\delta\beta}\big)
    &\widetilde{\Gamma}_{\psi_{l,\delta}(-p_1-p_2)\overline{\psi}_{i,\alpha}(p_2)\phi_a(p_1)}\\
    -  g \big({\hypL}{}_{,il}{\projR}{}_{,\alpha\delta}+{\hypR}{}_{,il}{\projL}{}_{,\alpha\delta}\big)
    &\widetilde{\Gamma}_{\psi_{j,\beta}(p_3)\overline{\psi}_{l,\delta}(-p_1-p_3)\phi_a(p_1)}\\
    &= i \big( \widehat{\Delta} \cdot \widetilde{\Gamma} \big)_{\psi_{j,\beta}(p_3)\overline{\psi}_{i,\alpha}(p_2)\phi_a(p_1)c(q)}
\end{split}\\[1.5ex]
\begin{split}\label{Eq:STI-QAP-FFbarSbarc}
    -g\hypS &\widetilde{\Gamma}_{\psi_{j,\beta}(p_3)\overline{\psi}_{i,\alpha}(p_2)\phi^{\dagger}_a(-p_2-p_3)}\\
    + g \big({\hypL}{}_{,lj}{\projL}{}_{,\delta\beta}+{\hypR}{}_{,lj}{\projR}{}_{,\delta\beta}\big)
    &\widetilde{\Gamma}_{\psi_{l,\delta}(-p_1-p_2)\overline{\psi}_{i,\alpha}(p_2)\phi^{\dagger}_a(p_1)}\\
    - g \big({\hypL}{}_{,il}{\projR}{}_{,\alpha\delta}+{\hypR}{}_{,il}{\projL}{}_{,\alpha\delta}\big)
    &\widetilde{\Gamma}_{\psi_{j,\beta}(p_3)\overline{\psi}_{l,\delta}(-p_1-p_3)\phi^{\dagger}_a(p_1)}\\
    &= i \big( \widehat{\Delta} \cdot \widetilde{\Gamma} \big)_{\psi_{j,\beta}(p_3)\overline{\psi}_{i,\alpha}(p_2)\phi^{\dagger}_a(p_1)c(q)}
\end{split}\\[1.5ex]
\begin{split}\label{Eq:STI-QAP-SSbarBc}
    q^{\nu}\,\widetilde{\Gamma}_{\phi_b(p_3)\phi_{a}^{\dagger}(p_2)B^{\mu}(p_1)B^{\nu}(q)}
    &+ g \hypS 
    \widetilde{\Gamma}_{\phi_b(-p_1-p_2)\phi^{\dagger}_{a}(p_2)B^{\mu}(p_1)}\\
    &- g \hypS
    \widetilde{\Gamma}_{\phi_b(p_3)\phi^{\dagger}_a(-p_1-p_3)B^{\mu}(p_1)}\\
    &= i \big( \widehat{\Delta} \cdot \widetilde{\Gamma} \big)_{\phi_b(p_3)\phi^{\dagger}_a(p_2)B^{\mu}(p_1)c(q)} 
\end{split}\\[1.5ex]
\begin{split}\label{Eq:STI-QAP-SSBc}
    q^{\nu}\,\widetilde{\Gamma}_{\phi_b(p_3)\phi_{a}(p_2)B^{\mu}(p_1)B^{\nu}(q)}
    &+ g \hypS 
    \widetilde{\Gamma}_{\phi_b(-p_1-p_2)\phi_{a}(p_2)B^{\mu}(p_1)}\\
    &+ g \hypS
    \widetilde{\Gamma}_{\phi_b(p_3)\phi_a(-p_1-p_3)B^{\mu}(p_1)}\\
    &= i \big( \widehat{\Delta} \cdot \widetilde{\Gamma} \big)_{\phi_b(p_3)\phi_a(p_2)B^{\mu}(p_1)c(q)}
\end{split}\\[1.5ex]
\begin{split}\label{Eq:STI-QAP-SbarSbarBc}
    q^{\nu}\,\widetilde{\Gamma}_{\phi_b^{\dagger}(p_3)\phi_{a}^{\dagger}(p_2)B^{\mu}(p_1)B^{\nu}(q)}
    &- g \hypS 
    \widetilde{\Gamma}_{\phi_b^{\dagger}(-p_1-p_2)\phi^{\dagger}_{a}(p_2)B^{\mu}(p_1)}\\
    &- g \hypS
    \widetilde{\Gamma}_{\phi_b^{\dagger}(p_3)\phi^{\dagger}_a(-p_1-p_3)B^{\mu}(p_1)}\\
    &= i \big( \widehat{\Delta} \cdot \widetilde{\Gamma} \big)_{\phi_b^{\dagger}(p_3)\phi^{\dagger}_a(p_2)B^{\mu}(p_1)c(q)}
\end{split}
\end{align}
\begin{align}
\begin{split}\label{Eq:STI-QAP-SSbarBBc}
    g \hypS 
    \widetilde{\Gamma}_{\phi_b(-p_1-p_2-p_3)\phi^{\dagger}_{a}(p_3)B^{\nu}(p_2)B^{\mu}(p_1)}
    &- g \hypS
    \widetilde{\Gamma}_{\phi_b(p_4)\phi^{\dagger}_a(-p_1-p_2-p_4)B^{\nu}(p_2)B^{\mu}(p_1)}\\
    &= i \big( \widehat{\Delta} \cdot \widetilde{\Gamma} \big)_{\phi_b(p_4)\phi^{\dagger}_a(p_3)B^{\nu}(p_2)B^{\mu}(p_1)c(q)} 
\end{split}\\[1.5ex]
\begin{split}\label{Eq:STI-QAP-SSBBc}
    g \hypS 
    \widetilde{\Gamma}_{\phi_b(-p_1-p_2-p_3)\phi_{a}(p_3)B^{\nu}(p_2)B^{\mu}(p_1)}
    &+ g \hypS
    \widetilde{\Gamma}_{\phi_b(p_4)\phi_a(-p_1-p_2-p_4)B^{\nu}(p_2)B^{\mu}(p_1)}\\
    &= i \big( \widehat{\Delta} \cdot \widetilde{\Gamma} \big)_{\phi_b(p_4)\phi_a(p_3)B^{\nu}(p_2)B^{\mu}(p_1)c(q)}
\end{split}\\[1.5ex]
\begin{split}\label{Eq:STI-QAP-SbarSbarBBc}
    - g \hypS 
    \widetilde{\Gamma}_{\phi_b^{\dagger}(-p_1-p_2-p_3)\phi^{\dagger}_{a}(p_3)B^{\nu}(p_2)B^{\mu}(p_1)}
    &- g \hypS
    \widetilde{\Gamma}_{\phi^{\dagger}_b(p_4)\phi^{\dagger}_a(-p_1-p_2-p_4)B^{\nu}(p_2)B^{\mu}(p_1)}\\
    &= i \big( \widehat{\Delta} \cdot \widetilde{\Gamma} \big)_{\phi^{\dagger}_b(p_4)\phi^{\dagger}_a(p_3)B^{\nu}(p_2)B^{\mu}(p_1)c(q)}
\end{split}\\[1.5ex]
\begin{split}\label{Eq:STI-QAP-SSSbarSbarc}
    g \hypS 
    \widetilde{\Gamma}_{\phi_d(-p_3-p_2-p_1)\phi_c(p_3)\phi_b^{\dagger}(p_2)\phi_a^{\dagger}(p_1)}
    &+ g \hypS 
    \widetilde{\Gamma}_{\phi_d(p_4)\phi_c(-p_4-p_2-p_1)\phi_b^{\dagger}(p_2)\phi_a^{\dagger}(p_1)}\\
    - g \hypS 
    \widetilde{\Gamma}_{\phi_d(p_4)\phi_c(p_3)\phi_b^{\dagger}(-p_4-p_3-p_1)\phi_a^{\dagger}(p_1)}
    &- g \hypS 
    \widetilde{\Gamma}_{\phi_d(p_4)\phi_c(p_3)\phi_b^{\dagger}(p_2)\phi_a^{\dagger}(-p_4-p_3-p_2)}\\
    &= i \big( \widehat{\Delta} \cdot \widetilde{\Gamma} \big)_{\phi_d(p_4)\phi_c(p_3)\phi_b^{\dagger}(p_2)\phi_a^{\dagger}(p_1)c(q)}
\end{split}\\[1.5ex]
\begin{split}\label{Eq:STI-QAP-SSSSc}
    + g \hypS 
    \widetilde{\Gamma}_{\phi_d(-p_3-p_2-p_1)\phi_c(p_3)\phi_b(p_2)\phi_a(p_1)}
    &+ g \hypS 
    \widetilde{\Gamma}_{\phi_d(p_4)\phi_c(-p_4-p_2-p_1)\phi_b(p_2)\phi_a(p_1)}\\
    + g \hypS 
    \widetilde{\Gamma}_{\phi_d(p_4)\phi_c(p_3)\phi_b(-p_4-p_3-p_1)\phi_a(p_1)}
    &+ g \hypS 
    \widetilde{\Gamma}_{\phi_d(p_4)\phi_c(p_3)\phi_b(p_2)\phi_a(-p_4-p_3-p_2)}\\
    &= i \big( \widehat{\Delta} \cdot \widetilde{\Gamma} \big)_{\phi_d(p_4)\phi_c(p_3)\phi_b(p_2)\phi_a(p_1)c(q)}
\end{split}\\[1.5ex]
\begin{split}\label{Eq:STI-QAP-SSSSbarc}
    g \hypS 
    \widetilde{\Gamma}_{\phi_d(-p_3-p_2-p_1)\phi_c(p_3)\phi_b(p_2)\phi_a^{\dagger}(p_1)}
    &+ g \hypS 
    \widetilde{\Gamma}_{\phi_d(p_4)\phi_c(-p_4-p_2-p_1)\phi_b(p_2)\phi_a^{\dagger}(p_1)}\\
    + g \hypS 
    \widetilde{\Gamma}_{\phi_d(p_4)\phi_c(p_3)\phi_b(-p_4-p_3-p_1)\phi_a^{\dagger}(p_1)}
    &- g \hypS 
    \widetilde{\Gamma}_{\phi_d(p_4)\phi_c(p_3)\phi_b(p_2)\phi_a^{\dagger}(-p_4-p_3-p_2)}\\
    &= i \big( \widehat{\Delta} \cdot \widetilde{\Gamma} \big)_{\phi_d(p_4)\phi_c(p_3)\phi_b(p_2)\phi_a^{\dagger}(p_1)c(q)}
\end{split}\\[1.5ex]
\begin{split}\label{Eq:STI-QAP-SSbarSbarSbarc}
    g \hypS 
    \widetilde{\Gamma}_{\phi_d(-p_3-p_2-p_1)\phi_c^{\dagger}(p_3)\phi_b^{\dagger}(p_2)\phi_a^{\dagger}(p_1)}
    &- g \hypS 
    \widetilde{\Gamma}_{\phi_d(p_4)\phi_c^{\dagger}(-p_4-p_2-p_1)\phi_b^{\dagger}(p_2)\phi_a^{\dagger}(p_1)}\\
    - g \hypS 
    \widetilde{\Gamma}_{\phi_d(p_4)\phi_c^{\dagger}(p_3)\phi_b^{\dagger}(-p_4-p_3-p_1)\phi_a^{\dagger}(p_1)}
    &- g \hypS 
    \widetilde{\Gamma}_{\phi_d(p_4)\phi_c^{\dagger}(p_3)\phi_b^{\dagger}(p_2)\phi_a^{\dagger}(-p_4-p_3-p_2)}\\
    &= i \big( \widehat{\Delta} \cdot \widetilde{\Gamma} \big)_{\phi_d(p_4)\phi_c^{\dagger}(p_3)\phi_b^{\dagger}(p_2)\phi_a^{\dagger}(p_1)c(q)}
\end{split}\\[1.5ex]
\begin{split}\label{Eq:STI-QAP-SbarSbarSbarSbarc}
    - g \hypS 
    \widetilde{\Gamma}_{\phi_d^{\dagger}(-p_3-p_2-p_1)\phi_c^{\dagger}(p_3)\phi_b^{\dagger}(p_2)\phi_a^{\dagger}(p_1)}
    &- g \hypS 
    \widetilde{\Gamma}_{\phi_d^{\dagger}(p_4)\phi_c^{\dagger}(-p_4-p_2-p_1)\phi_b^{\dagger}(p_2)\phi_a^{\dagger}(p_1)}\\
    - g \hypS 
    \widetilde{\Gamma}_{\phi_d^{\dagger}(p_4)\phi_c^{\dagger}(p_3)\phi_b^{\dagger}(-p_4-p_3-p_1)\phi_a^{\dagger}(p_1)}
    &- g \hypS 
    \widetilde{\Gamma}_{\phi_d^{\dagger}(p_4)\phi_c^{\dagger}(p_3)\phi_b^{\dagger}(p_2)\phi_a^{\dagger}(-p_4-p_3-p_2)}\\
    &= i \big( \widehat{\Delta} \cdot \widetilde{\Gamma} \big)_{\phi_d^{\dagger}(p_4)\phi_c^{\dagger}(p_3)\phi_b^{\dagger}(p_2)\phi_a^{\dagger}(p_1)c(q)}
\end{split}
\end{align}

% -----------------------------------------------------------------------------

% +++++++++++++++++++++++++++++++++++++++++++++++++++++++++++++++++++++++++++++
\section{One-Loop Renormalisation} \label{Sec:1-Loop-Ren-Results}

In this section we present the complete 1-loop renormalisation 
of the general abelian chiral gauge theory defined by Eq.\ 
\eqref{Eq:model-Lagrangian} in Sec.\ \ref{Sec:D-DimLagrangian}.
Here we focus on results on the generic level; the subsequent section presents results specialised to the models discussed in Sec.\ \ref{Sec:SpecialCases} and defined via Tab.\ \ref{Tab:SpecialCasesOverview}.
The results include the evanescent 
gauge interactions
governed by the evanescent hypercharges $\hypLR$ and $\hypRL$, as introduced in the second line of Eq.\ \eqref{Eq:Lfermion}. 
Furthermore, we work
in $R_{\xi}$-gauge, cf.\ Eq.\ (\ref{Eq:Ddim-LGaugeFixing+Ghost}), 
without restricting the gauge parameter $\xi$.

In order to obtain concise expressions, the results have been simplified
using the conditions on the  Yukawa couplings imposed by 4-dimensional gauge invariance given in Eqs.\ (\ref{Eq:YukawaHyperchargeBRSTCondition}), and relations derived from them, as well
as the commutation properties of the hypercharge matrices.

For simplicity, for the generic results in this
section we choose $\hypRL=\hypLR$, and choose them to be hermitian matrices that
do not only respect electric and colour charge conservation but also commute with
both $\hypL$ and $\hypR$. 
This restriction is satisfied by the SM evanescent hypercharges in Eq.\ 
\eqref{Eq:ASSM-YLR}, but not by those in Eq.\ \eqref{Eq:ASSM-YLR-YRL-Option2}, and the subsequent section will present appropriate generalisations.

The renormalisation procedure has been performed according to
our methodology described in the previous section.
An important and novel aspect of the renormalisation process in the BMHV scheme of the model
under consideration is that it cannot be restricted to hypercharge-conserving Green functions alone, 
due to the possible violation of global hypercharge conservation, as discussed in Sec.\ \ref{Sec:The-Model}.
Hence, Green functions of the kind $\phi\phi$ and related ones must be taken into account.

All results have been independently computed by 
two different computational setups. 
The first setup is the one which has been used and explained in Ref.\ \cite{Stockinger:2023ndm}
(cf.\ Ref.\ \cite{Kuhler:2024fak});
it employs a tadpole decomposition (see Refs.\ \cite{Misiak:1994zw,Chetyrkin:1997fm})
to extract UV divergences.
The second setup is also \texttt{Mathematica}-based, but works exclusively with
\texttt{FeynArts} \cite{Hahn:2000kx} and \texttt{FeynCalc} 
\cite{Mertig:1990an,Shtabovenko:2016sxi,Shtabovenko:2020gxv,Shtabovenko:2021hjx}.
This second setup has been developed entirely independently from the first one,
serving as a strong consistency check for our results.
Additionally, the UV-divergent BRST breaking contributions
can be obtained from both the standard 1PI Green functions and the 
single $\Delta$-operator inserted 1PI Green functions. 
Both results are in agreement across both setups
and satisfy the relations given in Sec.\ \ref{Sec:STI+QAP}, 
providing further confidence in the correctness
of the results below.

We express all results presented in this section in terms of coefficients defined
in App.\ \ref{App:Results-1LoopCoeffs} to allow for a more compact and clear presentation.
Subsections \ref{Sec:1-Loop-Div-GreenFuncs} and \ref{Sec:1-Loop-BRST-Breaking} contain the calculated
results of the Green functions, while the singular and finite part of the counterterm action is
to be found in subsections \ref{Sec:SingularCTAction-1Loop} and \ref{Sec:FiniteCTAction-1Loop}, respectively.
For each Green function, only brief comments are provided, whereas a more detailed discussion
is dedicated to both the singular and the finite contribution to the counterterm action.

\subsection{Divergent One-Loop Green Functions}\label{Sec:1-Loop-Div-GreenFuncs}

We start with all standard power-counting divergent 1PI Green functions, from which
all singular counterterms are derived. 
These singular counterterms are essential for rendering the theory finite.
Below, we provide a complete list of all relevant Green functions, excluding
those given in Tab.\ \ref{Tab:NonContributingOrdinaryGreenFunctions}.
\begin{table}[h!]
    \centering
    \begin{tabular}{|c|c|c|} \hline
         Degree of div. & Green function & Remark \\
         \hline\hline
         \rule{0pt}{2.3em} 2 & \makecell{$\phi$\,--\,$B^{\mu}$,\\ $\phi^{\dagger}$\,--\,$B^{\mu}$} & \makecell{Green func.\ $\sim \mathcal{O}(p^2)$ due to power counting.\\Field monomial has an open Lorentz index\\that needs to be saturated.} \\ 
         \hline
         \rule{0pt}{1.6em} 1 & \makecell{$\phi$\,--\,$B^{\mu}$\,--\,$B^{\nu}$,\\ $\phi^{\dagger}$\,--\,$B^{\mu}$\,--\,$B^{\nu}$} & \makecell{Green func.\ $\sim \mathcal{O}(p)$ due to power counting.\\Field monomial has no open Lorentz index.} \\ 
         \hline
         1 & \makecell{$\phi$\,--\,$\phi$\,--\,$\phi$,\\ $\phi^{\dagger}$\,--\,$\phi$\,--\,$\phi$,\\ $\phi^{\dagger}$\,--\,$\phi^{\dagger}$\,--\,$\phi$,\\ $\phi^{\dagger}$\,--\,$\phi^{\dagger}$\,--\,$\phi^{\dagger}$} & \makecell{Green func.\ $\sim \mathcal{O}(p)$ due to power counting.\\Field monomial has no open Lorentz index.} \\ 
         \hline
         \rule{0pt}{2.3em} 0 & \makecell{$\phi$\,--\,$B^{\mu}$\,--\,$B^{\nu}$\,--\,$B^{\rho}$,\\ $\phi^{\dagger}$\,--\,$B^{\mu}$\,--\,$B^{\nu}$\,--\,$B^{\rho}$} & \makecell{Green func.\ $\sim \mathcal{O}(p^0)$ due to power counting.\\Field monomial has an open Lorentz index\\that needs to be saturated.} \\ 
         \hline
         0 & \makecell{$\phi$\,--\,$\phi$\,--\,$\phi$\,--\,$B^{\mu}$,\\ $\phi^{\dagger}$\,--\,$\phi$\,--\,$\phi$\,--\,$B^{\mu}$,\\ $\phi^{\dagger}$\,--\,$\phi^{\dagger}$\,--\,$\phi$\,--\,$B^{\mu}$,\\ $\phi^{\dagger}$\,--\,$\phi^{\dagger}$\,--\,$\phi^{\dagger}$\,--\,$B^{\mu}$} & \makecell{Green func.\ $\sim \mathcal{O}(p^0)$ due to power counting.\\Field monomial has an open Lorentz index\\that needs to be saturated.} \\ 
         \hline
    \end{tabular}
    \caption{Power-counting divergent Green functions that vanish a priori, and thus do not contribute.
    The first column lists the superficial degree of divergence, the second column shows the fields associated with 
    a specific Green function, and the final column provides a remark implying the reason why the contribution
    from this particular Green function vanishes.
    Note that, considering a massless theory, the only scales of the theory are momenta, and divergences are restricted by the power-counting degree of divergence and must correspond to field monomials with saturated Lorentz indices.}
    \label{Tab:NonContributingOrdinaryGreenFunctions}
\end{table}

The Green functions listed in Tab.\ \ref{Tab:NonContributingOrdinaryGreenFunctions}, 
although power-counting divergent, 
do not give rise to non-vanishing UV divergences and can therefore be discarded.
The reason for this can be found in the remarks in the third column of Tab.\ 
\ref{Tab:NonContributingOrdinaryGreenFunctions}.

\subsubsection{2-Point Functions}

Starting with the $2$-point Green functions, there are five non-vanishing 
Green functions --- excluding the two listed in the first row of Tab.\ 
\ref{Tab:NonContributingOrdinaryGreenFunctions} ---
which can be categorised into three groups: those for the gauge boson, 
the fermions and the scalars.

\vspace{0.2cm}\noindent\textbf{(i) Gauge Boson Self Energy:}%\vspace{0.1cm}
\begin{equation}\label{Eq:GaugeBosonSelfEnergy_1-Loop}
    \begin{aligned}
        i \widetilde{\Gamma}&_{B_{\mu}(p)B_{\nu}(-p)}\big|_{\text{div}}^{1}\\
        &= 
        \frac{ig^2}{16 \pi^2} 
        \prescript{}{S}{\mathcal{A}}_{BB}^{1, \text{inv}} \, \frac{1}{\epsilon} \,
        \Big(p^{\mu} p^{\nu} - p^2 \eta^{\mu\nu}\Big) \, 
        + \frac{ig^2}{16 \pi^2} 
        \prescript{}{F}{\overline{\mathcal{A}}}_{BB}^{1, \text{inv}} \, \frac{1}{\epsilon} \,
        \Big(\overline{p}^{\mu} \overline{p}^{\nu} - \overline{p}^2 \overline{\eta}^{\mu\nu}\Big)\\
        &+ \frac{ig^2}{16 \pi^2} \, \frac{1}{\epsilon} \bigg\{
        \Big( 
        \prescript{}{F}{\widehat{\mathcal{A}}}_{BB,1}^{1, \text{break}}
        + \prescript{}{F}{\widehat{\mathcal{A}}}_{BB,3}^{1, \text{break}}
        \Big) \,
        \overline{p}^2 \widehat{\eta}^{\mu\nu}
        + \prescript{}{F}{\widehat{\mathcal{A}}}_{BB,4}^{1, \text{break}} \,
        \widehat{p}^2 \overline{\eta}^{\mu\nu}\\
        &\qquad\qquad + \frac{2}{3}
        \prescript{}{F}{\widehat{\mathcal{A}}}_{BB,3}^{1, \text{break}} \,
        \widehat{p}^2 \widehat{\eta}^{\mu\nu}
        +
        \prescript{}{F}{\widehat{\mathcal{A}}}_{BB,2}^{1, \text{break}} \,
        \big(\overline{p}^{\mu} \widehat{p}^{\nu} + \widehat{p}^{\mu} \overline{p}^{\nu} \big) 
        + 2 \prescript{}{F}{\widehat{\mathcal{A}}}_{BB,1}^{1, \text{break}} \,
        \widehat{p}^{\mu} \widehat{p}^{\nu}
        \bigg\}.
    \end{aligned}
\end{equation}
The first line of Eq.\ (\ref{Eq:GaugeBosonSelfEnergy_1-Loop}) displays the transversal part of the 
gauge boson self energy, while the remaining terms account for the divergent and BRST-breaking contributions.
As demonstrated in our previous studies on chiral QED  \cite{Belusca-Maito:2021lnk,Stockinger:2023ndm},
the fermionic contribution to the transversal part --- the second term of the first line
in Eq.\ (\ref{Eq:GaugeBosonSelfEnergy_1-Loop}) --- is purely $4$-dimensional, which is due to
the projectors in the fermion-gauge boson interaction currents.
In contrast, the scalar contribution in the first term of the first line is completely 
$D$-dimensional.
Among the divergent BRST breaking terms, the only term that is independent of the evanescent 
hypercharges is $\prescript{}{F}{\widehat{\mathcal{A}}}_{BB,4}^{1, \text{break}} \, \widehat{p}^2 \overline{\eta}^{\mu\nu}$,
cf.\ Eqs.\ (\ref{App-Eq:GaugeBosonCoeffsFermionic}) and (\ref{App-Eq:GaugeBosonCoeffsScalar}).
All other breaking terms are due to the evanescent gauge interactions $\hypLR$.
Thus, these additional terms introduce further Lorentz structures not present in the
1-loop results of chiral QED in previous publications and vanish for $\hypLR\equiv0$.

\vspace{0.2cm}\noindent\textbf{(ii) Fermion Self Energy:}%\vspace{0.1cm}
\begin{equation}\label{Eq:FermionSelfEnergy_1-Loop}
    \begin{aligned}
        i \widetilde{\Gamma}_{\psi_i(p)\overline{\psi}_j(-p)} \big|_{\text{div}}^{1}
        &= \frac{i}{16 \pi^2} \, \frac{1}{\epsilon} \, \overline{\slashed{p}} \, \bigg[
        \overline{\mathcal{A}}_{\psi\overline{\psi},\mathrm{R},ji}^{1,\text{inv}}
        \mathbb{P}_{\mathrm{R}} 
        +
        \overline{\mathcal{A}}_{\psi\overline{\psi},\mathrm{L},ji}^{1,\text{inv}}
        \mathbb{P}_{\mathrm{L}}
        \bigg]\\
        &+ \frac{i}{16 \pi^2} \, \frac{1}{\epsilon} \, \widehat{\slashed{p}} \, \bigg[
        \widehat{\mathcal{A}}_{\psi\overline{\psi},ji}^{1,\text{break}}
        \mathbb{P}_{\mathrm{R}} 
        + 
        {\widehat{\mathcal{A}}{}_{\psi\overline{\psi},ij}^{1,\text{break}}}^{\dagger}
        \mathbb{P}_{\mathrm{L}}
        \Big].
    \end{aligned}
\end{equation}
There are left- and right-handed contributions to both $4$-dimensional and evanescent terms.
Further, each coefficient contains a fermionic and a Yukawa contribution, see
Eq.\ (\ref{App-Eq:DivCoeffRelations}) in App.\ \ref{App:Results-1LoopCoeffs}, 
as well as Eq.\ (\ref{App-Eq:FermionCoeffFermionGauge})
for the fermionic and Eq.\ (\ref{App-Eq:FermionCoeffYukawa}) for the Yukawa contributions.
Note that this time, none of the coefficients vanishes entirely when setting $\hypLR\equiv0$,
and thus the evanescent interaction currents do not alter the Lorentz structure in this case.

\vspace{0.2cm}\noindent\textbf{(iii) Scalar Self Energy:}\vspace{0.1cm}

\begin{align}
    \begin{split}\label{Eq:ScalarSelfEnergy_1-Loop}
        i \widetilde{\Gamma}_{\phi_a(p)\phi^{\dagger}_b(-p)} \big|_{\text{div}}^{1} &= \frac{i}{16 \pi^2} \, \frac{1}{\epsilon} \, \Big(
        g^2
        \prescript{}{S}{\mathcal{A}}_{\phi\phi^{\dagger},ba}^{1,\text{inv}}
        + \prescript{}{Y}{\overline{\mathcal{A}}}_{\phi\phi^{\dagger},ba}^{1,\text{inv}}
        \Big) \, 
        \overline{p}^2\\ 
        &+ \frac{i}{16 \pi^2} \, \frac{1}{\epsilon} \, \Big(
        g^2
        \prescript{}{S}{\mathcal{A}}_{\phi\phi^{\dagger},ba}^{1,\text{inv}}
        + {\widehat{\mathcal{A}}}_{\phi\phi^{\dagger},ba}^{1, \text{break}}
        \Big) \, \widehat{p}^2,
    \end{split}\\[1.5ex]
    \begin{split}\label{Eq:phiphiSelfEnergy_1-Loop}
        i \widetilde{\Gamma}_{\phi_a(p)\phi_b(-p)} \big|_{\text{div}}^{1} &= 
        \frac{i}{16 \pi^2} \, \frac{1}{\epsilon} \,
        {\widehat{\mathcal{A}}}_{\phi\phi,ba}^{1,\text{break}}\,
        \widehat{p}^2,
    \end{split}\\[1.5ex]
    \begin{split}\label{Eq:phidaggerphidaggerSelfEnergy_1-Loop}
        i \widetilde{\Gamma}_{\phi^{\dagger}_a(p)\phi^{\dagger}_b(-p)} \big|_{\text{div}}^{1} &= 
        \frac{i}{16 \pi^2} \, \frac{1}{\epsilon} \,
        {\widehat{\mathcal{A}}{}_{\phi\phi,ba}^{1,\text{break}}}^{\dagger}\,
        \widehat{p}^2.
    \end{split}
\end{align}
For the scalars, we obtain contributions from the self energy given in 
Eq.\ (\ref{Eq:ScalarSelfEnergy_1-Loop}), as well as from the two 2-point 
Green functions that violate global hypercharge conservation provided in Eqs.\ (\ref{Eq:phiphiSelfEnergy_1-Loop})
and (\ref{Eq:phidaggerphidaggerSelfEnergy_1-Loop}).
As explained in Sec.\ \ref{Sec:The-Model}, the two sources for the violation of
global hypercharge conservation are the evanescent part of the fermion kinetic term and the
evanescent gauge interactions.
Their divergent 1-loop effects are purely evanescent.
Here, the contributions in Eqs.\ (\ref{Eq:phiphiSelfEnergy_1-Loop})
and (\ref{Eq:phidaggerphidaggerSelfEnergy_1-Loop}) stem entirely from
the first source, i.e.\ the fermion kinetic term, as they are independent of
the evanescent hypercharges.

In contrast, the ordinary scalar self energy in Eq.\ (\ref{Eq:ScalarSelfEnergy_1-Loop})
includes both $4$-dimensional and evanescent contributions.
Specifically, the terms arising solely from scalar interactions,
i.e.\ those terms $\sim \prescript{}{S}{\mathcal{A}}_{\phi\phi^{\dagger},ba}^{1,\text{inv}}$,
add up to a fully $D$-dimensional and BRST-invariant contribution.
This is not the case for the terms emerging from Yukawa interactions, where
the evanescent contribution differs from the $4$-dimensional one, and thus gives
rise to a BRST-breaking term.

\subsubsection{3-Point Functions}

Moving on to the 3-point Green functions, we find seven
power-counting divergent Green functions
--- excluding the two scalar-double gauge boson and the four
triple scalar self-interaction contributions listed in the second 
and third row of Tab.\ 
\ref{Tab:NonContributingOrdinaryGreenFunctions}, respectively ---
which are presented below.

\vspace{0.2cm}\noindent\textbf{(iv) Fermion-Gauge Boson Interaction:}%\vspace{0.1cm}
\begin{equation}\label{Eq:FFB_1-Loop}
    \begin{aligned}
        i \widetilde{\Gamma}_{\psi_i(p_2)\overline{\psi}_j(p_1)B_{\mu}(q)} \big|_{\text{div}}^{1} =
        &-\frac{ig}{16\pi^2} \, \frac{1}{\epsilon} \, \overline{\gamma}^{\mu} 
        \Big[
        \Big( \mathcal{Y}_{R}
        \overline{\mathcal{A}}_{\psi\overline{\psi},\mathrm{R}}^{1, \text{inv}}
        \Big)_{ji}
        \mathbb{P}_{\mathrm{R}}
        +
        \Big( \mathcal{Y}_{L}
        \overline{\mathcal{A}}_{\psi\overline{\psi},\mathrm{L}}^{1, \text{inv}}
        \Big)_{ji} 
        \mathbb{P}_{\mathrm{L}}
        \Big]\\
        &-\frac{ig}{16 \pi^2} \, \frac{1}{\epsilon} \, \widehat{\gamma}^{\mu} 
        \Big[
        \widehat{\mathcal{A}}_{\psi\overline{\psi}B,ji}^{1,\text{break}}\,
        \mathbb{P}_{\mathrm{R}}
        +
        {\widehat{\mathcal{A}}{}_{\psi\overline{\psi}B,ij}^{1,\text{break}}}^{\dagger}\,
        \mathbb{P}_{\mathrm{L}}
        \Big].
    \end{aligned}
\end{equation}
Again, we find fermionic and Yukawa contributions 
to both the $4$-dimensional
and the evanescent part of the Green function, as indicated by the coefficients
provided in App.\ \ref{App:Results-1LoopCoeffs} in Eq.\ (\ref{App-Eq:DivCoeffRelations})
as well as Eqs.\ (\ref{App-Eq:FermionCoeffFermionGauge}) to (\ref{App-Eq:FermionGaugeBosonCoeff}).
For the evanescent contributions, only the terms arising from Yukawa
interactions remain when setting $\hypLR\equiv0$.

\vspace{0.2cm}\noindent\textbf{(v) Triple Gauge Boson Interaction:}\vspace{0.1cm}

\noindent
As expected, and analogous to the case of chiral QED studied in Refs.\ \cite{Belusca-Maito:2021lnk,Stockinger:2023ndm},
the triple gauge boson interaction 
%in the anomaly-free case, cf.~\ref{Eq:AnomalyCancellationCondition}, 
does not lead to a UV-divergent contribution, and thus
\begin{equation}
    \begin{aligned}
        i \widetilde{\Gamma}_{B_{\mu}(q)B_{\nu}(p_1)B_{\rho}(p_2)} \big|_{\text{div}}^{1} = 0.
    \end{aligned}
\end{equation}

\vspace{0.2cm}\noindent\textbf{(vi) Yukawa Interaction:}\vspace{0.1cm}

\noindent
The Yukawa interactions are given by
\begin{align}
    \begin{split}\label{Eq:YukawaGreenFunc1-FFS}
        i \widetilde{\Gamma}_{\psi_i(p_2)\overline{\psi}_j(p_1)\phi_a(q)} \big|_{\text{div}}^{1} &=
        \frac{i}{16 \pi^2} \, \frac{1}{\epsilon} 
        \Big[
        \overline{\mathcal{A}}_{\psi\overline{\psi}\phi,\mathrm{R},ji}^{1,\text{inv},a}\,
        \mathbb{P}_{\mathrm{R}}
        + 
        \overline{\mathcal{A}}_{\psi\overline{\psi}\phi,\mathrm{L},ji}^{1,\text{inv},a}\,
        \mathbb{P}_{\mathrm{L}}
        \Big],
    \end{split}\\[1.5ex]
    \begin{split}\label{Eq:YukawaGreenFunc2-FFSbar}
        i \widetilde{\Gamma}_{\psi_i(p_2)\overline{\psi}_j(p_1)\phi^{\dagger}_a(q)} \big|_{\text{div}}^{1} &=
        \frac{i}{16 \pi^2} \, \frac{1}{\epsilon} 
        \Big[
        {\overline{\mathcal{A}}{}_{\psi\overline{\psi}\phi,\mathrm{L},ij}^{1,\text{inv},a}}^{\!\!\!\!\!\!\dagger}\,\,\,\,
        \mathbb{P}_{\mathrm{R}}
        + 
        {\overline{\mathcal{A}}{}_{\psi\overline{\psi}\phi,\mathrm{R},ij}^{1,\text{inv},a}}^{\!\!\!\!\!\!\!\dagger}\,\,\,\,
        \mathbb{P}_{\mathrm{L}}
        \Big],
    \end{split}
\end{align}
with coefficients provided in App.\ \ref{App:Results-1LoopCoeffs} in Eqs.\ (\ref{App-Eq:DivCoeffRelations})
and (\ref{App-Eq:DivYukawaCoeffs}).

\vspace{0.2cm}\noindent\textbf{(vii) Scalar-Gauge Boson Triple Interaction:}\vspace{0.1cm}

\begin{align}
    \begin{split}\label{Eq:phi-phiDagger-B-GreenFunc}
        i \widetilde{\Gamma}_{\phi_a(p_2)\phi^{\dagger}_b(p_1)B_{\mu}(q)} \big|_{\text{div}}^{1} &=
        \frac{ig}{16\pi^2} \, \frac{\mathcal{Y}_{S}}{\epsilon} \Big(
        g^2
        \prescript{}{S}{\mathcal{A}}_{\phi\phi^{\dagger},ba}^{1,\text{inv}}
        + \prescript{}{Y}{\overline{\mathcal{A}}}_{\phi\phi^{\dagger},ba}^{1,\text{inv}} 
        \Big) \Big( \overline{p}_1^{\mu} - \overline{p}_2^{\mu} \Big)\\
        &+ \frac{ig}{16 \pi^2} \, \frac{1}{\epsilon} \Big(
        g^2 \, \mathcal{Y}_{S}
        \prescript{}{S}{\mathcal{A}}_{\phi\phi^{\dagger},ba}^{1,\text{inv}}
        + {\widehat{\mathcal{A}}}_{\phi\phi^{\dagger}B,ba}^{1,\text{break}} 
        \Big) \Big( \widehat{p}_1^{\mu} - \widehat{p}_2^{\mu} \Big),
    \end{split}\\[1.5ex]
    \begin{split}\label{Eq:phi-phi-B-GreenFunc}
        i \widetilde{\Gamma}_{\phi_a(p_2)\phi_b(p_1) B_{\mu}(q)} \big|_{\text{div}}^{1} &=
        \frac{ig}{16 \pi^2} \, \frac{1}{\epsilon}
        {\widehat{\mathcal{A}}}_{\phi\phi B, ba}^{1,\text{break}} \, \Big( \widehat{p}_1^{\mu} - \widehat{p}_2^{\mu} \Big),
    \end{split}\\[1.5ex]
    \begin{split}\label{Eq:phiDagger-phiDagger-B-GreenFunc}
        i \widetilde{\Gamma}_{\phi^{\dagger}_a(p_2)\phi^{\dagger}_b(p_1) B_{\mu}(q)} \big|_{\text{div}}^{1} &=
        - \frac{ig}{16 \pi^2} \, \frac{1}{\epsilon} \,
        {\widehat{\mathcal{A}}{}_{\phi\phi B,ba}^{1,\text{break}}}^{\dagger} \Big( \widehat{p}_1^{\mu} - \widehat{p}_2^{\mu} \Big).
    \end{split}
\end{align}
Similar to the scalar 2-point Green functions, we again find purely evanescent contributions from two
global hypercharge violating Green functions, as given in Eqs.\
(\ref{Eq:phi-phi-B-GreenFunc}) and (\ref{Eq:phiDagger-phiDagger-B-GreenFunc}), 
in addition to the expected contribution from the 
scalar-gauge boson triple interaction in Eq.\ (\ref{Eq:phi-phiDagger-B-GreenFunc}).

However, this time the contributions that violate global hypercharge conservation vanish identically
for $\hypLR\equiv0$, as indicated by the associated coefficient provided in
App.\ \ref{App:Results-1LoopCoeffs} in Eq.\ (\ref{App-Eq:SSBCoeff}).
This is due to the fact that these contributions arise completely from the evanescent gauge interactions,
rather than from the evanescent part of the fermion kinetic term, as was the case for the 
2-point Green functions above.

\subsubsection{4-Point Functions}

Finally, for the 4-point Green functions, we identify nine
power-counting divergent Green functions
--- excluding the two scalar-triple gauge boson and the four
triple scalar-single gauge boson contributions listed in the penultimate
and the last row of Tab.\ 
\ref{Tab:NonContributingOrdinaryGreenFunctions}, respectively ---
which are provided below.

\vspace{0.2cm}\noindent\textbf{(viii) Quartic Gauge Boson Interaction:}\vspace{0.1cm}

\noindent
The 1-loop contribution to the quartic gauge boson self-interaction vanishes, analogous to
the case in chiral QED, and thus
\begin{equation}\label{Eq:BBBB-GreenFunc}
    \begin{aligned}
        i \widetilde{\Gamma}_{B_{\mu}(p_2)B_{\nu}(p_1)B_{\rho}(p_4)B_{\sigma}(p_3)} \big|_{\text{div}}^{1} = 0.
    \end{aligned}
\end{equation}
Note that this result holds true for $\hypRL=\hypLR$.
However, in the case where the two evanescent hypercharges are unequal, only satisfying the
necessary hermiticity requirement in Eq.\ \eqref{Eq:Requirement-YLR-YRL}
--- such as in the case for the hypercharges in Eq.\ \eqref{Eq:ASSM-YLR-YRL-Option2} --- 
the quartic gauge boson Green function yields a non-vanishing divergent contribution, as seen below in
Sec.\ \ref{Sec:AnalysisOfASSMResults}.

\vspace{0.2cm}\noindent\textbf{(ix) Scalar-Gauge Boson Quartic Interaction:}\vspace{0.1cm}

\begin{align}
    \begin{split}\label{Eq:phi-phiDagger-B-B}
        i \widetilde{\Gamma}_{\phi_a(p_2)\phi^{\dagger}_b(p_1) B_{\mu}(p_4)B_{\nu}(p_3)}\big|_{\text{div}}^{1} &=
        \frac{ig^2}{16 \pi^2} \, \frac{2\,\mathcal{Y}_{S}^2}{\epsilon} 
        \Big( 
        g^2 \prescript{}{S}{\mathcal{A}}_{\phi\phi^{\dagger},ba}^{1,\text{inv}}
        + \prescript{}{Y}{\overline{\mathcal{A}}}_{\phi\phi^{\dagger},ba}^{1,\text{inv}}
        \Big) \,
        \overline{\eta}^{\mu\nu}\\
        &+ \frac{ig^2}{16 \pi^2} \, \frac{2}{\epsilon} \, 
        \Big( 
        g^2 \, \mathcal{Y}_{S}^2 \prescript{}{S}{\mathcal{A}}_{\phi\phi^{\dagger},ba}^{1,\text{inv}}
        + \frac{1}{2} {\widehat{\mathcal{A}}}_{\phi\phi^{\dagger} BB,ba}^{1,\text{break}}
        \Big) \,
        \widehat{\eta}^{\mu\nu},
    \end{split}\\[1.5ex]
    \begin{split}\label{Eq:phi-phi-B-B}
        i \widetilde{\Gamma}_{\phi_a(p_2)\phi_b(p_1) B_{\mu}(p_4)B_{\nu}(p_3)} \big|_{\text{div}}^{1} &=
        \frac{ig^2}{16 \pi^2} \, \frac{1}{\epsilon}
        {\widehat{\mathcal{A}}}_{\phi\phi BB, ba}^{1, \text{break}} \, \widehat{\eta}^{\mu\nu},
    \end{split}\\[1.5ex]
    \begin{split}\label{Eq:phiDagger-phiDagger-B-B}
        i \widetilde{\Gamma}_{\phi^{\dagger}_a(p_2)\phi^{\dagger}_b(p_1) B_{\mu}(p_4)B_{\nu}(p_3)} \big|_{\text{div}}^{1} &=
        \frac{ig^2}{16 \pi^2} \, \frac{1}{\epsilon}
        {\widehat{\mathcal{A}}{}_{\phi\phi BB,ba}^{1, \text{break}}}^{\!\!\!\!\!\dagger} \,\,\, \widehat{\eta}^{\mu\nu}.     
    \end{split}
\end{align}
Once again, there are two fully evanescent global hypercharge violating Green functions for the 
scalar-gauge boson quartic interaction illustrated in Eqs.\ (\ref{Eq:phi-phi-B-B})
and (\ref{Eq:phiDagger-phiDagger-B-B}), alongside the expected contribution
in Eq.\ (\ref{Eq:phi-phiDagger-B-B}).

Their contributions also arise solely from the evanescent gauge interactions,
and thus vanish identically for $\hypLR\equiv0$, as can be seen from
the coefficients provided in App.\ \ref{App:Results-1LoopCoeffs},
Eq.\ (\ref{App-Eq:SSBBCoeffs}).

\vspace{0.2cm}\noindent\textbf{(x) Quartic Scalar Interaction:}%\vspace{0.1cm}
\begin{align}
    \begin{split}\label{Eq:SSSbarSbar-GreenFunc}
        i \widetilde{\Gamma}_{\phi_a(p_2)\phi_b(p_1)\phi^{\dagger}_c(p_4)\phi^{\dagger}_d(p_3)}\big|_{\text{div}}^{1}
        &=
        \frac{i}{16 \pi^2} \, \frac{1}{\epsilon} \, 
        \mathcal{A}_{\phi\phi^{\dagger}\phi\phi^{\dagger},cadb}^{1,\text{inv}},
    \end{split}\\[1.5ex]
    \begin{split}\label{Eq:SSSSbar-GreenFunc}
        i \widetilde{\Gamma}_{\phi_a(p_2)\phi_b(p_1)\phi_c(p_4)\phi^{\dagger}_d(p_3)} \big|_{\text{div}}^{1} &= 0,
    \end{split}\\[1.5ex]
    \begin{split}\label{Eq:SSbarSbarSbar-GreenFunc}
        i \widetilde{\Gamma}_{\phi_a(p_2)\phi^{\dagger}_b(p_1)\phi^{\dagger}_c(p_4)\phi^{\dagger}_d(p_3)} \big|_{\text{div}}^{1} &= 0,
    \end{split}\\[1.5ex]
    \begin{split}\label{Eq:SSSS-GreenFunc}
        i \widetilde{\Gamma}_{\phi_a(p_2)\phi_b(p_1)\phi_c(p_4)\phi_d(p_3)} \big|_{\text{div}}^{1} &= 0,
    \end{split}\\[1.5ex]
    \begin{split}\label{Eq:SbarSbarSbarSbar-GreenFunc}
        i \widetilde{\Gamma}_{\phi^{\dagger}_a(p_2)\phi^{\dagger}_b(p_1)\phi^{\dagger}_c(p_4)\phi^{\dagger}_d(p_3)} \big|_{\text{div}}^{1} &= 0.
    \end{split}
\end{align}
Evidently, only the hypercharge-conserving Green function yields a non-vanishing
contribution, while all hypercharge-violating Green functions vanish identically.
This is because any hypercharge-violating divergence at the 1-loop level
must be evanescent as it stems from the evanescent $\DeltaHat$-operator 
shown in Eq.\ (\ref{Eq:TreeLevelBreaking}).
However, the four scalar interaction Green functions at hand have a power counting
of zero, indicating that these Green functions do not involve any momenta.
Further, they do not contain any metric tensors or $\gamma^{\mu}$-matrices,
due to the absence of gauge bosons and fermions, respectively.
As a result, no evanescent divergences can emerge from these Green functions.
The situation becomes more intricate and might change at higher orders,
as then also $4$-dimensional hypercharge-breaking contributions can occur,
cf.\ e.g.\ \cite{Stockinger:2023ndm}.

\subsection{One-Loop Breaking of BRST Symmetry}\label{Sec:1-Loop-BRST-Breaking}

In this subsection, we continue with all single $\DeltaHat$-operator inserted,
power-counting divergent, 1PI Green functions. They determine the 
breaking of the regularised Slavnov-Taylor identity and allow to derive the  symmetry-restoring counterterms  via Eqs.~(\ref{Eq:QAPofDReg}) and (\ref{Eq:DefDeltaBreaking}).
As as cross-check, the divergent symmetry-violating contributions align perfectly
with those obtained from the standard Green functions discussed in 
Sec.\ \ref{Sec:1-Loop-Div-GreenFuncs}.

Again, we provide a comprehensive list of all relevant operator-inserted
Green functions, suppressing all finite evanescent terms, as they
vanish in the limit $\mathop{\text{LIM}}_{D\,\to\,4}$ after renormalisation, and
excluding those Green functions that vanish in the first place,
listed in Tab.\ \ref{Tab:NonContributingDeltaGreenFunctions}.
The reasons for them to vanish is indicated in the third column
of this table.
\begin{table}[h!]
    \centering
    \begin{tabular}{|c|c|c|} \hline
         Degree of div. & Green function & Remark \\ 
         \hline\hline
         \rule{0pt}{1.6em} 3 & \makecell{$c$\,--\,$\phi$,\\$c$\,--\,$\phi^{\dagger}$} & \makecell{Green func.\ $\sim \mathcal{O}(p^3)$ due to power counting.\\Field monomial has no open Lorentz index.} \\ 
         \hline
         \rule{0pt}{2.25em} 2 & \makecell{$c$\,--\,$B^{\mu}$\,--\,$\phi$,\\$c$\,--\,$B^{\mu}$\,--\,$\phi^{\dagger}$} & \makecell{Green func.\ $\sim \mathcal{O}(p^2)$ due to power counting.\\Field monomial has an open Lorentz index\\that needs to be saturated.} \\ 
         \hline
         1 & \makecell{$c$\,--\,$\phi$\,--\,$\phi$\,--\,$\phi$,\\$c$\,--\,$\phi^{\dagger}$\,--\,$\phi$\,--\,$\phi$,\\$c$\,--\,$\phi^{\dagger}$\,--\,$\phi^{\dagger}$\,--\,$\phi$,\\$c$\,--\,$\phi^{\dagger}$\,--\,$\phi^{\dagger}$\,--\,$\phi^{\dagger}$} & \makecell{Green func.\ $\sim \mathcal{O}(p)$ due to power counting.\\Field monomial has no open Lorentz index.} \\ 
         \hline
         1 & \makecell{$c$\,--\,$B^{\mu}$\,--\,$B^{\nu}$\,--\,$\phi$,\\$c$\,--\,$B^{\mu}$\,--\,$B^{\nu}$\,--\,$\phi^{\dagger}$} & \makecell{Green func.\ $\sim \mathcal{O}(p)$ due to power counting.\\Field monomial has no open Lorentz index.} \\ 
         \hline
         0 & \makecell{$c$\,--\,$B^{\mu}$\,--\,$\phi$\,--\,$\phi$\,--\,$\phi$,\\$c$\,--\,$B^{\mu}$\,--\,$\phi^{\dagger}$\,--\,$\phi$\,--\,$\phi$,\\$c$\,--\,$B^{\mu}$\,--\,$\phi^{\dagger}$\,--\,$\phi^{\dagger}$\,--\,$\phi$,\\$c$\,--\,$B^{\mu}$\,--\,$\phi^{\dagger}$\,--\,$\phi^{\dagger}$\,--\,$\phi^{\dagger}$} & \makecell{Green func.\ $\sim \mathcal{O}(p^0)$ due to power counting.\\Field monomial has an open Lorentz index\\that needs to be saturated.} \\ 
         \hline
         \rule{0pt}{2.25em} 0 & \makecell{$c$\,--\,$B^{\mu}$\,--\,$B^{\nu}$\,--\,$B^{\rho}$\,--\,$\phi$,\\$c$\,--\,$B^{\mu}$\,--\,$B^{\nu}$\,--\,$B^{\rho}$\,--\,$\phi^{\dagger}$} & \makecell{Green func.\ $\sim \mathcal{O}(p^0)$ due to power counting.\\Field monomial has an open Lorentz index\\that needs to be saturated.} \\ 
         \hline
         \rule{0pt}{1.6em} 0 & \makecell{$c$\,--\,$B^{\mu}$\,--\,$B^{\nu}$\,--\,$B^{\rho}$\,--\,$B^{\sigma}$} & 
         \makecell{Could only emerge from $s(BBBBB)$,\\which is non-renormalisable.} \\ 
         \hline
    \end{tabular}
    \caption{Power-counting divergent, single $\Delta$-operator inserted Green functions that vanish a priori.
    The first column lists the superficial degree of divergence, the second column shows the fields associated with 
    a specific Green function, and the final column provides a remark implying the reason why the contribution
    from this particular Green function vanishes.
    Note that, considering a massless theory, the only scales of the theory are momenta.
    Field monomials are restricted by power-counting and Lorentz invariance.}
    \label{Tab:NonContributingDeltaGreenFunctions}
\end{table}
The reasons are similar to the ones discussed for Tab.\ \ref{Tab:NonContributingOrdinaryGreenFunctions}, with the exception of the last line.
The Green function presented in the last row of Tab.\ \ref{Tab:NonContributingDeltaGreenFunctions}
might initially appear to be allowed.
However, in an abelian gauge theory, such a term in $\Delta\cdot\Gamma$ could only originate from 
the BRST transformation of the field monomial $\sim BBBBB$, which is non-renormalisable, i.e., not allowed in any renormalisable quantum field theory.
Note that this situation changes in the framework of non-abelian gauge theories,
where more complex BRST transformations, especially those involving the gauge bosons, come into play.

\subsubsection{2-Point Functions}

Beginning with the $2$-point Green functions, we are left with a single 
non-vanishing Green function, excluding the two listed in the first row of Tab.\ 
\ref{Tab:NonContributingDeltaGreenFunctions}, which is displayed below.

\vspace{0.2cm}\noindent\textbf{(xi) Ghost-Gauge Boson Contribution:}%\vspace{0.1cm}
\begin{equation}\label{Eq:GhostGaugeBoson_1-Loop}
    \begin{aligned}
        i \Big( \widehat{\Delta} \cdot \widetilde{\Gamma} \Big)_{B_{\mu}(-p) c(p)}^{1} 
        &= \frac{g^2}{16 \pi^2} \, \frac{1}{\epsilon} \, 
        \bigg[
        \Big( 
        \prescript{}{F}{\widehat{\mathcal{A}}}_{BB,1}^{1, \text{break}} 
        + \prescript{}{F}{\widehat{\mathcal{A}}}_{BB,2}^{1, \text{break}}
        + \prescript{}{F}{\widehat{\mathcal{A}}}_{BB,3}^{1, \text{break}}
        \Big) \,
        \overline{p}^2 \widehat{p}^{\mu}\\
        &\qquad\qquad + \Big( 
        \prescript{}{F}{\widehat{\mathcal{A}}}_{BB,2}^{1, \text{break}} 
        + \prescript{}{F}{\widehat{\mathcal{A}}}_{BB,4}^{1, \text{break}}
        \Big) \,
        \widehat{p}^2 \overline{p}^{\mu} \bigg]\\
        &+ \frac{g^2}{16 \pi^2}
        {\mathcal{F}}_{BB}^{1, \text{break}} \, 
        \overline{p}^2 \, \overline{p}^{\mu}.
    \end{aligned}
\end{equation}
It can be seen that contracting Eq.\ (\ref{Eq:GaugeBosonSelfEnergy_1-Loop}) 
with $p_{\nu}$ reproduces the divergent part of the result above in Eq.\
(\ref{Eq:GhostGaugeBoson_1-Loop}), and thus the divergences satisfy 
the relation of Eq.\ (\ref{Eq:STI-QAP-Bc}).

\subsubsection{3-Point Functions}

For the 3-point Green functions, we have five
contributing Green functions,
excluding the two listed in the second row of Tab.\ 
\ref{Tab:NonContributingDeltaGreenFunctions},
which are illustrated below.

\vspace{0.2cm}\noindent\textbf{(xii) Ghost-Fermion-Fermion Contribution:}\vspace{0.1cm}

\noindent
The BRST breaking attributed to the fermion self energy and the fermion-gauge boson
interaction current according to Eq.\ (\ref{Eq:STI-QAP-FFc}) is given by
\begin{equation}\label{Eq:FbarFc-GreenFunc}
    \begin{aligned}
        &i \Big( \widehat{\Delta} \cdot \widetilde{\Gamma} \Big)_{\psi_i(p_2) \overline{\psi}_j(p_1) c(q)}^{1} =\\
        &- \frac{g}{16 \pi^2} \, \frac{1}{\epsilon} 
        \bigg\{
        \widehat{\slashed{p}}_1
        \bigg[
        \Big(
        \Big(\widehat{\mathcal{A}}_{\psi\overline{\psi}}^{1, \text{break}}\mathcal{Y}_{R}\Big)_{ji}
        - \widehat{\mathcal{A}}_{\psi\overline{\psi}B,ji}^{1, \text{break}}
        \Big) 
        \mathbb{P}_{\mathrm{R}}
        +
        \Big(
        \Big(\mathcal{Y}_{L}\widehat{\mathcal{A}}_{\psi\overline{\psi}}^{1, \text{break}}\Big)^{\dagger}_{ij}
        - {\widehat{\mathcal{A}}{}_{\psi\overline{\psi}B,ij}^{1, \text{break}}}^{\dagger}
        \Big) 
        \mathbb{P}_{\mathrm{L}}
        \bigg]\\
        &\qquad\qquad + 
        \widehat{\slashed{p}}_2
        \bigg[
        \Big(
        \Big(\mathcal{Y}_{L}\widehat{\mathcal{A}}_{\psi\overline{\psi}}^{1, \text{break}}\Big)_{ji}
        - \widehat{\mathcal{A}}_{\psi\overline{\psi}B,ji}^{1, \text{break}}
        \Big)
        \mathbb{P}_{\mathrm{R}}
        +
        \Big(
        \Big(\widehat{\mathcal{A}}_{\psi\overline{\psi}}^{1, \text{break}}\mathcal{Y}_{R}\Big)^{\dagger}_{ij}
        - {\widehat{\mathcal{A}}{}_{\psi\overline{\psi}B,ij}^{1, \text{break}}}^{\dagger}
        \Big) 
        \mathbb{P}_{\mathrm{L}}
        \bigg] \bigg\}\\
        &+ \frac{g}{16 \pi^2} \, \Big( \overline{\slashed{p}}_1 + \overline{\slashed{p}}_2 \Big) 
        \bigg\{
        \mathcal{F}_{\psi\overline{\psi}B,\mathrm{R},ji}^{1, \text{break}}
        \mathbb{P}_{\mathrm{R}}
        + \mathcal{F}_{\psi\overline{\psi}B,\mathrm{L},ji}^{1, \text{break}} 
        \mathbb{P}_{\mathrm{L}}
        \bigg\}.
    \end{aligned}
\end{equation}

\vspace{0.2cm}\noindent\textbf{(xiii) Ghost-double Gauge Boson Contribution:}\vspace{0.1cm}

\noindent
The ghost-double gauge boson 3-point Green function, associated with the well-known ABJ-anomaly 
via the Slavnov-Taylor identity in Eq.\ (\ref{Eq:STI-QAP-BBc}), yields a vanishing contribution upon applying  the anomaly cancellation condition from Eq.\ (\ref{Eq:AnomalyCancellationCondition}),
as expected. 
Hence, we have
\begin{equation}\label{Eq:GhostDoubleGaugeBoson_1-Loop}
    \begin{aligned}
        i \Big( \widehat{\Delta} \cdot \widetilde{\Gamma} \Big)_{B_{\nu}(p_2)B_{\mu}(p_1)c(q)}^{1} &=
        - \frac{ig^3}{16\pi^2} \, \frac{4}{3} \, \Big(\mathrm{Tr}\big(\mathcal{Y}_R^3\big)-\mathrm{Tr}\big(\mathcal{Y}_L^3\big)\Big) \, \overline{\varepsilon}^{\mu\nu\rho\sigma} \, \overline{p}_{1,\rho} \, \overline{p}_{2,\sigma} = 0.
    \end{aligned}
\end{equation}

\vspace{0.2cm}\noindent\textbf{(xiv) Ghost-double Scalar Contribution:}\vspace{0.1cm}

\noindent
For the BRST breaking attributed to the scalar 2-point and the scalar-gauge boson
3-point Green functions according to Eqs.\ (\ref{Eq:STI-QAP-SSbarc}), (\ref{Eq:STI-QAP-SSc})
and (\ref{Eq:STI-QAP-SbarSbarc}), we find
\begin{align}
    \begin{split}\label{Eq:GhostDoubleScalar_1-Loop}
        i \Big( &\widehat{\Delta} \cdot \widetilde{\Gamma} \Big)_{\phi_b(p_2)\phi^{\dagger}_a(p_1) c(q)}^{1}\\
        = &- \frac{g}{16 \pi^2} \, \frac{1}{\epsilon} \, 
        \Big(
        {\widehat{\mathcal{A}}}_{\phi\phi^{\dagger}B,ab}^{1, \text{break}} 
        - \mathcal{Y}_{S} 
        {\widehat{\mathcal{A}}}_{\phi\phi^{\dagger},ab}^{1, \text{break}}
        \Big)
        \Big( \widehat{p}_1^2 - \widehat{p}_2^2 \Big)\\
        &+ \frac{g}{16 \pi^2} \, {\mathcal{F}}_{\phi\phi^{\dagger}B,ab}^{1,\text{break}} \, \Big( \overline{p}_1^2 - \overline{p}_2^2 \Big),
    \end{split}\\[1.5ex]
    \begin{split}\label{Eq:Ghostphiphi_1-Loop}
        i \Big( &\widehat{\Delta} \cdot \widetilde{\Gamma} \Big)_{\phi_b(p_2)\phi_a(p_1)c(q)}^{1}\\
        &= \frac{g}{16 \pi^2} \, \frac{1}{\epsilon} \,
        \Big[
        \Big(
        \mathcal{Y}_{S} 
        {\widehat{\mathcal{A}}}_{\phi\phi,ab}^{1, \text{break}}
        - {\widehat{\mathcal{A}}}_{\phi\phi B,ab}^{1, \text{break}}
        \Big) \, \widehat{p}_1^2
        +
        \Big(
        \mathcal{Y}_{S} 
        {\widehat{\mathcal{A}}}_{\phi\phi,ab}^{1, \text{break}}
        + {\widehat{\mathcal{A}}}_{\phi\phi B,ab}^{1, \text{break}}
        \Big) \, \widehat{p}_2^2
        \Big]\\
        &+ \,
        \frac{g}{16 \pi^2} \, 4 \, \mathcal{Y}_{S} \, {\mathcal{F}}_{\phi\phi,ab}^{1, \text{break}}
        \Big(\overline{p}_1^2 + \overline{p}_2^2 + \frac{3}{2} \, \overline{p}_1 \cdot \overline{p}_2\Big),
    \end{split}\\[1.5ex]
    \begin{split}\label{Eq:Ghostphibarphibar_1-Loop}
        i \Big( &\widehat{\Delta} \cdot \widetilde{\Gamma} \Big)_{\phi^{\dagger}_b(p_2)\phi^{\dagger}_a(p_1)c(q)}^{1}\\
        = &-\frac{g}{16 \pi^2} \, \frac{1}{\epsilon} \,
        \Big[
        \Big(
        \mathcal{Y}_{S} \,
        {\widehat{\mathcal{A}}{}_{\phi\phi,ab}^{1, \text{break}}}^{\dagger}
        - {\widehat{\mathcal{A}}{}_{\phi\phi B,ab}^{1, \text{break}}}^{\dagger}
        \Big) \, \widehat{p}_1^2
        +
        \Big(
        \mathcal{Y}_{S} \, 
        {\widehat{\mathcal{A}}{}_{\phi\phi,ab}^{1, \text{break}}}^{\dagger}
        + {\widehat{\mathcal{A}}{}_{\phi\phi B,ab}^{1, \text{break}}}^{\dagger}
        \Big) \, \widehat{p}_2^2
        \Big]\\
        &-\frac{g}{16 \pi^2} \, 4 \, \mathcal{Y}_{S} \, {{\mathcal{F}}_{\phi\phi,ab}^{1, \text{break}}}^{\dagger}
        \Big(\overline{p}_1^2 + \overline{p}_2^2 + \frac{3}{2} \, \overline{p}_1 \cdot \overline{p}_2\Big),
    \end{split}
\end{align}
where the latter two, shown in Eqs.\ (\ref{Eq:Ghostphiphi_1-Loop}) and (\ref{Eq:Ghostphibarphibar_1-Loop}), 
break global hypercharge conservation, cf.\ Sec.\ \ref{Sec:1-Loop-Div-GreenFuncs}.
Further, we see the first finite contributions that violate global hypercharge conservation, governed 
by the coefficient ${\mathcal{F}}_{\phi\phi,ab}^{1, \text{break}}$.

\subsubsection{4-Point Functions}

There are seven 4-point Green functions,
provided in the following, that yield a
non-vanishing contribution.
The six Green functions provided in the third and fourth row of Tab.\ 
\ref{Tab:NonContributingDeltaGreenFunctions} are excluded from this list.

\vspace{0.2cm}\noindent\textbf{(xv) Ghost-triple Gauge Boson Contribution:}\vspace{0.1cm}

\noindent
We obtain
\begin{equation}\label{Eq:BBBc-GreenFunc}
    \begin{aligned}
        i \Big( \widehat{\Delta} \cdot \widetilde{\Gamma}& \Big)_{B_{\rho}(p_3)B_{\nu}(p_2)B_{\mu}(p_1)c(q)}^{1}\\
        &= \frac{g^4}{16 \pi^2} 
        {\mathcal{F}}_{BBBB}^{1,\text{break}}\,
        \big(\overline{p}_1 + \overline{p}_2 + \overline{p}_3\big)_{\sigma} \, 
        \Big( \overline{\eta}^{\rho\sigma} \, \overline{\eta}^{\mu\nu}
        + \overline{\eta}^{\nu\sigma} \, \overline{\eta}^{\mu\rho}
        + \overline{\eta}^{\mu\sigma} \, \overline{\eta}^{\nu\rho} \Big)
    \end{aligned}
\end{equation}
for the BRST breaking associated with the quartic 
gauge boson Green function, cf.\ Eq.\ (\ref{Eq:STI-QAP-BBBc}).
Analogous to the case of chiral QED, the entire contribution to the quartic 
gauge boson Green function in case of $\hypRL=\hypLR$ is finite and BRST-breaking,
with no contributions from scalars, as seen above in Eq.\
(\ref{Eq:BBBc-GreenFunc}) and Eq.\ (\ref{Eq:BBBB-GreenFunc}).
In general, e.g.\ for evanescent hypercharges such as those given in Eq.\ \eqref{Eq:ASSM-YLR-YRL-Option2},
there is an additional divergent, BRST-breaking contribution governed by these evanescent hypercharges,
as demonstrated below in Sec.\ \ref{Sec:AnalysisOfASSMResults}.

\vspace{0.2cm}\noindent\textbf{(xvi) Ghost-Gauge Boson-Fermion-Fermion Contribution:}\vspace{0.1cm}

\noindent
There is another divergent BRST breaking associated to the fermion-gauge boson interaction, as indicated in Eq.\ (\ref{Eq:STI-QAP-FFbarBc}) and given by
\begin{equation}\label{Eq:FFbarBc-GreenFunc}
    \begin{aligned}
        i \Big( \widehat{\Delta} \cdot \widetilde{\Gamma} \Big)_{\psi_i(p_2) \overline{\psi}_j(p_3) B_{\mu}(p_1) c(q)}^{1}
        = \frac{g^2}{16 \pi^2} \, \frac{1}{\epsilon} \, \widehat{\gamma}^{\mu} 
        \bigg[
        &\Big(
        \mathcal{Y}_{L}\,\widehat{\mathcal{A}}_{\psi\overline{\psi}B}^{1,\text{break}}
        - \widehat{\mathcal{A}}_{\psi\overline{\psi}B}^{1,\text{break}}\,\mathcal{Y}_{R}
        \Big)_{ji} \,
        \mathbb{P}_{\mathrm{R}}\\
        - &\Big(
        \mathcal{Y}_{L}\,\widehat{\mathcal{A}}_{\psi\overline{\psi}B}^{1,\text{break}}
        - \widehat{\mathcal{A}}_{\psi\overline{\psi}B}^{1,\text{break}}\,\mathcal{Y}_{R}
        \Big)_{ij}^{\dagger} \,
        \mathbb{P}_{\mathrm{L}}
        \bigg],
    \end{aligned}
\end{equation}
which is purely evanescent.
Based on the coefficients provided in  Eqs.\
(\ref{App-Eq:DivCoeffRelations}), (\ref{App-Eq:FermionCoeffFermionGauge}) 
and (\ref{App-Eq:FermionGaugeBosonCoeff}),
we observe that there are both fermionic and Yukawa contributions arising from diagrams with the
$\widehat{\Delta}_2\big[c,B,\overline{\psi},\psi\big]$-vertex, while there
are only Yukawa contributions originating from diagrams containing the 
$\widehat{\Delta}_1\big[c,\overline{\psi},\psi\big]$-vertex, having
used the notation established in Eq.\ (\ref{Eq:TreeLevelBreaking}).
Therefore, this Green function vanishes in absence of scalars and with 
$\hypLR\equiv0$, as is the case of chiral QED.

\vspace{0.2cm}\noindent\textbf{(xvii) Ghost-Yukawa Contribution:}\vspace{0.1cm}

\noindent
The Yukawa interaction only gives rise to a finite BRST breaking, 
taking the form
\begin{align}
    \begin{split}\label{Eq:YukawaDeltaGreenFunc1-cSFF}
        i \Big( \widehat{\Delta} \cdot \widetilde{\Gamma} \Big)_{\psi_i(p_2) \overline{\psi}_j(p_3) \phi_{a}(p_1) c(q)}^{1}&\\ 
        = - \frac{g}{16 \pi^2} \, 
        \bigg[
        &\Big(
        \mathcal{Y}_{L}\mathcal{F}_{\psi\overline{\psi}\phi}^{1,\text{break}}
        -\mathcal{F}_{\psi\overline{\psi}\phi}^{1,\text{break}}\mathcal{Y}_{R}
        -\mathcal{Y}_{S}\mathcal{F}_{\psi\overline{\psi}\phi}^{1,\text{break}}
        \Big)^{a}_{ji} \, \mathbb{P}_{\mathrm{R}}\\
        -
        &\Big(
        \mathcal{Y}_{L}\mathcal{F}_{\psi\overline{\psi}\phi^{\dagger}}^{1,\text{break}}
        -\mathcal{F}_{\psi\overline{\psi}\phi^{\dagger}}^{1,\text{break}}\mathcal{Y}_{R}
        +\mathcal{Y}_{S}\mathcal{F}_{\psi\overline{\psi}\phi^{\dagger}}^{1,\text{break}}
        \Big)^{a \, \dagger}_{ij} \, \mathbb{P}_{\mathrm{L}}
        \bigg],
    \end{split}\\[1.5ex]
    \begin{split}\label{Eq:YukawaDeltaGreenFunc2-cSbarFF}
        i \Big( \widehat{\Delta} \cdot \widetilde{\Gamma} \Big)_{\psi_i(p_2) \overline{\psi}_j(p_3) \phi^{\dagger}_{a}(p_1) c(q)}^{1}&\\
        = - \frac{g}{16 \pi^2} \, 
        \bigg[
        &\Big(
        \mathcal{Y}_{L}\mathcal{F}_{\psi\overline{\psi}\phi^{\dagger}}^{1,\text{break}}
        -\mathcal{F}_{\psi\overline{\psi}\phi^{\dagger}}^{1,\text{break}}\mathcal{Y}_{R}
        +\mathcal{Y}_{S}\mathcal{F}_{\psi\overline{\psi}\phi^{\dagger}}^{1,\text{break}}
        \Big)^{a}_{ji} \, \mathbb{P}_{\mathrm{R}}\\
        -
        &\Big(
        \mathcal{Y}_{L}\mathcal{F}_{\psi\overline{\psi}\phi}^{1,\text{break}}
        -\mathcal{F}_{\psi\overline{\psi}\phi}^{1,\text{break}}\mathcal{Y}_{R}
        -\mathcal{Y}_{S}\mathcal{F}_{\psi\overline{\psi}\phi}^{1,\text{break}}
        \Big)^{a\,\dagger}_{ij} \, \mathbb{P}_{\mathrm{L}}
        \bigg].
    \end{split}
\end{align}
The coefficients used in Eqs.\ (\ref{Eq:YukawaDeltaGreenFunc1-cSFF})
and (\ref{Eq:YukawaDeltaGreenFunc2-cSbarFF}) and displayed in Eq.\ (\ref{App-Eq:FiniteYukawaCoeffs}), are 
chosen in such a way that the terms in the finite counterterm action, cf.\ Eq.\ 
(\ref{Eq:Sfct_1-Loop}), whose BRST transformation produce the corresponding 
terms in $\DeltaHat\cdot\widetilde{\Gamma}$, take a simple and compact form, which 
can directly be read off from Eqs.\ (\ref{Eq:YukawaDeltaGreenFunc1-cSFF})
and (\ref{Eq:YukawaDeltaGreenFunc2-cSbarFF}).

\vspace{0.2cm}\noindent\textbf{(xviii) Ghost-Gauge boson-double Scalar Contribution:}\vspace{0.1cm}

\noindent
The BRST breakings associated with the triple and quartic scalar-gauge boson interactions
according to Eqs.\ (\ref{Eq:STI-QAP-SSbarBc}) to (\ref{Eq:STI-QAP-SbarSbarBc}) are given by
\begin{equation}
    \begin{aligned}\label{Eq:GreenFunc-c-B-phiDagger-phi}
        i \Big( \widehat{\Delta} &\cdot \widetilde{\Gamma} \Big)_{\phi_b(p_3) \phi_a^{\dagger}(p_2) B_{\mu}(p_1) c(q)}^{1}\\
        &= \frac{g^2}{16 \pi^2} \,
        \frac{1}{\epsilon}
        \Big(
        2 \,\mathcal{Y}_{S} {\widehat{\mathcal{A}}}_{\phi\phi^{\dagger}B,ab}^{1,\text{break}} - {\widehat{\mathcal{A}}}_{\phi\phi^{\dagger}BB,ab}^{1,\text{break}}\Big) 
        \Big(\widehat{p}_1^{\mu} + \widehat{p}_2^{\mu} + \widehat{p}_3^{\mu}\Big)\\
        &- \frac{g^2}{16 \pi^2}
        \Big(2\,\mathcal{Y}_{S}\, {\mathcal{F}}_{\phi\phi^{\dagger}B,ab}^{1, \text{break}} 
        - {\mathcal{F}}_{\phi\phi^{\dagger}BB,ab}^{1, \text{break}}\Big) 
        \Big(\overline{p}_1^{\mu} + \overline{p}_2^{\mu} + \overline{p}_3^{\mu}\Big),
    \end{aligned}
\end{equation}
\begin{equation}
    \begin{alignedat}{2}
        i \Big( \widehat{\Delta} &\cdot \widetilde{\Gamma} \Big)_{\phi_b(p_3) \phi_a(p_2) B_{\mu}(p_1) c(q)}^{1}
        &&\\
        &= \frac{g^2}{16 \pi^2} \, 
        \frac{1}{\epsilon}
        \bigg[
        - {\widehat{\mathcal{A}}}_{\phi\phi BB,ab}^{1, \text{break}} \, \widehat{p}_1^{\mu}
        &&+ \Big(
        2 \, \mathcal{Y}_{S} {\widehat{\mathcal{A}}}_{\phi\phi B,ab}^{1, \text{break}}
        - {\widehat{\mathcal{A}}}_{\phi\phi BB,ab}^{1, \text{break}}
        \Big) \, \widehat{p}_2^{\mu}\\
        &
        &&- \Big(
        2 \, \mathcal{Y}_{S} {\widehat{\mathcal{A}}}_{\phi\phi B,ab}^{1, \text{break}}
        + {\widehat{\mathcal{A}}}_{\phi\phi BB,ab}^{1, \text{break}}
        \Big) \, \widehat{p}_3^{\mu}
        \bigg]\\
        &-\mathrlap{\frac{g^2}{16 \pi^2} \, 6 \, \mathcal{Y}_{S}^2
        \, {\mathcal{F}}_{\phi\phi,ab}^{1,\text{break}} \,
        \Big(\overline{p}_2^{\mu} + \overline{p}_3^{\mu}\Big),}
        &&
    \end{alignedat}
\end{equation}
\begin{equation}
    \begin{alignedat}{2}
        i \Big( \widehat{\Delta} &\cdot \widetilde{\Gamma} \Big)_{\phi_b^{\dagger}(p_3) \phi_a^{\dagger}(p_2) B_{\mu}(p_1) c(q)}^{1}
        &&\\
        &= \frac{g^2}{16 \pi^2} \, 
        \frac{1}{\epsilon}
        \bigg[
        - {\widehat{\mathcal{A}}{}_{\phi\phi BB,ab}^{1, \text{break}}}^{\!\!\!\!\!\dagger} \,\,\, \widehat{p}_1^{\mu}
        &&+ \Big(
        2 \, \mathcal{Y}_{S} \, {\widehat{\mathcal{A}}{}_{\phi\phi B,ab}^{1, \text{break}}}^{\dagger}
        - {\widehat{\mathcal{A}}{}_{\phi\phi BB,ab}^{1, \text{break}}}^{\!\!\!\!\!\dagger} \,\,
        \Big) \, \widehat{p}_2^{\mu}\\
        &
        &&- \Big(
        2 \, \mathcal{Y}_{S} \, {\widehat{\mathcal{A}}{}_{\phi\phi B,ab}^{1, \text{break}}}^{\dagger}
        + {\widehat{\mathcal{A}}{}_{\phi\phi BB,ab}^{1, \text{break}}}^{\!\!\!\!\!\dagger} \,\,
        \Big) \, \widehat{p}_3^{\mu}
        \bigg]\\
        &- \mathrlap{\frac{g^2}{16 \pi^2} \, 6 \, \mathcal{Y}_{S}^2
        \, {{\mathcal{F}}_{\phi\phi,ab}^{1,\text{break}}}^{\dagger} \,
        \Big(\overline{p}_2^{\mu} + \overline{p}_3^{\mu}\Big).}
        &&
    \end{alignedat}
\end{equation}
Evidently, these contributions again come along with both finite and 
divergent breakings of not only BRST invariance but also global hypercharge conservation.

\subsubsection{5-Point Functions}

Finally, because the ghost has mass dimension zero,
there are also power-counting divergent 5-point Green functions
that provide non-vanishing contributions. 
Excluding the seven Green functions in the last three rows of Tab.\
\ref{Tab:NonContributingDeltaGreenFunctions}, we list
the remaining eight 5-point Green functions below.

\vspace{0.2cm}\noindent\textbf{(ixx) Ghost-double Gauge boson-double Scalar Contribution:}\vspace{0.1cm}

\noindent
Next to item (xviii), further BRST-breaking contributions associated to the quartic scalar-gauge boson
interaction, as described by the relations in Eqs.\ (\ref{Eq:STI-QAP-SSbarBBc}) to (\ref{Eq:STI-QAP-SbarSbarBBc}),
are provided by
\begin{align}
    \begin{split}\label{Eq:SSbarBBc-DeltaGreenFunc}
        i \Big( \widehat{\Delta} \cdot \widetilde{\Gamma} &\Big)_{\phi_b(p_4) \phi_a^{\dagger}(p_3) B_{\nu}(p_2) B_{\mu}(p_1) c(q)}^{1}
        = 0,
    \end{split}\\[1.5ex]
    \begin{split}\label{Eq:SSBBc-DeltaGreenFunc}
        i \Big( \widehat{\Delta} \cdot \widetilde{\Gamma} &\Big)_{\phi_b(p_4) \phi_a(p_3) B_{\nu}(p_2) B_{\mu}(p_1) c(q)}^{1}\\
        &= \frac{g^3}{16 \pi^2}
        \bigg[
        \frac{2\,\mathcal{Y}_{S}}{\epsilon}
        {\widehat{\mathcal{A}}}_{\phi\phi BB,ab}^{1, \text{break}} \, \widehat{\eta}^{\mu\nu}
        + 
        12 \, \mathcal{Y}_{S}^3 \,
        {\mathcal{F}}_{\phi\phi,ab}^{1,\text{break}} \, \overline{\eta}^{\mu\nu}
        \bigg],
    \end{split}\\[1.5ex]
    \begin{split}\label{Eq:SbarSbarBBc-DeltaGreenFunc}
        i \Big( \widehat{\Delta} \cdot \widetilde{\Gamma} &\Big)_{\phi_b^{\dagger}(p_4) \phi_a^{\dagger}(p_3) B_{\nu}(p_2) B_{\mu}(p_1) c(q)}^{1}\\
        &= - \frac{g^3}{16 \pi^2}
        \bigg[
        \frac{2\,\mathcal{Y}_{S}}{\epsilon}\,
       {\widehat{\mathcal{A}}{}_{\phi\phi BB,ab}^{1, \text{break}}}^{\!\!\!\!\!\dagger} \,\,\, \widehat{\eta}^{\mu\nu}
        + 
        12 \, \mathcal{Y}_{S}^3 \,
        {{\mathcal{F}}_{\phi\phi,ab}^{1,\text{break}}}^{\dagger} \, \overline{\eta}^{\mu\nu}
        \bigg].
    \end{split}
\end{align}
Since the local part of the Green functions on the LHS of Eq.\ (\ref{Eq:STI-QAP-SSbarBBc}) is
independent of momenta, the associated breaking on the RHS must vanish.
This is confirmed by the result given in Eq.\ (\ref{Eq:SSbarBBc-DeltaGreenFunc}),
obtained by direct calculation.
In contrast, the two Green functions that violate global hypercharge conservation
may generally result in non-vanishing divergent and finite contributions, as provided in
Eqs.\ \eqref{Eq:SSBBc-DeltaGreenFunc} and \eqref{Eq:SbarSbarBBc-DeltaGreenFunc}.

\vspace{0.2cm}\noindent\textbf{(xx) Ghost-quartic Scalar Contribution:}\vspace{0.1cm}

\noindent
In contrast to the divergent part of the quartic scalar interaction illustrated in
Eqs.\ (\ref{Eq:SSSbarSbar-GreenFunc}) to (\ref{Eq:SbarSbarSbarSbar-GreenFunc}), the 
associated BRST breaking vanishes for the hypercharge-conserving Green function, while there
are non-zero finite hypercharge-violating contributions, as demonstrated below.
\begin{align}
    \begin{split}\label{Eq:SSSbarSbarc-DeltaGreenFunc}
        i \Big( \widehat{\Delta} \cdot \widetilde{\Gamma} \Big)_{\phi_d(p_4) \phi_c(p_3) \phi_b^{\dagger}(p_2) \phi_a^{\dagger}(p_1) c(q)}^{1}
        &= 0,
    \end{split}\\[1.5ex]
    \begin{split}\label{Eq:SSSS-DeltaGreenFunc}
        i \Big( \widehat{\Delta} \cdot \widetilde{\Gamma} \Big)_{\phi_d(p_4) \phi_c(p_3) \phi_b(p_2) \phi_a(p_1) c(q)}^{1}
        &= \frac{g}{16 \pi^2} \, \mathcal{Y}_{S} \, {\mathcal{F}}_{\phi\phi\phi\phi,abcd}^{1,\text{break}},
    \end{split}\\[1.5ex]
    \begin{split}
        i \Big( \widehat{\Delta} \cdot \widetilde{\Gamma} \Big)_{\phi_d(p_4) \phi_c(p_3) \phi_b(p_2) \phi_a^{\dagger}(p_1) c(q)}^{1}
        &= \frac{g}{16 \pi^2} \, \mathcal{Y}_{S} \, {\mathcal{F}}_{\phi^{\dagger}\phi\phi\phi,abcd}^{1,\text{break}},
    \end{split}\\[1.5ex]
    \begin{split}
        i \Big( \widehat{\Delta} \cdot \widetilde{\Gamma} \Big)_{\phi_d(p_4) \phi_c^{\dagger}(p_3) \phi_b^{\dagger}(p_2) \phi_a^{\dagger}(p_1) c(q)}^{1}
        &= - \frac{g}{16 \pi^2} \, \mathcal{Y}_{S} \, \Big({{\mathcal{F}}_{\phi^{\dagger}\phi\phi\phi}^{1,\text{break}}}^{\dagger}\Big)_{dbca},
    \end{split}\\[1.5ex]
    \begin{split}\label{Eq:SbarSbarSbarSbar-DeltaGreenFunc}
        i \Big( \widehat{\Delta} \cdot \widetilde{\Gamma} \Big)_{\phi_d^{\dagger}(p_4) \phi_c^{\dagger}(p_3) \phi_b^{\dagger}(p_2) \phi_a^{\dagger}(p_1) c(q)}^{1}
        &= - \frac{g}{16 \pi^2} \, \mathcal{Y}_{S} \, \Big({{\mathcal{F}}_{\phi\phi\phi\phi}^{1,\text{break}}}^{\dagger}\Big)_{abcd}.
    \end{split}
\end{align}
The global hypercharge-conserving $\Delta$-inserted Green function
in Eq.\ (\ref{Eq:SSSbarSbarc-DeltaGreenFunc}) vanishes for the same reason as
the ghost-double scalar-double gauge boson Green function in Eq.\ (\ref{Eq:SSbarBBc-DeltaGreenFunc})
discussed above, cf.\ Eq.\ (\ref{Eq:STI-QAP-SSSbarSbarc}).

\subsection{One-Loop Singular Counterterm Action}\label{Sec:SingularCTAction-1Loop}

Here we present the 1-loop  singular counterterm action which exactly cancels 
all divergent contributions of the Green functions given in Sec.\ \ref{Sec:1-Loop-Div-GreenFuncs}.
 We decompose it into a BRST-invariant and
a non-invariant, symmetry-restoring part as
\begin{equation}\label{Eq:Ssct_1-Loop}
    \begin{aligned}
        S^1_{\mathrm{sct}} = S^1_{\mathrm{sct,inv}} + S^1_{\mathrm{sct,break}}.
    \end{aligned}
\end{equation}
The BRST-invariant part is found to be
\begin{equation}\label{Eq:Ssct_1-Loop_inv}
    \begin{aligned}
        S^1_{\mathrm{sct,inv}} = 
        - \frac{1}{16\pi^2} \, \frac{1}{\epsilon} &\Dintx 
        \bigg[  
        g^2 \! \prescript{}{F}{\overline{\mathcal{A}}}_{BB}^{1,\text{inv}}
        \Big(-\frac{1}{4}\overline{F}^{\mu\nu}\overline{F}_{\mu\nu}\Big)
        + g^2
        \prescript{}{S}{\mathcal{A}}_{BB}^{1,\text{inv}}
        \Big(-\frac{1}{4}F^{\mu\nu}F_{\mu\nu}\Big)\\
        &+ \overline{\mathcal{A}}_{\psi\overline{\psi},\mathrm{R},kj}^{1,\text{inv}} \overline{\psi}_i i \overline{\slashed{D}}_{\mathrm{R},ik} \psi_j
        + \overline{\mathcal{A}}_{\psi\overline{\psi},\mathrm{L},kj}^{1,\text{inv}} \overline{\psi}_i i \overline{\slashed{D}}_{\mathrm{L},ik} \psi_j\\
        &+ \Big( \overline{\psi}_i 
        \Big[
        \overline{\mathcal{A}}_{\psi\overline{\psi}\phi,\mathrm{R},ij}^{1,\text{inv},a} \,
        \phi_a
        + 
        {\overline{\mathcal{A}}_{\psi\overline{\psi}\phi,\mathrm{L},ji}^{1,\text{inv},a}}^{\hspace{-0.29cm}\dagger} \hspace{0.2cm} 
        \phi_a^{\dagger}
        \Big]
        \mathbb{P}_{\mathrm{R}}
        \psi_j 
        + \mathrm{h.c.}\Big)\\
        &+ g^2 \! \prescript{}{S}{\mathcal{A}}_{\phi\phi^{\dagger},ab}^{1,\text{inv}} 
        \big(D^{\mu}\phi_a\big)^{\dagger}\big(D_{\mu}\phi_b\big)
        + \prescript{}{Y}{\overline{\mathcal{A}}}_{\phi\phi^{\dagger},ab}^{1,\text{inv}}
        \big(\overline{D}^{\mu}\phi_a\big)^{\dagger}\big(\overline{D}_{\mu}\phi_b\big)\\
        &+ \frac{1}{4} \, \mathcal{A}_{\phi\phi^{\dagger}\phi\phi^{\dagger},abcd}^{1,\text{inv}}
        \, \phi_a^{\dagger} \phi_b \phi_c^{\dagger} \phi_d
        \bigg],
    \end{aligned}
\end{equation}
with covariant derivatives
\begin{equation}
    \begin{aligned}
        \overline{D}^{\mu}_{\mathrm{L/R},ij} &=
        \Big( \overline{\partial}^{\mu} \delta_{ij} + i g \mathcal{Y}_{L/R,ij} \overline{B}^{\mu} \Big) \mathbb{P}_{\mathrm{L/R}},\\
        D^{\mu} \phi_a &=
        \Big( \partial^{\mu} + i g \hypS B^{\mu} \Big) \phi_a.
    \end{aligned}
\end{equation}
This part of the counterterm action in Eq.\ (\ref{Eq:Ssct_1-Loop_inv}) is clearly invariant under the 
BRST transformations  Eq.\ (\ref{Eq:Ddim-BRSTTrafos}), as it only consists of the original
 symmetric field combinations appearing in the tree-level Lagrangian discussed in Sec.\ \ref{Sec:D-DimLagrangian}.
In other words, Eq.\ (\ref{Eq:Ssct_1-Loop_inv}) comprises standard counterterms that can be obtained 
via a multiplicative renormalisation transformation and which are also expected in a naive calculation
that avoids the spurious breaking of BRST invariance at intermediate steps.

The non-symmetric, symmetry-restoring part of the singular counterterm action is given by
\begin{align}\label{Eq:Ssct_1-Loop_break}
        S&^1_{\mathrm{sct,break}} = \frac{1}{16\pi^2} \frac{1}{\epsilon} \Dintx
        \bigg[
        \frac{g^2}{2}
        \bigg\{ \!
        \Big(
        \prescript{}{F}{\widehat{\mathcal{A}}}_{BB,1}^{1,\text{break}}
        + \prescript{}{F}{\widehat{\mathcal{A}}}_{BB,3}^{1,\text{break}}
        \Big)
        \widehat{B}_{\mu}\overline{\Box}\widehat{B}^{\mu}
        + 
        \prescript{}{F}{\widehat{\mathcal{A}}}_{BB,4}^{1,\text{break}}
        \overline{B}_{\mu}\widehat{\Box}\overline{B}^{\mu}
        \nonumber\\
        &+
        \frac{2}{3} \prescript{}{F}{\widehat{\mathcal{A}}}_{BB,3}^{1,\text{break}}
        \widehat{B}_{\mu}\widehat{\Box}\widehat{B}^{\mu}
        - 2 \prescript{}{F}{\widehat{\mathcal{A}}}_{BB,2}^{1,\text{break}}
        \big(\overline{\partial}\cdot\overline{B}\big) \big(\widehat{\partial}\cdot\widehat{B}\big)
        - 2 \prescript{}{F}{\widehat{\mathcal{A}}}_{BB,1}^{1,\text{break}}
        \big(\widehat{\partial}\cdot\widehat{B}\big)^2
        \bigg\}
        \nonumber\\
        &- \Big[ \overline{\psi}_i i
        \Big(
        \widehat{\slashed{\partial}} \widehat{\mathcal{A}}_{\psi\overline{\psi},ij}^{1,\text{break}}
        + i g \widehat{\slashed{B}} \widehat{\mathcal{A}}_{\psi\overline{\psi}B,ij}^{1,\text{break}}
        \Big) \projR
        \psi_j + \mathrm{h.c.} \Big]
        \\
        &+
        {\widehat{\mathcal{A}}}_{\phi\phi^{\dagger},ab}^{1,\text{break}} \phi_{a}^{\dagger} \widehat{\Box} \phi_b
        - i g {\widehat{\mathcal{A}}}_{\phi\phi^{\dagger}B,ab}^{1,\text{break}}
        \Big[ \big(\widehat{\partial}^{\mu}\phi_{a}^{\dagger}\big)\phi_b 
        - \phi_{a}^{\dagger}\big(\widehat{\partial}^{\mu}\phi_b\big) \Big] \widehat{B}_{\mu}
        - \frac{g^2}{2} {\widehat{\mathcal{A}}}_{\phi\phi^{\dagger}BB,ab}^{1,\text{break}}
        \phi_{a}^{\dagger}\phi_{b} \widehat{B}^{\mu} \widehat{B}_{\mu}
        \nonumber\\
        &+
        \frac{1}{2} \bigg\{ \!
        {\widehat{\mathcal{A}}}_{\phi\phi,ab}^{1,\text{break}} \phi_{a} \widehat{\Box} \phi_b
        - i g {\widehat{\mathcal{A}}}_{\phi\phi B,ab}^{1,\text{break}}
        \Big[ \big(\widehat{\partial}^{\mu}\phi_{a}\big)\phi_b
        - \phi_{a}\big(\widehat{\partial}^{\mu}\phi_b\big) \Big] \widehat{B}_{\mu}
        \nonumber\\
        &
        - \frac{g^2}{2} {\widehat{\mathcal{A}}}_{\phi\phi BB,ab}^{1,\text{break}}
        \phi_{a} \phi_{b} \widehat{B}^{\mu} \widehat{B}_{\mu} + \mathrm{h.c.}
        \bigg\}
        \bigg].\nonumber
\end{align}
All these singular non-symmetric 1-loop counterterms are evanescent, as expected.
The first two lines correspond to bilinear terms in the gauge boson. Compared to chiral QED as treated in Refs.~\cite{Belusca-Maito:2021lnk,Stockinger:2023ndm} these terms involve several new structures. All of these new structures vanish for $\hypLR\equiv0$, reducing the first two lines to only the term $\sim\overline{B}_{\mu}\widehat{\Box}\overline{B}^{\mu}$.

The fermionic terms in the third line of Eq.\ (\ref{Eq:Ssct_1-Loop_break})  cannot be written as a covariant derivative, because the kinetic term and the term with the gauge boson carry different and independent coefficients.
For $\hypLR\equiv0$ these terms do not vanish, but they simplify significantly, as can be seen 
from looking at the corresponding coefficients given in 
Eqs.\ (\ref{App-Eq:DivCoeffRelations}) and (\ref{App-Eq:FermionCoeffFermionGauge}) to
(\ref{App-Eq:FermionGaugeBosonCoeff}).

The last three lines of Eq.\ (\ref{Eq:Ssct_1-Loop_break}) contain scalar and scalar-gauge boson contributions
to the BRST breaking.
Again, these terms cannot be combined to
covariant derivatives and thus break BRST invariance.
The terms in the last two lines even violate
global hypercharge conservation, which is a new phenomenon not present in Refs.~\cite{Belusca-Maito:2020ala,Belusca-Maito:2021lnk,Stockinger:2023ndm,Belusca-Maito:2023wah}.

From Sec.\ \ref{Sec:D-DimLagrangian}, we know that these global hypercharge violating contributions 
have two different origins: the evanescent part of the fermion kinetic term combining physical left- and right-handed fermions (Option \ref{Opt:Option1} as defined in Sec.~\ref{Sec:FermionsInGeneral}) and 
the evanescent gauge interactions.
The contributions to the scalar kinetic terms,
governed by ${\widehat{\mathcal{A}}}_{\phi\phi,ab}^{1,\text{break}}$,
emerge entirely from the evanescent part of the fermion kinetic term, as they are independent of $\hypLR$.
Thus, these terms would vanish only if one follows the approach of Option 
\hyperref[Opt:Option2a-2b-GroupReference]{2}, which introduces a sterile partner field for
each fermion to avoid the violation of global hypercharge that is induced by the fermion kinetic term
in $D$ dimensions.
In contrast, the contributions to the scalar-gauge boson terms, governed by 
${\widehat{\mathcal{A}}}_{\phi\phi B,ab}^{1,\text{break}}$ and ${\widehat{\mathcal{A}}}_{\phi\phi BB,ab}^{1,\text{break}}$,
arise solely from the evanescent gauge interactions, since
every term in those coefficients depends on $\hypLR$, see Eqs.\ (\ref{App-Eq:Div2ScalarCoeffs}), (\ref{App-Eq:SSBCoeff}) and (\ref{App-Eq:SSBBCoeffs}).
Therefore, these contributions would vanish for $\hypLR\equiv0$. 

In general, the divergent symmetry-restoring part of the counterterm action Eq.\ \eqref{Eq:Ssct_1-Loop_break}, 
simplifies significantly for vanishing evanescent gauge interactions.

\subsection{One-Loop Finite Symmetry-Restoring Counterterm Action}\label{Sec:FiniteCTAction-1Loop}

As discussed in Sec.\ \ref{Sec:STI+QAP}, the ultimate symmetry requirement is the validity of the Slavnov-Taylor identity after renormalisation, Eq.~(\ref{Eq:UltimateSymmetryRequirement}). It determines in particular the
finite symmetry-restoring counterterm action  $S^1_{\mathrm{fct}}$.
In an abelian gauge theory, the requirement amounts to the  
simple, implicit relation $\DeltaHat\cdot\widetilde{\Gamma}\big|_{\mathrm{fin}}^{1}=-sS^1_{\mathrm{fct}}$,
with the BRST-operator $s$, for the finite symmetry-restoring counterterm action.
Hence, the following 1-loop result for $S^1_{\mathrm{fct}}$ is obtained 
from the $\Delta$-inserted Green functions provided
in Sec.\ \ref{Sec:1-Loop-BRST-Breaking},
\begin{equation}\label{Eq:Sfct_1-Loop}
    \begin{aligned}
        S^1_{\mathrm{fct}} &= \frac{1}{16\pi^2} \intx \bigg[
        \frac{g^2}{2} \mathcal{F}_{BB}^{1,\text{break}} \, \overline{B}_{\mu} \overline{\Box} \, \overline{B}^{\mu}
        + \frac{g^4}{8} \mathcal{F}_{BBBB}^{1,\text{break}} \, \overline{B}_{\mu} \overline{B}^{\mu} \overline{B}_{\nu} \overline{B}^{\nu}\\
        &+ g \, \overline{\psi}_i \overline{\slashed{B}} \Big[
        \mathcal{F}_{\psi\overline{\psi}B,\mathrm{R},ij}^{1,\text{break}}
        \mathbb{P}_{\mathrm{R}}
        +
        \mathcal{F}_{\psi\overline{\psi}B,\mathrm{L},ij}^{1,\text{break}}
        \mathbb{P}_{\mathrm{L}}
        \Big] \psi_j\\
        &-
        \Big(
        \overline{\psi}_i 
        \Big[ 
        \mathcal{F}_{\psi\overline{\psi}\phi,ij}^{1,\text{break},a} \phi_{a}
        + \mathcal{F}_{\psi\overline{\psi}\phi^{\dagger},ij}^{1,\text{break},a} \phi_{a}^{\dagger}
        \Big]
        \projR
        {\psi}_j
        + \mathrm{h.c.} \Big)
        \\ 
        &+ ig\, \mathcal{F}_{\phi\phi^{\dagger}B,ab}^{1,\text{break}}
        \Big[
        \big(\overline{\partial}^{\mu}\phi_{a}^{\dagger}\big)\phi_b
        - \phi_{a}^{\dagger}\big(\overline{\partial}^{\mu}\phi_b\big) 
        \Big] 
        \overline{B}_{\mu}
        + \frac{g^2}{2} \mathcal{F}_{\phi\phi^{\dagger}BB,ab}^{1,\text{break}}
        \phi_{a}^{\dagger}\phi_{b}\overline{B}^{\mu}\overline{B}_{\mu}\\
        &- \frac{1}{2} 
        \Big(
        \mathcal{F}_{\phi\phi,ab}^{1,\text{break}}
        \Big[
        \overline{\partial}_{\mu}\phi_{a}\overline{\partial}^{\mu}\phi_{b}
        - 3 i g \mathcal{Y}_{S} \overline{\partial}_{\mu}\big(\phi_a\phi_b\big)\overline{B}^{\mu}
        + 3 g^2 \mathcal{Y}_{S}^2 \phi_a \phi_b \overline{B}_{\mu} \overline{B}^{\mu}
        \Big]
        + \mathrm{h.c.} \Big)\\
        &-
        \Big(
        \frac{1}{96} \, \mathcal{F}_{\phi\phi\phi\phi,abcd}^{1,\text{break}} \,
        \phi_{a}\phi_{b}\phi_{c}\phi_{d}
        + \frac{1}{12} \, \mathcal{F}_{\phi^{\dagger}\phi\phi\phi,abcd}^{1,\text{break}} \,
        \phi_{a}^{\dagger}\phi_{b}\phi_{c}\phi_{d}
        + \mathrm{h.c.} \Big)
        \bigg].
    \end{aligned}
\end{equation}
In total there are eleven relevant coefficients that appear in $S^1_{\mathrm{fct}}$, 
which are given in 
Eqs.\ (\ref{App-Eq:FiniteFermionGaugeBosonCoeffsFull}), (\ref{App-Eq:FiniteGaugeBosonCoeffs}) and
(\ref{App-Eq:FiniteYukawaCoeffs}) to (\ref{App-Eq:FiniteScalarGaugeBosonCoeffs}).
In contrast to the divergent, symmetry-violating part of the counterterm action shown in Eq.\
(\ref{Eq:Ssct_1-Loop_break}), none of these coefficients vanishes completely for the case $\hypLR\equiv0$.

As explained e.g.\ in the review  Ref.\ \cite{Belusca-Maito:2023wah}, we have the 
freedom to add any finite, but symmetric counterterm to this finite counterterm action.
Here, we have chosen to attribute the breaking associated to the fermion self-energy and the
fermion-gauge boson interaction current completely to the fermion-gauge boson vertex rather than to
the fermion self-energy (as done in Refs.\ \cite{Belusca-Maito:2021lnk,Stockinger:2023ndm}), or to a combination of both. 
This is due to the fact that, in this general model, it is simply not possible
to attribute the breaking fully to the fermion self-energy, because of the additional breaking from
the Green function $cB\overline{\psi}\psi$, shown 
in Eq.\ \eqref{Eq:FFbarBc-GreenFunc}, that is associated exclusively to the gauge interaction vertex and 
not to the fermion self-energy.
This can directly be seen from the Slavnov-Taylor identity in Eq.\ (\ref{Eq:STI-QAP-FFbarBc}),
and contrasts the breaking from the Green function $c\overline{\psi}\psi$ in Eq.\ (\ref{Eq:FbarFc-GreenFunc}), 
which is associated to both the fermion self-energy and the
fermion-gauge boson interaction vertex, as seen in Eq.\ \eqref{Eq:STI-QAP-FFc}.
The additional BRST breaking in Eq.\ \eqref{Eq:FFbarBc-GreenFunc} emerges, on the one hand,
from the evanescent gauge interactions, governed by $\hypLR$, 
and, on the other hand, through the traditional
$\widehat{\Delta}_1\big[c,\overline{\psi},\psi\big]$-interaction,
see Eq.\ (\ref{Eq:TreeLevelBreaking}), and
scalar fields coupling to the external fermions and gauge boson.

Similarly, for the scalar-gauge boson breaking, we have chosen to fully attribute it to
the two scalar-gauge boson vertices and not to the scalar self-energy, the reason being analogous to the previous one.
From Eqs.\ (\ref{Eq:GhostDoubleScalar_1-Loop}) and (\ref{Eq:GreenFunc-c-B-phiDagger-phi})
we see that we have two non-vanishing breaking contributions from
the Green functions $c\phi^{\dagger}\phi$ and $cB\phi^{\dagger}\phi$, respectively, 
associated to the scalar self-energy and the scalar-gauge boson interactions according to 
Eqs.\ (\ref{Eq:STI-QAP-SSbarc}) and (\ref{Eq:STI-QAP-SSbarBc}).
The contribution from the Green function $cBB\phi^{\dagger}\phi$ vanishes identically,
as shown in Eq.\ (\ref{Eq:SSbarBBc-DeltaGreenFunc}).
Since the breaking from $cB\phi^{\dagger}\phi$ is solely associated with 
the scalar-gauge boson interactions, cf.\ Eq.\ (\ref{Eq:STI-QAP-SSbarBc}), 
it is again impossible to attribute the breaking to the scalar self-energy alone.
Thus, we decided to exclusively attribute it to the scalar-gauge boson interaction terms,
omitting the scalar self-energy.

In the finite symmetry-restoring counterterm action, we also identify
contributions that violate global hypercharge conservation,
similar to the divergent breakings presented in Eq.\ \eqref{Eq:Ssct_1-Loop_break}.
At first glance, the last two lines of Eq.\ \eqref{Eq:Sfct_1-Loop} stand out,
as they clearly break global hypercharge conservation.
From Eqs.\ \eqref{App-Eq:Finite2ScalarCoeff}, \eqref{App-Eq:Finite4ScalarCoeff-SSSS}
and \eqref{App-Eq:Finite4ScalarCoeff-SdaggerSSS}
in App.\ \ref{App:Results-1LoopCoeffs}, we observe that the associated coefficients
do not depend on the evanescent hypercharges.
Consequently, these terms would only vanish if we employ Option \hyperref[Opt:Option2a-2b-GroupReference]{2}
for the treatment of fermions in $D$ dimensions and introduce a sterile partner field for each fermion.
Further, note that here in this case,
it is not possible to add a symmetric counterterm to shift the breaking 
to preferred terms, as done before.
This is because all of these terms break global hypercharge conservation,
and thus necessarily also BRST invariance.

Focusing on the quartic scalar contributions, seen in the last line of
Eq.\ \eqref{Eq:Sfct_1-Loop}, the term $\phi^{\dagger}\phi\phi^{\dagger}\phi$ 
is BRST invariant by its own, and thus the only BRST-breaking contributions
necessarily violate global hypercharge conservation.

In addition to these obvious terms, the finite Yukawa breakings in the third line of 
Eq.\ \eqref{Eq:Sfct_1-Loop} also violate global hypercharge conservation.
This is because the corresponding coefficients, shown in 
Eq.\ \eqref{App-Eq:FiniteYukawaCoeffs}, combine fermions and scalars in such 
a way that either the assigned hypercharges do not add up to zero or the resulting
term is BRST-invariant, and can thus be subtracted by an appropriately chosen
finite, symmetric counterterm.
In contrast to the scalar terms in the last two lines of Eq.\ \eqref{Eq:Sfct_1-Loop},
discussed in the passage above, the coefficients in Eq.\ \eqref{App-Eq:FiniteYukawaCoeffs} 
reveal that the Yukawa breakings receive contributions from both
the evanescent part of the fermion kinetic term and from the evanescent gauge interactions.
Hence, they can only entirely be avoided if both sources of global hypercharge violation
are tackled simultaneously, i.e.\ employing Option \hyperref[Opt:Option2a-2b-GroupReference]{2}
and setting the evanescent hypercharges to zero.

% -----------------------------------------------------------------------------

% +++++++++++++++++++++++++++++++++++++++++++++++++++++++++++++++++++++++++++++
\section{Impact of Evanescent Choices: Fermions \& Hypercharges}
\label{Sec:AnalysisOfResults}

Recall that the  spurious  breaking of BRST invariance and global symmetry  
has two sources, cf.\ Eq.\ \eqref{Eq:TreeLevelBreaking}:
\begin{itemize}
    \item 
    Evanescent parts of fermion kinetic terms, which depend on details how the fermion fields are defined in $D$ dimensions (see Sec.\ \ref{Sec:FermionsInGeneral}).
    \item 
    Evanescent gauge interactions, governed by evanescent hypercharges 
    that can be set to zero or to different non-vanishing values (see  Sec.\ \ref{Sec:FermionsInGeneral} and Eq.~(\ref{Eq:Lfermion})).
\end{itemize}
Both of them mix fields of different chiralities and gauge quantum numbers, thereby introducing potential sources of symmetry breaking. In this section we explore the impact of different choices for these terms in detail.

To focus the attention on all crucial aspects and provide detailed results we specialise
the general model in several specific ways. First, in Sec.\ \ref{Sec:FermionicSectorNoScalars} we consider the model in the absence of scalar fields, 
focusing solely on the fermionic sector, and especially on the first source of symmetry breaking; the fermion kinetic term. 
This also serves as a consistency check for our results, as we are able to reproduce 
known results from the literature, as outlined in Sec.\ \ref{Sec:ModelsFromLiterature}.
Second, in Sec.~\ref{Sec:AnalysisOfASSMResults} we specialise to the abelian sector of the SM 
as illustrated in Sec.\ \ref{Sec:ASSM} because of its phenomenological relevance.
Here, we consider both sources of symmetry breaking and allow for general non-universal evanescent hypercharges.
In particular we work with Eq.\ \eqref{Eq:ASSM-YLR} for Option \ref{Opt:Option1} 
and Eq.\ \eqref{Eq:ASSM-YLR-YRL-Option2} for Option \hyperref[Opt:Option2a-2b-GroupReference]{2}.
Finally, in Sec.~\ref{Sec:TheRoleOf-YLR-YRL} we shift our focus to the chiral gauge interactions in $D$ dimensions and discuss various concrete 
choices for the evanescent hypercharges using the abelian sector of the SM as a case study.

\subsection{Exploring Options for the Fermionic Sector}\label{Sec:FermionicSectorNoScalars}

We begin by comparing with Refs.~\cite{Martin:1999cc}
and \cite{Cornella:2022hkc} in the appropriate special cases. We discard all scalar contributions 
and set the evanescent hypercharges to zero, i.e.\ $\hypLR=\hypRL\equiv0$,
such that the  symmetry breaking originates only from the fermion kinetic term.
By doing so, our results for the singular and finite counterterm action in 
Eqs.\ \eqref{Eq:Ssct_1-Loop_inv}, \eqref{Eq:Ssct_1-Loop_break} 
and \eqref{Eq:Sfct_1-Loop}, respectively, reduce to 
\begin{align}
    \begin{split}\label{Eq:Ssct_YLYR_1-Loop}
        S^1_{\mathrm{sct,\mathrm{R+L}}} &= - \frac{g^2}{16\pi^2} \, \frac{1}{\epsilon} \Dintx 
        \bigg[
        \frac{2}{3} \mathrm{Tr}\big(\hypR^2+\hypL^2\big)
        \Big(-\frac{1}{4}\overline{F}^{\mu\nu}\overline{F}_{\mu\nu}\Big)\\
        &\qquad\qquad\qquad\;\;\;
        + \xi (\hypR^2)_{kj}
        \overline{\psi}_i i \overline{\slashed{D}}_{\mathrm{R},ik} \psi_j
        + \xi (\hypL^2)_{kj}
        \overline{\psi}_i i \overline{\slashed{D}}_{\mathrm{L},ik} \psi_j\\
        &\qquad\qquad\qquad\;\;\;
        + \frac{1}{3} \mathrm{Tr}\big((\hypR+\hypL)^2\big) \frac{1}{2} \overline{B}_{\mu}\widehat{\Box}\overline{B}^{\mu}
        + 2 \frac{2+\xi}{3} (\hypR\hypL)_{ij} \overline{\psi}_i i \widehat{\slashed{\partial}} \psi_j
        \bigg],
    \end{split}\\[1.5ex]
    \begin{split}\label{Eq:Sfct_YLYR_1-Loop}
        S^1_{\mathrm{fct,\mathrm{R+L}}} &=
        \frac{1}{16\pi^2} \intx 
        \bigg[-\frac{g^2}{3} \mathrm{Tr}\big((\hypR-\hypL)^2\big) \frac{1}{2} \overline{B}_{\mu} \overline{\Box} \, \overline{B}^{\mu}\\
        &\qquad\qquad\qquad\;\;\;\;
        + \frac{2 g^4}{3} \mathrm{Tr}\big((\hypR-\hypL)^4\big) \frac{1}{8} \overline{B}_{\mu} \overline{B}^{\mu} \overline{B}_{\nu} \overline{B}^{\nu}\\
        &\qquad\qquad\qquad\;\;\;\;
        + \frac{5+\xi}{6} g^3 (\hypR-\hypL)_{ik} \overline{\psi}_i \overline{\slashed{B}} 
        \Big( (\hypR^2)_{kj} \mathbb{P}_{\mathrm{R}}
        - (\hypL^2)_{kj} \mathbb{P}_{\mathrm{L}} \Big) \psi_j
        \bigg],
    \end{split}
\end{align}
where only the physical fermionic hypercharges $\hypL$ and $\hypR$ are left.
Evidently, Eq.\ (\ref{Eq:Sfct_YLYR_1-Loop}) precisely reproduces the result 
for the finite symmetry-restoring counterterms
presented in Ref.\ \cite{Cornella:2022hkc} for the abelian special case 
in which the considered gauge group is $U(1)$.
In general, these results align with those from Refs.\ \cite{Martin:1999cc}
and \cite{Cornella:2022hkc}, in the cases where the hypercharges exhibit a 
block-diagonal structure or not, corresponding to the third and fourth row of
Tab.\ \ref{Tab:SpecialCasesOverview}, respectively.

We can now use Eqs.\ \eqref{Eq:Ssct_YLYR_1-Loop} and \eqref{Eq:Sfct_YLYR_1-Loop}
to examine the impact of the different options to treat fermions in $D$ dimensions discussed in Sec.~\ref{Sec:FermionsInGeneral}.
A first difference arises from the block structure, cf.\ Eq.\ \eqref{Eq:ASSM-YLYR-Option2}, 
that the hypercharges $\hypL$ and $\hypR$
exhibit in Option \hyperref[Opt:Option2a-2b-GroupReference]{2},
where a non-interacting, sterile partner field is introduced for each fermion.
In particular, this block structure results in vanishing 
``left- and right-mixing'' contributions, i.e.\ $\hypL\hypR=0$, which can be considered an advantage of this option.
Clearly, this is not the case for Option \ref{Opt:Option1}, cf.\ Eq.\ \eqref{Eq:ASSM-YLYR}, where
we combine the two chiral components of a fermion to the natural Dirac spinor
without introducing sterile partners.

As a result of this difference,  the symmetry-breaking counterterm contributions (appearing in the last line of Eq.\ \eqref{Eq:Ssct_YLYR_1-Loop} and in Eq.\ \eqref{Eq:Sfct_YLYR_1-Loop})
vary depending on which option is chosen.
Specifically,  in the last line of Eq.\ \eqref{Eq:Ssct_YLYR_1-Loop}, the gauge boson term only changes its
prefactor, whereas the fermion term vanishes entirely in the case of 
Option \hyperref[Opt:Option2a-2b-GroupReference]{2},
giving it a slight advantage over Option \ref{Opt:Option1}.
In the finite symmetry-restoring counterterm action shown in
Eq.\ \eqref{Eq:Sfct_YLYR_1-Loop}, only the prefactors change,
and none of the terms vanish for either option.
In contrast, 
we find that the symmetric part of
the singular counterterm action, shown in the first two lines of Eq.\ \eqref{Eq:Ssct_YLYR_1-Loop},
leads to the same result for both options, as none of these terms emerged from
a chirality mixing source.

Next we consider a further specialised case, 
and set all left-handed hypercharges to zero, $\hypL=0$, cf.\ Tab.\ \ref{Tab:SpecialCasesOverview}. This corresponds to recasting all left-handed fermions as right-handed spinor fields with opposite charge
and introducing appropriate left-handed sterile partner fields,
in line with Option \ref{Opt:Option2b}, and results in the following counterterm action
\begin{align}
    \begin{split}\label{Eq:Ssct_chiralQED_1-Loop}
        S^1_{\mathrm{sct,\chi QED}} &= - \frac{g^2}{16\pi^2} \, \frac{1}{\epsilon} \Dintx 
        \bigg[
        \frac{2}{3} \mathrm{Tr}\big(\hypR^2\big)
        \Big(-\frac{1}{4}\overline{F}^{\mu\nu}\overline{F}_{\mu\nu}\Big)
        + \xi (\hypR^2)_{kj}
        \overline{\psi}_i i \overline{\slashed{D}}_{\mathrm{R},ik} \psi_j\\
        &\qquad\qquad\qquad\quad\;\;\;
        + \frac{1}{3} \mathrm{Tr}\big(\hypR^2\big) \frac{1}{2} \overline{B}_{\mu}\widehat{\Box}\overline{B}^{\mu}
        \bigg],
    \end{split}\\[1.5ex]
    \begin{split}\label{Eq:Sfct_chiralQED_1-Loop}
        S^1_{\mathrm{fct,\chi QED}} &=
        \frac{1}{16\pi^2} \intx \bigg[
        -\frac{g^2}{3} \mathrm{Tr}\big(\hypR^2\big) \frac{1}{2} \overline{B}_{\mu} \overline{\Box} \, \overline{B}^{\mu}
        + \frac{2 g^4}{3} \mathrm{Tr}\big(\hypR^4\big) \frac{1}{8} \overline{B}_{\mu} \overline{B}^{\mu} \overline{B}_{\nu} \overline{B}^{\nu}\\
        &\qquad\qquad\qquad\;
        + \frac{5+\xi}{6} g^3 (\hypR^3)_{ij} \overline{\psi}_i \overline{\slashed{B}} \mathbb{P}_{\mathrm{R}} \psi_j
        \bigg].
    \end{split}
\end{align}
This special case corresponds to
right-handed chiral QED, as studied in the previous publications
\cite{Belusca-Maito:2021lnk,Belusca-Maito:2023wah,Stockinger:2023ndm}, and the result is in agreement with these references.
Note that here, we attributed the finite breaking of the fermion-gauge boson interaction to the vertex,
as seen in the last line of Eq.\ (\ref{Eq:Sfct_chiralQED_1-Loop}), which is in contrast
to Refs.\ \cite{Belusca-Maito:2021lnk,Belusca-Maito:2023wah,Stockinger:2023ndm}.
However, making use of the freedom of adding an appropriate symmetric finite counterterm
of the form $\frac{5+\xi}{6}g^2(\hypR^2)_{ik}\overline{\psi}_{i}i\overline{\slashed{D}}_{\mathrm{R},kj}\psi_j$,
we can shift this breaking to the fermion self-energy and identically recover the results
from Refs.\ \cite{Belusca-Maito:2021lnk,Belusca-Maito:2023wah,Stockinger:2023ndm}. 
Further, in line with earlier results and with the discussion above, the divergent, symmetry-breaking and evanescent fermion term 
from Eq.\ \eqref{Eq:Ssct_YLYR_1-Loop}
has vanished since $\hypL\hypR=0$ in the considered special case of Option \ref{Opt:Option2b} 
(or equivalently Option \ref{Opt:Option2} in the absence of left-handed fermions).

As another check we specialise our general setup to standard QED. Indeed, if we reintroduce evanescent gauge interactions and set all
hypercharges of the fermions equal to the electromagnetic charge, as shown in
the first row of Tab.\ \ref{Tab:SpecialCasesOverview}, we obtain
\begin{equation}\label{Eq:Sct_QED_1-Loop}
    \begin{aligned}
        S^1_{\mathrm{ct,QED}} = 
        - \frac{g^2}{16\pi^2} \, \frac{1}{\epsilon} \Dintx 
        \bigg[  
        \frac{4}{3} \mathrm{Tr}\big(Q^2\big)
        \Big(-\frac{1}{4}F^{\mu\nu}F_{\mu\nu}\Big)
        + \xi (Q^2)_{kj}
        \overline{\psi}_i i \slashed{D}_{ik} \psi_j
        \bigg],
    \end{aligned}
\end{equation}
with the well-known covariant derivative $D_{\mu}=\partial_{\mu}+igQB_{\mu}$,
which is the counterterm action of standard QED.
Note that this result requires non-zero values of the evanescent hypercharges $\hypLR$ and $\hypRL$.

Clearly, this counterterm action is completely symmetric. 
Moreover, the result is the same for both options for the $D$-dimensional treatment of fermions. 
This is analogous to the previous case discussed above, 
where the symmetric part of the counterterm
action in Eq.~\eqref{Eq:Ssct_YLYR_1-Loop} also remained unchanged under different options to treat the fermions.

Finally, we discuss the scenario where $\hypR=\hypL=\mathcal{Y}$ while 
$\hypLR=\hypRL=0$, which corresponds to QED, however not treated in the standard way but rather by keeping the photon couplings 4-dimensional. Clearly, this treatment of QED breaks gauge invariance on the regularised level. Accordingly, the operator $\widehat{\Delta}$ is non-zero and some parts of the divergent symmetry-breaking contributions from the 1-loop Green functions 
$cB$, $c\overline{\psi}\psi$ and $c\phi^{\dagger}\phi$ 
--- in Eqs.\ \eqref{Eq:GhostGaugeBoson_1-Loop}, \eqref{Eq:FbarFc-GreenFunc} 
and \eqref{Eq:GhostDoubleScalar_1-Loop}, respectively ---
persist. All other
symmetry breakings vanish. In particular, the finite symmetry breaking vanishes completely, as pointed out by the authors of 
Ref.\ \cite{Cornella:2022hkc}, who focused exclusively on finite breakings.\footnote{%
Note that the vanishing of the coefficients for the symmetry-breaking contributions
presented in App.\ \ref{App:Results-1LoopCoeffs} is obvious for some of them, as
they directly depend on $\sim(\hypR-\hypL)$.
However, the situation is less straightforward for 
others, such as e.g.\ the Yukawa coefficients in Eq.\ \eqref{App-Eq:FiniteYukawaCoeffs} and the scalar coefficients displayed in Eqs.\ 
\eqref{App-Eq:Finite2ScalarCoeff} to \eqref{App-Eq:Finite4ScalarCoeff-SdaggerSSS}, some of 
which are, in fact, independent of the hypercharges.
This issue is resolved by BRST invariance of the Yukawa sector, as expressed in the 
requirement of Eq.\ \eqref{Eq:YukawaHyperchargeBRSTCondition}, which relates  
the Yukawa and hypercharge matrices. Using such relations it is possible to show that all coefficients either vanish for $\hypL=\hypR$ or the counterterms reduce to finite symmetric counterterms.
% For instance, if we keep $\hypS\neq0$, none of the Yukawa matrices can have diagonal entries
% due to hypercharge conservation.
% Therefore, these coefficients also vanish.
% Moreover, most coefficients provided in App.\ \ref{App:Results-1LoopCoeffs} 
% have been simplified using relations such as
% Eq.\ \eqref{Eq:YukawaHyperchargeBRSTCondition}. 
% The purpose of this is that the coefficients of the associated finite
% symmetry-restoring counterterm action take a concise form and can directly be read off 
% from the $\Delta$-operator inserted Green functions.
% For example, this concerns the ghost-Yukawa contributions in Eqs.\ 
% \eqref{Eq:YukawaDeltaGreenFunc1-cSFF} and \eqref{Eq:YukawaDeltaGreenFunc2-cSbarFF},
% which initially vanish for $\hypR=\hypL$.
% However, by employing Eq.\ \eqref{Eq:YukawaHyperchargeBRSTCondition} to bring them into the form
% presented in Eqs.\ \eqref{Eq:YukawaDeltaGreenFunc1-cSFF} and \eqref{Eq:YukawaDeltaGreenFunc2-cSbarFF},
% with coefficients given in Eq.\ \eqref{App-Eq:FiniteYukawaCoeffs},
% a non-vanishing contribution remains in the finite counterterm action
% if $\hypR=\hypL$ and $\hypLR=\hypRL=0$ is just applied to Eq.\ \eqref{Eq:Sfct_1-Loop}.
% This contribution, however, is finite and symmetric for $\hypR=\hypL$, and
% can simply be removed by adding an appropriate finite symmetric counterterm,
% resulting indeed in the absence of any finite symmetry breaking.
}
Moreover, as anticipated due to $\hypR=\hypL$, there is no violation of global hypercharge.

To understand why
all $4$-dimensional finite symmetry-breaking contributions vanish in the case where $\hypR=\hypL$ and $\hypLR=\hypRL=0$, while some
of the evanescent, divergent symmetry-breaking contributions persist,
we examine the tree-level breaking Eq.\ \eqref{Eq:TreeLevelBreaking}
for this scenario. It simplifies to
\begin{equation}\label{Eq:BRST-Breaking-YL=YR}
        \begin{aligned}
            \widehat{\Delta}_{\mathrm{L}=\mathrm{R}}
            &= - g \mathcal{Y}_{ij} \Dintx \, c \widehat{\partial}_{\mu} \Big( \overline{\psi}_i \widehat{\gamma}^{\mu} \psi_j \Big)
            \,\,\,\,\Longrightarrow
        \end{aligned}
    \begin{tabular}{rl}
        \raisebox{-37pt}{\includegraphics[scale=0.5]{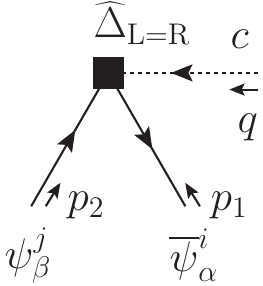}}&
        $\begin{aligned}
            &= g \mathcal{Y}_{ij} \widehat{\slashed{q}}.
        \end{aligned}$
    \end{tabular}
\end{equation}
Evidently, the Feynman rule reduces to $\sim\widehat{\slashed{q}}$, 
with $q$ being the inflowing ghost momentum.
Hence, for $q_{\mu}=\text{4-dimensional}$, all $\DeltaHat$-inserted diagrams vanish. 
This observation explains the complete absence of any 4-dimensional and finite symmetry-restoring counterterms 
at the 1-loop level.
Whether this situation changes at higher loop orders due to subrenormalisation  requires further investigations.
In particular, in non-abelian gauge theories the ghost is not always external, and thus
needs to carry $D$-dimensional momentum in these cases.
Even in the abelian 1-loop case considered here, 
divergent symmetry-restoring counterterms exist.
Examples are divergent breaking contributions $\sim\widehat{q}^2\overline{q}^{\mu}$ from
$cB$ and $\sim\widehat{\slashed{q}}$ from $c\overline{\psi}\psi$,
with $q$ being the ghost momentum.

An important takeaway from this subsection is that 
only symmetry breaking terms 
--- specifically, those terms that emerged from a chirality mixing source of symmetry breaking ---
are affected by a different treatment of fermions in $D$ dimensions.

\subsection{The Abelian Sector of the Standard Model}\label{Sec:AnalysisOfASSMResults}

In this section, the general model defined in Sec.\ \ref{Sec:D-DimLagrangian} is specialised 
to the abelian sector of the SM, as outlined in Sec.\ \ref{Sec:ASSM}. 
We consider both approaches for the treatment of fermions in $D$ dimensions 
and a general realisation of evanescent gauge interactions,
as introduced in Sections \ref{Sec:FermionsInGeneral} and \ref{Sec:D-DimLagrangian}, respectively.
In both cases we provide the full counterterm actions including divergent and finite symmetry-restoring counterterms, for general values of the evanescent hypercharges.

First, we focus on the ``natural'' approach to fermions in $D$ dimensions, corresponding to
Option \ref{Opt:Option1}, where we
combine the two chiral components of each fermion into one Dirac spinor.
In this case, the fermionic content is given by the multiplets in Eq.\
\eqref{Eq:ASSM-FermionMultiplets}, and evanescent gauge interactions are implemented via 
the evanescent hypercharges in Eq.\ \eqref{Eq:ASSM-YLR}.
Next, we transition to Option \ref{Opt:Option2}, where each fermion
is paired with a sterile partner field in order to avoid breaking of global hypercharge conservation.
The fermions in this approach are given via the multiplets in Eq.\ 
\eqref{Eq:ASSM-FermionMultiplets-Option2}, and evanescent gauge interactions are
incorporated via the evanescent hypercharges shown in Eq.\ \eqref{Eq:ASSM-YLR-YRL-Option2}.

\paragraph{ASM with fermion multiplets according to Option \ref{Opt:Option1}:}
As announced, we begin by considering the ``natural'' Option \ref{Opt:Option1} for the fermions, allowing diagonal evanescent hypercharges $\hypLR=\hypRL$. We first provide the explicit results for all three parts of the counterterm action including explicit definitions of coefficients, and thereafter discuss the impact of the evanescent couplings $\hypLR$.

The invariant part of the divergent counterterm action for the abelian sector 
of the SM is given by
\begin{equation}\label{Eq:Ssct_1-Loop_inv_ASSM_Option1}
    \begin{aligned}
        S^1_{\mathrm{sct,inv}} &= 
        - \frac{1}{16\pi^2} \, \frac{1}{\epsilon} \Dintx 
        \bigg[  
        \frac{20 g^2}{9}
        \Big(-\frac{1}{4}\overline{F}^{\mu\nu}\overline{F}_{\mu\nu}\Big)
        + \frac{g^2}{6}
        \Big(-\frac{1}{4}F^{\mu\nu}F_{\mu\nu}\Big)\\
        &+ \overline{\psi}_f i \overline{\slashed{\partial}} 
        \Big(\delta Z^{f}_{\overline{\psi}\psi,\mathrm{R}} \projR
        + \delta Z^{f}_{\overline{\psi}\psi,\mathrm{L}} \projL \Big) \psi_f
        - g \overline{\psi}_f \overline{\slashed{B}} 
        \Big( \delta Z^{f}_{\overline{\psi}B\psi,\mathrm{R}} \projR
        + \delta Z^{f}_{\overline{\psi}B\psi,\mathrm{L}} \projL \Big) \psi_f\\
        &- \Big\{
        \delta Z_{Y}^{e} y_e \big(\overline{\nu_{L}} \phi_{1} e_{R} 
        + \overline{e_{L}} \phi_{2} e_{R} \big)
        + \delta Z_{Y}^{d} y_d \big( \overline{u_{L}} \phi_{1} d_{R}
        + \overline{d_{L}} \phi_{2} d_{R} \big)\\
        &- \delta Z_{Y}^{u} y_u \big( \overline{d_{L}} \phi_{1}^{\dagger} u_{R}
        - \overline{u_{L}} \phi_{2}^{\dagger} u_{R} \big)
        +\mathrm{h.c.} \Big\}\\
        &- \frac{3-\xi}{4} g^2 
        \big(D^{\mu}\phi_a\big)^{\dagger}\big(D_{\mu}\phi_a\big)
        + \delta y_2
        \big(\overline{D}^{\mu}\phi_a\big)^{\dagger}\big(\overline{D}_{\mu}\phi_a\big)\\
        &+ \frac{\delta \lambda}{4} \Big( \phi_1^{\dagger} \phi_1 \phi_1^{\dagger} \phi_1
        + 2 \, \phi_1^{\dagger} \phi_1 \phi_2^{\dagger} \phi_2 + \phi_2^{\dagger} \phi_2 \phi_2^{\dagger} \phi_2 \Big)
        \bigg],
    \end{aligned}
\end{equation}
where we define fermionic counterterm coefficients
\begin{equation}\label{Eq:ASSM-Ssctinv-Coeff-FF}
    \begin{aligned}
        \delta Z^{f}_{\overline{\psi}\psi,\mathrm{R}} =
        \begin{cases}
            0, &f=\nu\\
            \xi g^2 + y_e^2, &f=e\\
            \frac{4\xi}{9} g^2 + y_u^2, &f=u\\
            \frac{\xi}{9} g^2 + y_d^2, &f=d
        \end{cases},
        \quad
        \delta Z^{f}_{\overline{\psi}\psi,\mathrm{L}} =
        \begin{cases}
            \frac{\xi}{4} g^2 + \frac{y_e^2}{2}, &f\in\{\nu,e\}\\
            \frac{\xi}{36} g^2 + \frac{y_u^2+y_d^2}{2}, &f\in\{u,d\},
        \end{cases}
    \end{aligned}
\end{equation}
and, for the fermion-gauge interactions, 
\begin{equation}\label{Eq:ASSM-Ssctinv-Coeff-FBF}
    \begin{aligned}
        \delta Z^{f}_{\overline{\psi}B\psi,\mathrm{R}} =
        \begin{cases}
            0, &f=\nu\\
            - \xi g^2 - y_e^2, &f=e\\
            \frac{8\xi}{27} g^2 + \frac{2 y_u^2}{3}, &f=u\\
            -\frac{\xi}{27} g^2 - \frac{y_d^2}{3}, &f=d
        \end{cases},
        \quad
        \delta Z^{f}_{\overline{\psi}B\psi,\mathrm{L}} =
        \begin{cases}
            -\frac{\xi}{8} g^2 - \frac{y_e^2}{4}, &f\in\{\nu,e\}\\
            \frac{\xi}{216} g^2 + \frac{y_u^2+y_d^2}{12}, &f\in\{u,d\}.
        \end{cases}
    \end{aligned}
\end{equation}
The Yukawa counterterms are governed by the coefficients
\begin{equation}\label{Eq:ASSM-Ssctinv-Coeff-Yukawa}
    \begin{aligned}
        \delta Z_{Y}^{e} =\frac{6+3\xi}{4} g^2, \qquad
        \delta Z_{Y}^{u} =\frac{12+13\xi}{36} g^2 + y_d^2, \qquad
        \delta Z_{Y}^{d} =-\frac{6-7\xi}{36} g^2 + y_u^2,
    \end{aligned}
\end{equation}
whereas we introduce
\begin{equation}\label{Eq:ASSM-Ssctinv-Coeff-YukawaCombinations}
    \begin{aligned}
        \delta y_2 = y_e^2 + 3 \big(y_u^2+y_d^2\big), \qquad
        \delta y_4 = y_e^4 + 3 \big(y_u^4+y_d^4\big)
    \end{aligned}
\end{equation}
for counterterm coefficients emerging from Yukawa interactions.
Finally, we introduce
\begin{equation}\label{Eq:ASSM-Ssctinv-Coeff-Scalar}
    \begin{aligned}
        \delta \lambda &= 48\lambda_{SM}^2 - 2\xi g^2 \lambda_{SM} + \frac{3}{4} g^4 - 4 \delta y_4
    \end{aligned}
\end{equation}
for the counterterm of the scalar self-interaction. 
The result in Eq.\ \eqref{Eq:Ssct_1-Loop_inv_ASSM_Option1} 
is precisely the counterterm action that we expect from 
renormalisation transformations in the ASM.

The divergent symmetry-restoring counterterm action
for the ASM, with fermions treated according to Option \ref{Opt:Option1}, 
is provided by
\begin{equation}\label{Eq:Ssct_1-Loop_break_ASSM_Option1}
    \begin{aligned}
        S^1_{\mathrm{sct,break}} &= \frac{1}{16\pi^2} \, \frac{1}{\epsilon} \Dintx
        \bigg[
        - \frac{7g^2}{9}
        \overline{B}_{\mu}\widehat{\Box}\overline{B}^{\mu}\\
        &+ \Big( \frac{2}{3} \delta y_2 - 2 y_u y_d \Big) \phi_{1}^{\dagger} \widehat{\Box} \phi_1
        + \frac{2}{3} \delta y_2 \phi_{2}^{\dagger} \widehat{\Box} \phi_2
        + \frac{1}{6} \Big( \delta y_2 \phi_{2} \widehat{\Box} \phi_2 + \mathrm{h.c.} \Big)\\
        &+ \delta\widehat{X}^{\text{opt1}}_{\psi,f} \overline{\psi}_f i \widehat{\slashed{\partial}} \psi_f
        - g \, \delta\widehat{X}^{\text{opt1}}_{G,f} \overline{\psi}_f \widehat{\slashed{B}} \psi_f\\
        &+ \frac{g^2}{3} \delta \widehat{H}_1
        \big(\overline{\partial}\cdot\overline{B}\big) \big(\widehat{\partial}\cdot\widehat{B}\big)
        -\frac{2g^2}{3} \delta \widehat{H}_{2} \Big(
        \widehat{B}_{\mu}\overline{\Box}\widehat{B}^{\mu}
        +\widehat{B}_{\mu}\widehat{\Box}\widehat{B}^{\mu}
        +\big(\widehat{\partial}\cdot\widehat{B}\big)^2
        \Big)\\
        &+ i \frac{2g}{3} \delta \widehat{S}^{\text{opt1}}_3 
        \Big[ \big(\widehat{\partial}^{\mu}\phi_{1}^{\dagger}\big)\phi_1 
        - \phi_{1}^{\dagger}\big(\widehat{\partial}^{\mu}\phi_1\big) \Big] \widehat{B}_{\mu}
        + \frac{2g^2}{3} \delta \widehat{S}^{\text{opt1}}_4
        \phi_{1}^{\dagger}\phi_{1} \widehat{B}^{\mu} \widehat{B}_{\mu}
        \bigg],
    \end{aligned}
\end{equation}
with $\delta y_2$ as defined in Eq.\ (\ref{Eq:ASSM-Ssctinv-Coeff-YukawaCombinations}), 
as well as fermion kinetic and fermion-gauge interaction coefficients provided by
\begin{equation}\label{Eq:ASSM-Ssctbreak-Coeff-FF}
    \begin{aligned}
        \delta\widehat{X}^{\text{opt1}}_{\psi,f} &=
        \begin{cases}
            0, &f=\nu\\
            - g^2 \Big( \frac{2+\xi}{3} + \frac{1-\xi}{2} \hypLR^e - \frac{2+\xi}{3} (\hypLR^e)^2 \Big), &f=e\\
            - g^2 \Big( \frac{4+2\xi}{27} - 5\frac{1-\xi}{18} \hypLR^u - \frac{2+\xi}{3} (\hypLR^u)^2 \Big) + \frac{y_u y_d}{2}, &f=u\\
            g^2 \Big( \frac{2+\xi}{27} - \frac{1-\xi}{18} \hypLR^d + \frac{2+\xi}{3} (\hypLR^d)^2 \Big) + \frac{y_u y_d}{2}, &f=d,
        \end{cases}
    \end{aligned}
\end{equation}
and
\begin{equation}\label{Eq:ASSM-Ssctbreak-Coeff-FBF}
    \begin{aligned}
        \delta\widehat{X}^{\text{opt1}}_{G,f} &=
        \begin{cases}
            0, &f=\nu\\
            - g^2 \hypLR^e \Big( \frac{2+\xi}{3} + \frac{1-\xi}{2} \hypLR^e - \frac{2+\xi}{3} (\hypLR^e)^2 \Big), &f=e\\
            - g^2 \hypLR^u \Big( \frac{4+2\xi}{27} - 5\frac{1-\xi}{18} \hypLR^u - \frac{2+\xi}{3} (\hypLR^u)^2 \Big) + \frac{y_u y_d}{2} \Big( \frac{1}{2} + \hypLR^d \Big), &f=u\\
            g^2 \hypLR^d \Big( \frac{2+\xi}{27} - \frac{1-\xi}{18} \hypLR^d + \frac{2+\xi}{3} (\hypLR^d)^2 \Big) - \frac{y_u y_d}{2} \Big( \frac{1}{2} - \hypLR^u \Big), &f=d,
        \end{cases}
    \end{aligned}
\end{equation}
respectively.
The divergent, symmetry-breaking and purely evanescent gauge boson contributions are
governed by the counterterm coefficients
\begin{equation}\label{Eq:ASSM-Ssctbreak-Coeff-BB}
    \begin{aligned}
        \delta \widehat{H}_1 &= 3 \hypLR^e - 5 \hypLR^u + \hypLR^d,\\
        \delta \widehat{H}_2 &= \lVert \hypLR \rVert^2 = (\hypLR^e)^2 + 3 \big[(\hypLR^u)^2+(\hypLR^d)^2\big],
    \end{aligned}
\end{equation}
which only consist of the evanescent hypercharges, 
while the scalar-gauge interactions of the same kind come with the coefficients
\begin{equation}\label{Eq:ASSM-Ssctbreak-Coeff-SSB-SSBB}
    \begin{aligned}
        \delta \widehat{S}^{\text{opt1}}_3 &= y_e^2 \hypLR^e - 3 \big( y_u^2+y_d^2-y_uy_d \big) \big( \hypLR^u-\hypLR^d \big),\\
        \delta \widehat{S}^{\text{opt1}}_4 &= - y_e^2 (\hypLR^e)^2 - 3 \big( y_u^2+y_d^2-y_uy_d \big) \big( \hypLR^u-\hypLR^d \big)^2.
    \end{aligned}
\end{equation}

From all these definitions, it becomes evident that the evanescent hypercharges $\hypLR$ have a significant impact on the divergent  symmetry-restoring counterterm action Eq.\ (\ref{Eq:Ssct_1-Loop_break_ASSM_Option1}).
While the first two lines of  Eq.\ (\ref{Eq:Ssct_1-Loop_break_ASSM_Option1})
are independent of evanescent hypercharges $\hypLR$, 
the third line contains a mix of terms --- some involve evanescent hypercharges
and others do not ---
and the last two lines would vanish entirely for $\hypLR=0$.
Hence, the divergent contribution of the symmetry-restoring counterterms
significantly simplifies for vanishing evanescent hypercharges.

Finally, the finite symmetry-restoring counterterm action in the considered case reads
\begin{equation}\label{Eq:Sfct_1-Loop_ASSM_Option1}
    \begin{aligned}
        S^1_{\mathrm{fct}} &= \frac{1}{16\pi^2} \intx \bigg[
        -\frac{g^2}{3} \overline{B}_{\mu} \overline{\Box} \, \overline{B}^{\mu}
        + \frac{g^4}{24} \overline{B}_{\mu} \overline{B}^{\mu} \overline{B}_{\nu} \overline{B}^{\nu}\\
        &- i \frac{g}{3} \delta y_2
        \Big[
        \big(\overline{\partial}^{\mu}\phi_{2}^{\dagger}\big)\phi_2
        - \phi_{2}^{\dagger}\big(\overline{\partial}^{\mu}\phi_2\big) 
        \Big] 
        \overline{B}_{\mu}
        - \frac{g^2}{4} \delta y_2 \Big[
        \phi_{1}^{\dagger}\phi_{1}\overline{B}^{\mu}\overline{B}_{\mu}
        + \frac{5}{3} \phi_{2}^{\dagger}\phi_{2}\overline{B}^{\mu}\overline{B}_{\mu} \Big]\\
        &+ g \, \overline{\psi}_f \overline{\slashed{B}} \Big[
        \delta F_{\mathrm{R},f}^{\text{opt1}}
        \mathbb{P}_{\mathrm{R}}
        +
        \delta F_{\mathrm{L},f}^{\text{opt1}}
        \mathbb{P}_{\mathrm{L}}
        \Big] \psi_f\\
        &- \Big\{
        \delta Y_{e,2}^{\text{opt1}} y_e \overline{e_{L}} \phi_{2}^{\dagger} e_{R}
        + \delta Y_{u,2}^{\text{opt1}} y_u \overline{u_{L}} \phi_{2} u_{R}
        + \delta Y_{d,2}^{\text{opt1}} y_d \overline{d_{L}} \phi_{2}^{\dagger} d_{R}
        +\mathrm{h.c.} \Big\}\\
        &+ \frac{1}{6} \delta y_2
        \Big[
        \overline{\partial}_{\mu}\phi_{2}\overline{\partial}^{\mu}\phi_{2}
        - \frac{3}{2} i g \overline{\partial}_{\mu}\big(\phi_2\phi_2\big)\overline{B}^{\mu}
        + \frac{3}{4} g^2 \phi_2 \phi_2 \overline{B}_{\mu} \overline{B}^{\mu} + \mathrm{h.c.}
        \Big]\\
        &- \Big(
        \frac{1}{12} \delta y_4 \,
        \phi_{2}\phi_{2}\phi_{2}\phi_{2}
        + \frac{2}{3} \big( \delta y_4 - \delta y_{ud} \big)
        \phi_{1}^{\dagger}\phi_{1}\phi_{2}\phi_{2}
        + \frac{2}{3} \delta y_4 \, 
        \phi_{2}^{\dagger}\phi_{2}\phi_{2}\phi_{2}
        + \mathrm{h.c.} \Big)
        \bigg],
    \end{aligned}
\end{equation}
with coefficients $\delta y_2$ and $\delta y_4$ as in 
Eq.\ (\ref{Eq:ASSM-Ssctinv-Coeff-YukawaCombinations}) and
\begin{equation}\label{Eq:ASSM-Sfct-Coeff-YukawaCombinations}
    \begin{aligned}
        \delta y_{ud} &= \frac{3}{2} y_u y_d \big( y_u^2 + y_d^2 \big).
    \end{aligned}
\end{equation}
Further, the coefficients of the finite fermion-gauge interaction counterterms 
are given by
\begin{equation}\label{Eq:ASSM-Sfct-Coeff-FBF-R-Option1}
    \begin{aligned}
        \delta F_{\mathrm{R},f}^{\text{opt1}} &=
        \begin{cases}
            0, &f=\nu\\
            - g^2 \Big( \frac{5+\xi}{12} - \frac{1-\xi}{6} \hypLR^e + \frac{5+\xi}{12} (\hypLR^e)^2 \Big), &f=e\\
            g^2 \Big( \frac{5+\xi}{27} + \frac{1-\xi}{9} \hypLR^u + \frac{5+\xi}{12} (\hypLR^u)^2 \Big), &f=u\\
            - g^2 \Big( \frac{5+\xi}{108} - \frac{1-\xi}{18} \hypLR^d + \frac{5+\xi}{12} (\hypLR^d)^2 \Big), &f=d,
        \end{cases}
    \end{aligned}
\end{equation}
and
\begin{equation}\label{Eq:ASSM-Sfct-Coeff-FBF-L-Option1}
    \begin{aligned}
        \delta F_{\mathrm{L},f}^{\text{opt1}} &=
        \begin{cases}
            - g^2 \frac{5+\xi}{48} - \frac{y_e^2}{4}, &f=\nu\\
            g^2 \Big( \frac{5+\xi}{48} - \frac{1-\xi}{12} \hypLR^e + \frac{5+\xi}{12} (\hypLR^e)^2 \Big) - \frac{y_e^2}{4}, &f=e\\
            - g^2 \Big( \frac{5+\xi}{432} + \frac{1-\xi}{36} \hypLR^u + \frac{5+\xi}{12} (\hypLR^u)^2 \Big) + \frac{y_u^2 - y_d^2}{4}, &f=u\\
            g^2 \Big( \frac{5+\xi}{432} + \frac{1-\xi}{36} \hypLR^d + \frac{5+\xi}{12} (\hypLR^d)^2 \Big) + \frac{y_u^2 - y_d^2}{4}, &f=d,
        \end{cases}
    \end{aligned}
\end{equation}
while the finite Yukawa interaction counterterms are governed by the coefficients
\begin{equation}\label{Eq:ASSM-Sfct-Coeff-Yukawas-Option1}
    \begin{aligned}
        \delta Y_{e,2}^{\text{opt1}} &=g^2 \Big( \frac{5+\xi}{6} - \frac{1-\xi}{2} \hypLR^e + \frac{5+\xi}{3} (\hypLR^e)^2 \Big),\\
        \delta Y_{u,2}^{\text{opt1}} &=g^2 \Big( \frac{5+\xi}{27} + 5 \frac{1-\xi}{18} \hypLR^u + \frac{5+\xi}{3} (\hypLR^u)^2 \Big) + \frac{y_d^2}{2},\\
        \delta Y_{d,2}^{\text{opt1}} &=-g^2 \Big( \frac{5+\xi}{54} + \frac{1-\xi}{18} \hypLR^d - \frac{5+\xi}{3} (\hypLR^d)^2 \Big) + \frac{y_u^2}{2}.
    \end{aligned}
\end{equation}

Again the choice of $\hypLR$ significantly impacts the form of these symmetry-restoring counterterms, although we can observe that none of the terms in Eq.\ (\ref{Eq:Sfct_1-Loop_ASSM_Option1})
vanishes entirely when the evanescent hypercharges are set to zero, i.e.\
for $\hypLR=\hypRL=0$.
However, the counterterms involving fermions simplify significantly
for vanishing evanescent hypercharges, which can be seen from each of the coefficients in Eqs.\ (\ref{Eq:ASSM-Sfct-Coeff-FBF-R-Option1})
to (\ref{Eq:ASSM-Sfct-Coeff-Yukawas-Option1}).

As anticipated when using Option \ref{Opt:Option1} for the 
$D$-dimensional treatment of fermions, global hypercharge conservation is violated on the regularised level. Correspondingly, the divergent and finite 
symmetry-restoring counterterm actions in Eqs.\ 
(\ref{Eq:Ssct_1-Loop_break_ASSM_Option1}) and
 (\ref{Eq:Sfct_1-Loop_ASSM_Option1}) contain terms that,
while preserving electric and colour charge,
break global hypercharge conservation.

These global hypercharge-violating terms are the evanescent term 
$\phi_2\widehat{\Box}\phi_2+\mathrm{h.c.}$ in Eq.\ 
(\ref{Eq:Ssct_1-Loop_break_ASSM_Option1}) and 
the $4$-dimensional terms with finite coefficients in the last three lines of
Eq.\ (\ref{Eq:Sfct_1-Loop_ASSM_Option1}).
Among these terms there are  Yukawa
counterterms with the ``wrong'' scalar field. These
involve both terms that depend on the evanescent hypercharges
and those that do not, as seen in Eq.\ \eqref{Eq:ASSM-Sfct-Coeff-Yukawas-Option1}.
The further hypercharge-violating counterterms in the last two lines of Eq.~(\ref{Eq:Sfct_1-Loop_ASSM_Option1}) are purely bosonic and contain products of scalar fields such as $\phi_2\phi_2$, etc. These arise solely from the chirality-violating evanescent part 
of the fermion kinetic term Eq.\ (\ref{Eq:Lkin-Electron-D-dim-Decomposed}) and are independent of the evanescent hypercharges.
In total, none of the global hypercharge-violating terms vanish entirely for the choice $\hypLR=\hypRL=0$.

This highlights a major drawback of Option \ref{Opt:Option1}, mentioned already in Sec.\ \ref{Sec:FermionsInGeneral}.
The approach is natural since it combines the two chiral components of each fermion into one Dirac spinor, but it introduces global hypercharge
violation through the evanescent part of the fermion kinetic term.

\paragraph{ASM with fermion multiplets according to Option \ref{Opt:Option2}:}
Now, we focus on the abelian SM treated according to Option \ref{Opt:Option2}, where we introduce a sterile partner 
field for each fermion and thus eliminate the fermion kinetic term as a potential source 
of global hypercharge violation. As in the previous case we provide the explicit form of the counterterms and discuss the impact of the choice of the $\hypLR$ and $\hypRL$. 
As observed  in Sec.\ \ref{Sec:FermionicSectorNoScalars}, 
the symmetric singular counterterm action is the same as for Option \ref{Opt:Option1} 
discussed above, i.e., corresponds to Eq.\ (\ref{Eq:Ssct_1-Loop_inv_ASSM_Option1}).
In contrast the divergent and the finite symmetry-restoring counterterms are different.

In particular, for the divergent symmetry-restoring counterterms  we find 
in this case of Option \ref{Opt:Option2},
\begin{equation}\label{Eq:Ssct_1-Loop_break_ASSM_Option2}
    \begin{aligned}
        &S^1_{\mathrm{sct,break}} = \frac{1}{16\pi^2} \, \frac{1}{\epsilon} \Dintx
        \bigg[
        - \frac{5g^2}{9}
        \overline{B}_{\mu}\widehat{\Box}\overline{B}^{\mu}
        + \frac{2}{3} \delta y_2 \phi_{a}^{\dagger} \widehat{\Box} \phi_a\\
        &+ \delta\widehat{X}^{\text{opt2}}_{\psi,f} \Big(
        \overline{\psi_{2}}_{f} i \widehat{\slashed{\partial}} \projL {\psi_{1}}_{f}
        + \overline{\psi_{1}}_{f} i \widehat{\slashed{\partial}} \projR {\psi_{2}}_{f}
        \Big)
        - g \delta\widehat{X}^{\text{opt2}}_{G,f} \Big(
        \overline{\psi_{2}}_{f} \widehat{\slashed{B}} \projL {\psi_{1}}_{f}
        + \overline{\psi_{1}}_{f} \widehat{\slashed{B}} \projR {\psi_{2}}_{f}
        \Big)\\
        &- g^2 \delta \widehat{H}_2 \Big( \widehat{B}^{\mu}\overline{\Box}\widehat{B}_{\mu}
        + \frac{2}{3} \widehat{B}^{\mu}\widehat{\Box}\widehat{B}_{\mu} \Big)
        + g^4 \delta \widehat{H}_2 \Big( 
        \frac{1}{4} \overline{B}^{\mu}\overline{B}_{\mu}\widehat{B}^{\nu}\widehat{B}_{\nu}
        + \widehat{B}^{\mu}\widehat{B}_{\mu}\widehat{B}^{\nu}\widehat{B}_{\nu}
        \Big)\\
        &+ \frac{g^2}{2} \Big(\delta \widehat{S}_{4a}^{\text{opt2}} \phi_1^{\dagger}\phi_1 \widehat{B}^{\mu}\widehat{B}_{\mu}
        + \delta \widehat{S}_{4b}^{\text{opt2}} \phi_2^{\dagger}\phi_2 \widehat{B}^{\mu}\widehat{B}_{\mu}\Big)
        - \frac{g^2}{4} \Big( \frac{\delta \widehat{S}_{4b}^{\text{opt2}}}{2} \phi_2\phi_2 \widehat{B}^{\mu}\widehat{B}_{\mu} +\mathrm{h.c.}\Big)
        \bigg],
    \end{aligned}
\end{equation}
where we employed the notation of Eq.\ (\ref{Eq:Fermions-psi1-psi2-Option2})
for $\psi_1$ and $\psi_2$, as well as with
$\delta y_2$ and $\delta \widehat{H}_2$ as defined in 
Eqs.\ (\ref{Eq:ASSM-Ssctinv-Coeff-YukawaCombinations}) and
(\ref{Eq:ASSM-Ssctbreak-Coeff-BB}), respectively.
The coefficients for the fermionic counterterms are given by
\begin{equation}\label{Eq:ASSM-Ssctbreak-Coeff-FF-Option2}
    \begin{aligned}
        \delta\widehat{X}^{\text{opt2}}_{\psi,f} &=
        \begin{cases}
            0, &f=\nu\\
            - g^2 \frac{1-\xi}{2} \hypLR^e, &f=e\\
            g^2 \frac{5-5\xi}{18} \hypLR^u, &f=u\\
            - g^2 \frac{1-\xi}{18} \hypLR^d, &f=d,
        \end{cases}
    \end{aligned}
\end{equation}
and
\begin{equation}\label{Eq:ASSM-Ssctbreak-Coeff-FBF-Option2}
    \begin{aligned}
        \delta\widehat{X}^{\text{opt2}}_{G,f} &=
        \begin{cases}
            0, &f=\nu\\
            - g^2 \hypLR^e \Big( \frac{3+\xi}{2} - \frac{5+\xi}{3} (\hypLR^e)^2 \Big), &f=e\\
            - g^2 \hypLR^u \Big( \frac{3+\xi}{9} - \frac{5+\xi}{3} (\hypLR^u)^2 \Big) + \hypLR^d y_u y_d, &f=u\\
            g^2 \hypLR^d \Big( \frac{3+\xi}{18} + \frac{5+\xi}{3} (\hypLR^d)^2 \Big) + \hypLR^u y_u y_d, &f=d.
        \end{cases}
    \end{aligned}
\end{equation}
For the evanescent, divergent scalar-gauge interactions under Option \ref{Opt:Option2},
the coefficients are found to be
\begin{equation}\label{Eq:ASSM-Ssctbreak-Coeff-SSBB-Option2}
    \begin{aligned}
        \delta \widehat{S}_{4a}^{\text{opt2}} &= 4 y_e^2 (\hypLR^e)^2 + 12 \Big[
        \big( y_u^2 + y_d^2 \big) \big( (\hypLR^u)^2 + (\hypLR^d)^2 \big) + 2 y_u y_d \hypLR^u \hypLR^d \big) \Big],\\
        \delta \widehat{S}_{4b}^{\text{opt2}} &= 8 y_e^2 (\hypLR^e)^2 + 24 \Big[ y_u^2 (\hypLR^u)^2 + y_d^2 (\hypLR^d)^2\Big].
    \end{aligned}
\end{equation}
In Eq.\ (\ref{Eq:Ssct_1-Loop_break_ASSM_Option2}), with the additional choice  $\hypLR=\hypRL=0$ only the first line remains non-zero while all other terms vanish.  In particular in this case Eq.\ \eqref{Eq:Ssct_1-Loop_break_ASSM_Option2} is free of any 
fermionic contributions.
This represents a significant simplification of the divergent symmetry-restoring 
counterterm action compared to Option \ref{Opt:Option1} analysed above
(cf.\ Eq.\ \eqref{Eq:Ssct_1-Loop_break_ASSM_Option1}).
However, when evanescent hypercharges are non-zero, the difference between the two options is not substantial,
primarily reflected by different coefficients and the exchange of some field monomials with
others.

Furthermore, note that Eq.\ \eqref{Eq:Ssct_1-Loop_break_ASSM_Option2} is not entirely free
of global hypercharge violation either, due to evanescent hypercharges as a source of
chirality-mixing.
Specifically, the last term of the final line in Eq.\ \eqref{Eq:Ssct_1-Loop_break_ASSM_Option2} of the form $\phi_2\phi_2\widehat{B}^{\mu}\widehat{B}_{\mu}+\mathrm{h.c.}$
clearly violates global hypercharge conservation.
However, in contrast to the $\phi_2\widehat{\Box}\phi_2+\mathrm{h.c.}$ term in Option \ref{Opt:Option1}, 
it vanishes identically for $\hypLR=\hypRL=0$.

In terms of the gauge boson contributions, 
Option \ref{Opt:Option2}, Eq.\ \eqref{Eq:Ssct_1-Loop_break_ASSM_Option2},
features fewer bilinear contributions compared to 
Option \ref{Opt:Option1}, Eq.\ \eqref{Eq:Ssct_1-Loop_break_ASSM_Option1},
but it introduces entirely new quartic gauge boson contributions, 
as seen in the second term of line three in Eq.\ \eqref{Eq:Ssct_1-Loop_break_ASSM_Option2}.
All of these gauge boson counterterms are fully governed by the evanescent hypercharges
and the reason for the differences arises from the different structure of the evanescent
hypercharge matrices, cf.\ Eqs.\ \eqref{Eq:ASSM-YLR} and \eqref{Eq:ASSM-YLR-YRL-Option2},
as previously noted in Eq.\ \eqref{Eq:BBBB-GreenFunc}.

The finite symmetry-restoring counterterm action under Option \ref{Opt:Option2}
is found to be
\begin{equation}\label{Eq:Sfct_1-Loop_ASSM_Option2}
    \begin{aligned}
        S^1_{\mathrm{fct}} &= \frac{1}{16\pi^2} \intx \bigg[
        -\frac{5g^2}{9} \overline{B}_{\mu} \overline{\Box} \, \overline{B}^{\mu}
        + \frac{95g^4}{648} \overline{B}_{\mu} \overline{B}^{\mu} \overline{B}_{\nu} \overline{B}^{\nu}\\
        &- i \frac{g}{6} \delta y_2
        \Big[
        \big(\overline{\partial}^{\mu}\phi_{a}^{\dagger}\big)\phi_a
        - \phi_{a}^{\dagger}\big(\overline{\partial}^{\mu}\phi_a\big) 
        \Big] 
        \overline{B}_{\mu}
        - \frac{g^2}{4} \delta y_2^{\text{opt2}}
        \phi_{a}^{\dagger}\phi_{a}\overline{B}^{\mu}\overline{B}_{\mu}\\
        &+ g \Big( \delta F_{\mathrm{R},f}^{\text{opt2}} \, \overline{\psi_2}_f \overline{\slashed{B}} \projR {\psi_2}_f 
        + \delta F_{\mathrm{L},f}^{\text{opt2}} \, \overline{\psi_1}_f \overline{\slashed{B}} \projL {\psi_1}_f \Big)\\
        &- \Big\{
        \delta Y_{e,2}^{\text{opt2}} y_e \overline{e_{L}} \phi_{2}^{\dagger} e_{R}
        + \delta Y_{u,2}^{\text{opt2}} y_u \overline{u_{L}} \phi_{2} u_{R} 
        + \delta Y_{d,2}^{\text{opt2}} y_d \overline{d_{L}} \phi_{2}^{\dagger} d_{R}
        +\mathrm{h.c.} \Big\}
        \bigg],
    \end{aligned}
\end{equation}
with coefficients $\delta y_2$ as defined in 
Eqs.\ (\ref{Eq:ASSM-Ssctinv-Coeff-YukawaCombinations}) and
\begin{equation}\label{Eq:ASSM-Sfct-Coeff-YukawaCombinations-Option2}
    \begin{aligned}
        \delta y_2^{\text{opt2}} = 3 y_e^2 + \frac{13}{3} y_u^2 + \frac{7}{3} y_d^2.
    \end{aligned}
\end{equation}
The coefficients of the finite fermion-gauge interactions under Option \ref{Opt:Option2} are given by
\begin{equation}\label{Eq:ASSM-Sfct-Coeff-FBF-R-Option2}
    \begin{aligned}
        \delta F_{\mathrm{R},f}^{\text{opt2}} &=
        \begin{cases}
            0, &f=\nu\\
            - g^2 \Big( \frac{5+\xi}{6} + \frac{1+\xi}{4} (\hypLR^e)^2 \Big) - \frac{y_e^2}{2}, &f=e\\
            g^2 \Big( \frac{20+4\xi}{81} + \frac{11+7\xi}{36} (\hypLR^u)^2 \Big) + \frac{y_u^2}{6}, &f=u\\
            - g^2 \Big( \frac{5+\xi}{162} + \frac{13+5\xi}{36} (\hypLR^d)^2 \Big) + \frac{y_d^2}{6}, &f=d,
        \end{cases}
    \end{aligned}
\end{equation}
\begin{equation}\label{Eq:ASSM-Sfct-Coeff-FBF-L-Option2}
    \begin{aligned}
        \delta F_{\mathrm{L},f}^{\text{opt2}} &=
        \begin{cases}
            - g^2 \frac{5+\xi}{48} - \frac{y_e^2}{2}, &f=\nu\\
            - g^2 \Big( \frac{5+\xi}{48} - \frac{1}{2} (\hypLR^e)^2 \Big) - \frac{y_e^2}{2}, &f=e\\
            g^2 \Big( \frac{5+\xi}{1296} - \frac{8+\xi}{18} (\hypLR^u)^2 \Big) + \frac{2 y_u^2 - y_d^2}{6}, &f=u\\
            g^2 \Big( \frac{5+\xi}{1296} + \frac{7+2\xi}{18} (\hypLR^d)^2 \Big) + \frac{2 y_u^2 - y_d^2}{6}, &f=d,
        \end{cases}
    \end{aligned}
\end{equation}
whereas the finite Yukawa contributions come with the coefficients
\begin{equation}\label{Eq:ASSM-Sfct-Coeff-Yukawas-Option2}
    \begin{aligned}
        \delta Y_{f,2}^{\text{opt2}} &=g^2 \frac{5+\xi}{3} (\hypLR^f)^2, \quad &&\text{with} \quad f\in\{e,u,d\}.
    \end{aligned}
\end{equation}
In contrast to Eq.\ \eqref{Eq:Sfct_1-Loop_ASSM_Option1} for Option \ref{Opt:Option1},
the finite Yukawa contributions, that violate global hypercharge conservation and are
shown in the last line of Eq.\ \eqref{Eq:Sfct_1-Loop_ASSM_Option2},
vanish completely for $\hypLR=\hypRL=0$, as they are entirely governed by the 
evanescent hypercharges.
This is because the evanescent gauge interactions are the only source of potential
global hypercharge violation in Option \ref{Opt:Option2}.
Analogously, the global hypercharge violating contributions from the scalar fields,
present in the last two lines of Eq.\ \eqref{Eq:Sfct_1-Loop_ASSM_Option1}
in the case of Option \ref{Opt:Option1}, are completely avoided here.
As a result, Option \ref{Opt:Option2} yields a shorter finite symmetry-restoring 
counterterm action compared to Option \ref{Opt:Option1}.

\paragraph{Concluding Comments:}
As explained earlier, there are two sources of (spurious) symmetry breaking
induced by a non-anticommuting $\gamma_5$ within the BMHV scheme.
With this in mind the main results of this section can be summarised as follows:
\begin{itemize}
    \item Choosing  Option \ref{Opt:Option1} or  \ref{Opt:Option2} does not change the divergent symmetric counterterm action.
    \item The evanescent part of the fermion kinetic terms break global hypercharge invariance in the natural
        Option \ref{Opt:Option1} but not in Option \ref{Opt:Option2} with sterile fermion partners.
        Hence, in general 
        Option \ref{Opt:Option2} leads to a more concise finite symmetry-restoring 
        counterterm action.
    \item For general evanescent hypercharges $\hypLR\ne0$,
        the (structural) difference for the divergent symmetry-restoring part is mainly limited
        to the presence of different evanescent field monomials in certain cases.
        For example, in Option \ref{Opt:Option1}, there is a divergent counterterm
        of the form $\phi_2\widehat{\Box}\phi_2+\mathrm{h.c.}$, entirely 
        independent of evanescent hypercharges, whereas in
        Option \ref{Opt:Option2}, there is one of the form
        $\phi_2\phi_2\widehat{B}^{\mu}\widehat{B}_{\mu}+\mathrm{h.c.}$,
        fully governed by evanescent hypercharges.
        Both of these violate global hypercharge conservation.
    \item For vanishing 
        evanescent gauge interactions $\hypLR=0$, both options show significant simplifications.
        However, Option \ref{Opt:Option2} has the additional advantage that in this case
        global hypercharge violation is completely eliminated, thereby removing both sources for global hypercharge violation
        and rendering the symmetry-restoring parts of the counterterm action even more concise.
\end{itemize}
While these remarks suggest
that Option \ref{Opt:Option2} is advantageous compared to Option \ref{Opt:Option1},
it is crucial to consider the significant problem associated with 
fermion propagators in the massive case within the framework of Option \ref{Opt:Option2},
as discussed in Sec.\ \ref{Sec:FermionsInGeneral}.
This complicates the decision of choosing one method in such cases.

In conclusion, analysing different concrete values for the evanescent hypercharges
is of great interest, with the case $\hypLR=\hypRL=0$ already standing out.
This investigation will be conducted in the following subsection for both options.

\subsection{The Impact of the Evanescent Hypercharges in the Abelian SM}
\label{Sec:TheRoleOf-YLR-YRL}

As discussed in Sec.\ \ref{Sec:The-Model}, the evanescent gauge interactions do not preserve chirality, 
and thus serve as a source of both local and global symmetry breaking, in addition to the
evanescent part of the fermion kinetic term.
The effects of this have explicitly been shown in the previous section for general evanescent hypercharges.
In this section, we study different choices for explicit values of these evanescent hypercharges.

We begin by recalling the fermion-gauge interactions in their most general form with arbitrary hypercharges
only constrained by hermiticity $\hypRL = \hypLR^{\dagger}$, see Eq.\ (\ref{Eq:Requirement-YLR-YRL}). The general ansatz in $D$ dimensions Eq.\ \eqref{Eq:Lfermion} contains four independent covariants,\footnote{
In contrast, there are only two such covariants in $4$ dimensions or in naive schemes with anticommuting $\gamma_5$.} and it is instructive to write the interaction terms in two equivalent ways as
\begin{equation}\label{Eq:GaugeInt-Vector-Axial-Current}
    \begin{aligned}
        &- g B_{\mu} \overline{\psi}_i \bigg[
        {\hypR}_{ij} \projL \overline{\gamma}^{\mu} \projR
        + {\hypL}_{ij} \projR \overline{\gamma}^{\mu} \projL
        + {\hypRL}_{ij} \projL \widehat{\gamma}^{\mu} \projL
        + {\hypLR}_{ij} \projR \widehat{\gamma}^{\mu} \projR 
        \bigg] \psi_j\\
        =&
        -g B_{\mu} \overline{\psi}_i \bigg[
        \frac{{\hypR}_{ij}+{\hypL}_{ij}}{2} \overline{\gamma}^{\mu}
        + \frac{{\hypLR}_{ij}+{\hypLR^{*}}_{ji}}{2} \widehat{\gamma}^{\mu}\\
        &\qquad\qquad\,\,\,\,
        + \frac{{\hypR}_{ij}-{\hypL}_{ij}}{2} \overline{\gamma}^{\mu}\gamma_5
        + \frac{{\hypLR}_{ij}-{\hypLR^{*}}_{ji}}{2} \widehat{\gamma}^{\mu}\gamma_5
        \bigg] \psi_j
    \end{aligned}
\end{equation}
The first version is expressed in terms of left- and right-handed currents, the second version exhibits the 4-dimensional vector and axial vector currents, and their evanescent analogs.

\begin{table}[t!]
    \centering
    \begin{tabular}{|c|c|c|} \hline 
         fermion current & $\mathrm{h.c.}$ & hermiticity \\ \hline \hline
         \rule{0pt}{1.1em}$\overline{\psi}\overline{\gamma}^{\mu}\psi$ & $\overline{\psi}\overline{\gamma}^{\mu}\psi$ & hermitian \\ \hline
         \rule{0pt}{1.1em}$\overline{\psi}\widehat{\gamma}^{\mu}\psi$ & $\overline{\psi}\widehat{\gamma}^{\mu}\psi$ & hermitian \\ \hline
         \rule{0pt}{1.1em}$\overline{\psi}\overline{\gamma}^{\mu}\gamma_5\psi$ & $\overline{\psi}\overline{\gamma}^{\mu}\gamma_5\psi$ & hermitian \\ \hline
         \rule{0pt}{1.1em}$\overline{\psi}\widehat{\gamma}^{\mu}\gamma_5\psi$ & $-\,\overline{\psi}\widehat{\gamma}^{\mu}\gamma_5\psi$ & anti-hermitian \\ \hline
         \rule{0pt}{1.1em}$\overline{\psi}\projR\overline{\gamma}^{\mu}\projL\psi$ & $\overline{\psi}\projR\overline{\gamma}^{\mu}\projL\psi$ & hermitian \\ \hline
         \rule{0pt}{1.1em}$\overline{\psi}\projL\overline{\gamma}^{\mu}\projR\psi$ & $\overline{\psi}\projL\overline{\gamma}^{\mu}\projR\psi$ & hermitian \\ \hline
         \rule{0pt}{1.1em}$\overline{\psi}\projL\widehat{\gamma}^{\mu}\projL\psi$ & $\overline{\psi}\projR\widehat{\gamma}^{\mu}\projR\psi$ & --- \\ \hline
         \rule{0pt}{1.1em}$\overline{\psi}\projR\widehat{\gamma}^{\mu}\projR\psi$ & $\overline{\psi}\projL\widehat{\gamma}^{\mu}\projL\psi$ & --- \\ \hline
    \end{tabular}
    \caption{Hermiticity properties of fermion currents appearing in $D$ dimensions.}
    \label{Tab:Currents-Hermiticity}
\end{table}
The hermiticity properties of the various
fermion currents appearing in 
Eq.\ \eqref{Eq:GaugeInt-Vector-Axial-Current} are summarised in Tab.\ \ref{Tab:Currents-Hermiticity}:
Accordingly,
the prefactors of $\overline{\gamma}^{\mu}$, $\widehat{\gamma}^{\mu}$ and $\overline{\gamma}^{\mu}\gamma_5$
are hermitian, whereas the prefactor of $\widehat{\gamma}^{\mu}\gamma_5$ is anti-hermitian. Hence the vector part can be uniformly extended to $D$ dimensions by assigning equal coefficients to $\widehat{\gamma}^\mu$ and $\overline{\gamma}^\mu$, as done in the standard treatment of QED discussed in Sec.~\ref{Sec:FermionicSectorNoScalars}. But the axial component of the gauge interaction
cannot be extended uniformly to $D$ dimensions,
because $\overline{\gamma}^{\mu}\gamma_5$ and $\widehat{\gamma}^{\mu}\gamma_5$ must
have different types of coefficients 
--- hermitian for the former and anti-hermitian for the latter.

\begin{table}[t!]
%    \scriptsize
    \tiny
    \centering
    \begin{tabular}{|c||c|c|c|c|c|} \hline
         & $0$ & $\frac{\hypR+\hypL}{2}$ & $Q=\hypR$ & $\hypL$  \\ \hline \hline
        $\delta \widehat{H}_1$ & $0$ & $-\frac{53}{12}$ & $-\frac{20}{3}$ & $-\frac{13}{6}$  \\ \hline
        $\delta \widehat{H}_2$ & $0$ & $+\frac{53}{48}$ & $+\frac{8}{3}$ & $+\frac{5}{12}$  \\ \hline\hline
        $\delta\widehat{X}^{\text{opt1}}_{\psi,\nu}$ & $0$ & $0$ & $0$ & $0$ \\ \hline
        $\delta\widehat{X}^{\text{opt1}}_{\psi,e}$ & $-g^2$ & $-\frac{7}{16}g^2$ & $0^*$ & $-\frac{3}{4}g^2$  \\ \hline
        $\delta\widehat{X}^{\text{opt1}}_{\psi,u}$ & $-\frac{2}{9}g^2 + \frac{y_u y_d}{2}$ & $-\frac{7}{144}g^2 + \frac{y_u y_d}{2}$ & $+\frac{2}{9}g^2 + \frac{y_u y_d}{2}$ & $-\frac{7}{36}g^2 + \frac{y_u y_d}{2}$ \\ \hline
        $\delta\widehat{X}^{\text{opt1}}_{\psi,d}$ & $+\frac{1}{9}g^2+\frac{y_u y_d}{2}$ & $+\frac{17}{144}g^2+\frac{y_u y_d}{2}$ & $+\frac29g^2+\frac{y_u y_d}{2}$ & $+\frac{5}{36}g^2+\frac{y_u y_d}{2}$  \\ \hline\hline
        $\delta\widehat{X}^{\text{opt1}}_{G,\nu}$ & $0$ & $0$ & $0$ & $0$  \\ \hline
        $\delta\widehat{X}^{\text{opt1}}_{G,e}$ & $0$ & $+\frac{21}{64}g^2$ & $0^*$ & $+\frac38 g^2$  \\ \hline
        $\delta\widehat{X}^{\text{opt1}}_{G,u}$ & $+\frac14{y_uy_d}$ & $-\frac{35}{1728}g^2 + \frac{5}{24}y_u y_d$ & $+ \frac{4}{27}g^2 + \frac{1}{12}y_u y_d$ & $-\frac{7}{216}g^2 + \frac{1}{3}y_u y_d$  \\ \hline
        $\delta\widehat{X}^{\text{opt1}}_{G,d}$ & $-\frac14{y_uy_d}$ & $- \frac{17}{1728}g^2 - \frac{1}{24}y_u y_d$ & $- \frac{2}{27}g^2 + \frac{1}{12}y_u y_d$ & $+ \frac{5}{216}g^2 - \frac{1}{6}y_u y_d$  \\ \hline\hline
        $\delta\widehat{S}^{\text{opt1}}_{3}$ & $0$ & $-\frac{3}{2}(\frac12 y_e^2 + y_u^2 + y_d^2 - y_u y_d )$ & $-y_e^2 -3(y_u^2 + y_d^2 - y_u y_d)$ & $- \frac12 y_e^2$  \\ \hline
        $\delta\widehat{S}^{\text{opt1}}_{4}$ & $0$ & $-\frac{3}{4}(\frac34 y_e^2 + y_u^2 + y_d^2 - y_u y_d)$ & $-y_e^2 -3(y_u^2 + y_d^2 - y_u y_d)$ & $- \frac14 y_e^2$  \\ \hline\hline
        $\delta F_{\mathrm{R},\nu}^{\text{opt1}}$ & $0$ & $0$ & $0$ & $0$  \\ \hline
        $\delta F_{\mathrm{R},e}^{\text{opt1}}$ & $-\frac12 g^2$ & $-\frac{25}{32}g^2$ & $-g^2$ & $-\frac{5}{8}g^2$  \\ \hline
        $\delta F_{\mathrm{R},u}^{\text{opt1}}$ & $+\frac29 g^2$ & $+\frac{89}{288}g^2$ & $+\frac49g^2$ & $+\frac{17}{72}g^2$ \\ \hline
        $\delta F_{\mathrm{R},d}^{\text{opt1}}$ & $-\frac{1}{18} g^2$ & $-\frac{17}{288}g^2$ & $-\frac{1}{9}g^2$ & $-\frac{5}{72}g^2$  \\ \hline\hline
        $\delta F_{\mathrm{L},\nu}^{\text{opt1}}$ & $- \frac18 g^2 - \frac14 y_e^2$ & $- \frac18 g^2 - \frac14 y_e^2$ & $-\frac18g^2 - \frac14 y_e^2$ & $-\frac18g^2 - \frac14 y_e^2$  \\ \hline
        $\delta F_{\mathrm{L},e}^{\text{opt1}}$ & $+ \frac18 g^2 - \frac14 y_e^2$ & $+ \frac{13}{32} g^2 - \frac14 y_e^2$  & $+\frac58g^2 - \frac14 y_e^2$ & $+\frac14g^2 - \frac14 y_e^2$ \\ \hline
        $\delta F_{\mathrm{L},u}^{\text{opt1}}$ & $-\frac{1}{72}g^2 + \frac14(y_u^2 - y_d^2)$ & $-\frac{29}{288}g^2 + \frac14(y_u^2 - y_d^2)$ & $-\frac{17}{72}g^2 + \frac14(y_u^2 - y_d^2)$ & $-\frac{1}{36}g^2 + \frac14(y_u^2 - y_d^2)$  \\ \hline
        $\delta F_{\mathrm{L},d}^{\text{opt1}}$ & $+\frac{1}{72}g^2 + \frac14(y_u^2 - y_d^2)$ & $+\frac{5}{288}g^2 + \frac14(y_u^2 - y_d^2)$ & $+\frac{5}{72}g^2 + \frac14(y_u^2 - y_d^2)$ & $+\frac{1}{36}g^2 + \frac14(y_u^2 - y_d^2)$  \\ \hline\hline
        $\delta Y_{e,2}^{\text{opt1}}$ & $+g^2$ & $+\frac{17}{8}g^2$ & $+3g^2$ & $+\frac32 g^2$  \\ \hline
        $\delta Y_{u,2}^{\text{opt1}}$ & $+\frac{2}{9}g^2 + \frac{y_d^2}{2}$ & $+\frac{41}{72}g^2 + \frac{y_d^2}{2}$ & $+\frac{10}{9}g^2 + \frac{y_d^2}{2}$ & $+\frac{5}{18}g^2 + \frac{y_d^2}{2}$  \\ \hline
        $\delta Y_{d,2}^{\text{opt1}}$ & $-\frac{1}{9}g^2 + \frac{y_u^2}{2}$ & $-\frac{7}{72}g^2 + \frac{y_u^2}{2}$ & $+\frac{1}{9}g^2 + \frac{y_u^2}{2}$ & $-\frac{1}{18}g^2 + \frac{y_u^2}{2}$ \\ \hline
    \end{tabular}
    \caption{Explicit results for the coefficients of the divergent and finite symmetry-restoring counterterms 
    for Option \ref{Opt:Option1},
    as shown in Eqs.\ \eqref{Eq:Ssct_1-Loop_break_ASSM_Option1} and \eqref{Eq:Sfct_1-Loop_ASSM_Option1},
    respectively.
    The results are given in Feynman gauge, i.e.\ for $\xi=1$, and the asterisk indicates cases where the coefficient vanishes
    only for $\xi=1$ but not for other gauge choices.
    % The different values of the evanescent hypercharges are presented across the columns, while the rows list
    % the various coefficients for the respective set of evanescent hypercharges.
    }
    \label{Tab:YLR-Specialisations-Option-1}
\end{table}
\begin{table}[t!]
%    \scriptsize
    \tiny
    \centering
    \begin{tabular}{|c||c|c|c|c|c|c|} \hline 
         & $0$ & $\frac{\hypR+\hypL}{2}$ & $Q=\hypR$ & $\hypL$ \\ \hline \hline
        $\delta \widehat{H}_1$ & $0$ & $-\frac{53}{12}$ & $-\frac{20}{3}$ & $-\frac{13}{6}$ \\ \hline
        $\delta \widehat{H}_2$ & $0$ & $\frac{53}{48}$ & $\frac{8}{3}$ & $\frac{5}{12}$ \\ \hline\hline
        $\delta\widehat{X}^{\text{opt2}}_{\psi,\nu}$ & $0$ & $0$ & $0$ & $0$ \\ \hline
        $\delta\widehat{X}^{\text{opt2}}_{\psi,e}$ & $0$ & $0^*$ & $0^*$ & $0^*$ \\ \hline
        $\delta\widehat{X}^{\text{opt2}}_{\psi,u}$ & $0$ & $0^*$ & $0^*$ & $0^*$ \\ \hline
        $\delta\widehat{X}^{\text{opt2}}_{\psi,d}$ & $0$ & $0^*$ & $0^*$ & $0^*$ \\ \hline\hline
        $\delta\widehat{X}^{\text{opt2}}_{G,\nu}$ & $0$ & $0$ & $0$ & $0$ \\ \hline
        $\delta\widehat{X}^{\text{opt2}}_{G,e}$ & $0$ & $+ \frac{21}{32}g^2$ & $0^*$ & $+\frac34 g^2$ \\ \hline
        $\delta\widehat{X}^{\text{opt2}}_{G,u}$ & $0$ & $- \frac{35}{864}g^2 - \frac{1}{12}y_u y_d$ & $+\frac{8}{27}g^2 - \frac{1}{3}y_uy_d$ & $-\frac{7}{108}g^2 + \frac{1}{6}y_uy_d$ \\ \hline
        $\delta\widehat{X}^{\text{opt2}}_{G,d}$ & $0$ & $- \frac{17}{864}g^2 + \frac{5}{12}y_u y_d$ & $-\frac{4}{27}g^2 + \frac{2}{3}y_uy_d$ & $+\frac{5}{108}g^2 + \frac{1}{6}y_uy_d$ \\ \hline\hline
        $\delta\widehat{S}^{\text{opt2}}_{4a}$ & $0$ & $\frac94 y_e^2 + \frac{13}{6}(y_u^2 + y_d^2) - \frac{5}{6} y_u y_d$ & $4 y_e^2 + \frac{20}{3}(y_u^2 +y_d^2) - \frac{16}{3} y_u y_d$ & $y_e^2 + \frac23(y_u^2 + y_d^2 + y_u y_d)$ \\ \hline
        $\delta\widehat{S}^{\text{opt2}}_{4b}$ & $0$ & $\frac92 y_e^2 + \frac16(25 y_u^2 + y_d^2)$ & $8y_e^2 + \frac13(32y_u^2 + 8y_d^2)$ & $2y_e^2 + \frac23(y_u^2 + y_d^2)$ \\ \hline\hline
        $\delta F_{\mathrm{R},\nu}^{\text{opt2}}$ & $0$ & $0$ & $0$ & $0$ \\ \hline
        $\delta F_{\mathrm{R},e}^{\text{opt2}}$ & $-g^2 -\frac12 y_e^2$ & $-\frac{41}{32}g^2 -\frac12 y_e^2$ & $-\frac32g^2 - \frac12 y_e^2$ & $-\frac98g^2 - \frac12 y_e^2$ \\ \hline
        $\delta F_{\mathrm{R},u}^{\text{opt2}}$ & $+\frac{8}{27} g^2 + \frac16 y_u^2$ & $+\frac{331}{864}g^2 +\frac16 y_u^2$ & $+\frac{14}{27}g^2 + \frac16 y_u^2$ & $+\frac{67}{216}g^2 + \frac16 y_u^2$ \\ \hline
        $\delta F_{\mathrm{R},d}^{\text{opt2}}$ & $-\frac{1}{27}g^2 + \frac16 y_d^2$ & $-\frac{35}{864}g^2 +\frac16 y_d^2$ & $-\frac{5}{54}g^2 + \frac16 y_d^2$ & $-\frac{11}{216}g^2 + \frac16 y_d^2$ \\ \hline\hline
        $\delta F_{\mathrm{L},\nu}^{\text{opt2}}$ & $-\frac18g^2 - \frac12y_e^2$ & $-\frac18g^2 - \frac12y_e^2$ & $-\frac18g^2 - \frac12y_e^2$ & $-\frac18g^2 - \frac12y_e^2$ \\ \hline
        $\delta F_{\mathrm{L},e}^{\text{opt2}}$ & $-\frac18g^2 - \frac12y_e^2$ & $+\frac{5}{32}g^2 - \frac12 y_e^2$ & $+\frac38g^2 - \frac12 y_e^2$ & $-\frac12 y_e^2$ \\ \hline
        $\delta F_{\mathrm{L},u}^{\text{opt2}}$ & $\frac{1}{216}g^2 +\frac13 y_u^2 - \frac16 y_d^2$ & $-\frac{71}{864}g^2 + \frac13y_u^2 - \frac16y_d^2$ & $-\frac{47}{216}g^2 +\frac13 y_u^2 - \frac16 y_d^2$ & $-\frac{1}{108}g^2 +\frac13 y_u^2 - \frac16 y_d^2$ \\ \hline
        $\delta F_{\mathrm{L},d}^{\text{opt2}}$ & $\frac{1}{216}g^2 +\frac13 y_u^2 - \frac16 y_d^2$ & $+\frac{7}{864}g^2 +\frac13 y_u^2 - \frac16 y_d^2$ & $+\frac{13}{216}g^2+\frac13 y_u^2 - \frac16 y_d^2$ & $+\frac{1}{54}g^2 +\frac13 y_u^2 - \frac16 y_d^2$ \\ \hline\hline
        $\delta Y_{e,2}^{\text{opt2}}$ & $0$ & $+\frac98 g^2$ & $+2g^2 $ & $+\frac12 g^2$ \\ \hline
        $\delta Y_{u,2}^{\text{opt2}}$ & $0$ & $+\frac{25}{72}g^2$ & $+\frac89g^2$ & $+\frac{1}{18}g^2$ \\ \hline
        $\delta Y_{d,2}^{\text{opt2}}$ & $0$ & $+\frac{1}{72}g^2$ & $+\frac29g^2$ & $+\frac{1}{18}g^2$ \\ \hline
    \end{tabular}
    \caption{Explicit results for the coefficients of the divergent and finite symmetry-restoring counterterms 
    for Option \ref{Opt:Option2}, as shown in Eqs.\ \eqref{Eq:Ssct_1-Loop_break_ASSM_Option2} and \eqref{Eq:Sfct_1-Loop_ASSM_Option2},
    respectively.
    The results are given in Feynman gauge, i.e.\ for $\xi=1$, and the asterisk indicates 
    cases where the coefficient vanishes only for $\xi=1$ but not for other gauge choices.
    % The different values of the evanescent hypercharges are presented across the columns, while the rows list
    % the various coefficients for the respective set of evanescent hypercharges.
    }
    \label{Tab:YLR-Specialisations-Option-2}
\end{table}

Based on the appearances of the evanescent hypercharges in the fermion--gauge interactions Eq.~(\ref{Eq:GaugeInt-Vector-Axial-Current}) we can identify several motivated choices of the $\hypLR$.
\begin{itemize}
\item $\hypLR=\hypRL=0$: In this simplest choice, the interactions become purely 4-dimensional,
\begin{align}
        -g B_{\mu}
            \overline{\psi}_i \bigg[
            \frac{{\hypR}_{ij}+{\hypL}_{ij}}{2} \overline{\gamma}^{\mu}
            + \frac{{\hypR}_{ij}-{\hypL}_{ij}}{2} \overline{\gamma}^{\mu}\gamma_5
            \bigg] \psi_j.  
\end{align}
We have already seen in the previous subsection that this choice simplifies the results for the counterterms and eliminates one source of global hypercharge violation.
\item $\hypLR=\hypRL=\frac{\hypR+\hypL}{2}$: In this case, the vector part of the interaction becomes fully $D$-dimensional,
\begin{align}
        -g B_{\mu}
            \overline{\psi}_i \bigg[
            \frac{{\hypR}_{ij}+{\hypL}_{ij}}{2} \gamma^{\mu}
            + \frac{{\hypR}_{ij}-{\hypL}_{ij}}{2} \overline{\gamma}^{\mu}\gamma_5
            \bigg] \psi_j.
\end{align}
This uniform extension of the vector interaction to $D$ dimensions is the standard treatment of QED discussed in Sec.~\ref{Sec:FermionicSectorNoScalars}. 
\item Other options: Interestingly, the axial component of the gauge interaction
cannot be extended uniformly to $D$-dimensions, as explained above.
For this reason no other choice of $\hypLR$ leads to results as simple as the ones above. 
Nevertheless, a noteworthy case is the choice $\hypLR=Q$, where $Q$ is the electric charge. 
In analogy to the previous case, such a choice might be of interest for the full SM, 
where one might want to treat the photon fully $D$-dimensionally. 
Since here, $Q=\hypR$, this choice amounts to the fermion--gauge interaction
\begin{align}
        -g B_{\mu}
            \overline{\psi}_i \bigg[
            \frac{-{\hypR}_{ij}+{\hypL}_{ij}}{2} \overline{\gamma}^{\mu} + Q_{ij}\gamma^\mu
            + \frac{{\hypR}_{ij}-{\hypL}_{ij}}{2} \overline{\gamma}^{\mu}\gamma_5
            \bigg] \psi_j.
\end{align}
This, and similar choices such as $\hypLR=\hypL$ or $\hypLR=(\hypR-\hypL)/2$, do not lead to significant simplifications compared to the generic case.
\end{itemize}
We have explicitly worked out all counterterm coefficients for the ASM results discussed in Sec.\ \ref{Sec:AnalysisOfASSMResults} for all
such motivated evanescent hypercharge values. The results for the most representative cases where $\hypLR$ is set to $0$, $\frac{\hypR+\hypL}{2}$, $Q$, $\hypL$, respectively, are shown in two tables:
in Tab.\ \ref{Tab:YLR-Specialisations-Option-1}
for Option \ref{Opt:Option1} and in Tab.\ \ref{Tab:YLR-Specialisations-Option-2} for Option \ref{Opt:Option2}.
The different values of the evanescent hypercharges are presented across the columns, while the rows list
the various coefficients for the respective set of evanescent hypercharges.

The tables confirm that setting $\hypLR=\hypRL=0$ leads to the simplest results. 
In particular, the bosonic contributions governed by $\delta\widehat{H}_1$,
$\delta\widehat{H}_2$, $\delta\widehat{S}^{\text{opt1}}_{3}$, $\delta\widehat{S}^{\text{opt1}}_{4}$,
$\delta\widehat{S}^{\text{opt2}}_{4a}$ and $\delta\widehat{S}^{\text{opt2}}_{4b}$ vanish
entirely for this choice.
In  case of Option \ref{Opt:Option1} with physical left- and right-handed fermions the simplicity affects mainly the divergent counterterms.  
The advantage of  setting $\hypLR=\hypRL=0$ is even more significant if Option \ref{Opt:Option2} with sterile fermions is used such that global hypercharge is manifestly preserved. 
None of the other settings for the evanescent hypercharges leads to noteworthy simplifications.

% -----------------------------------------------------------------------------

% +++++++++++++++++++++++++++++++++++++++++++++++++++++++++++++++++++++++++++++
\section{Conclusion}

We have worked out the impact of different implementations of the BMHV scheme for $\gamma_5$ 
in chiral gauge theories. 
For a Dirac fermion $\psi$, we allow different approaches for its $D$-dimensional extension,
resulting in different fermion kinetic terms. One possibility is the 
obvious $D$-dimensional kinetic term involving physical left-handed and right-handed fermions (Option \ref{Opt:Option1}), another possibility is to add fictitious sterile fields to the theory (Option \hyperref[Opt:Option2a-2b-GroupReference]{2}),
as discussed in Sec.\ \ref{Sec:FermionsInGeneral}. 
Both break local gauge and BRST invariance and Option  \ref{Opt:Option1} is more natural, but Option \hyperref[Opt:Option2a-2b-GroupReference]{2} has the advantage to
at least preserve global gauge invariance. 
For the gauge interaction we allow arbitrary contributions of interaction terms with evanescent currents of the form $\overline{\psi}\widehat{\gamma}^{\mu}\psi$, as shown in 
Sec.\ \ref{Sec:D-DimLagrangian}. 
We thus consider the most general $D$-dimensional interactions between gauge bosons and fermions for interactions, including the examples shown in Eq.\ (\ref{Eq:QED_vs_BMHV-Z-Boson}) 
as special cases.

The symmetry-restoring counterterms have been evaluated in general in Sec.\ \ref{Sec:1-Loop-Ren-Results}, 
and the results have also been specialised to a number of scenarios of 
interest such as the SM matter content, see Sec.\ \ref{Sec:AnalysisOfResults}.
As expected, the results demonstrate that the spurious symmetry breaking indeed depends on
both evanescent details of the regularisation reviewed above ---
the specifics of the
fermion kinetic term and of the gauge interactions.

In particular, we observe a significant proliferation of terms if the evanescent
gauge interactions $\sim\widehat{\gamma}^{\mu}$ are included. 
Even for plausible choices for the evanescent hypercharges such as $\hypL$, $\hypR$ or $(\hypR+\hypL)/2$,
the symmetry-restoring counterterms are significantly more involved than in the case where $\hypLR=\hypRL=0$,
as illustrated in Sec.\ \ref{Sec:TheRoleOf-YLR-YRL}.
For many applications it will therefore be advantageous to omit such evanescent interactions and keep $\hypLR=\hypRL=0$.
Even for QED, which emerges as a vector-like gauge theory from SSB in the EWSM, a 
purely $4$-dimensional treatment of the photon appears to be the most reasonable
compromise, particularly in the light of the discussions presented at the end of Sec.\ 
\ref{Sec:FermionicSectorNoScalars}.

The situation for the kinetic term is slightly different. 
We compared both options for the $D$-dimensional extension
of the fermions in the massless case with a general configuration of the evanescent hypercharges
in Sec.\ \ref{Sec:AnalysisOfASSMResults} in detail and refer to the end of Sec.\ \ref{Sec:AnalysisOfASSMResults} for a more detailed summary of our findings.
Evanescent kinetic terms of the form $\overline{\psi_L}i\widehat{\slashed{\partial}}\psi_R$ for a physical, chiral fermion inevitably break global 
gauge invariance which leads to additional symmetry-restoring counterterms. 
This could  be avoided if the field content is enlarged by  the additional, fictitious sterile fields according to our Option \hyperref[Opt:Option2a-2b-GroupReference]{2}. 
This in turn is not problematic in the general, purely massless calculations, as e.g.\ done in Ref.\ \cite{Martin:1999cc,Belusca-Maito:2020ala,Belusca-Maito:2021lnk,Stockinger:2023ndm,Kuhler:2024fak}. 
But in the context of massive fermions and spontaneous electroweak symmetry breaking the sterile fields will lead to complications 
such as propagator Feynman rules with combinations of $4$- and $D$-dimensional objects in the denominator, see Eq.\ \eqref{Eq:MassivePropagatorMatrix}.
This appears to be a more critical issue than the additional global symmetry breaking.

In conclusion, allowing global symmetry breaking in the kinetic terms while avoiding
additional breaking from evanescent gauge interactions emerges as the most 
straightforward and promising BMHV implementation for key applications, such as 
multi-loop analyses in the EWSM.

% -----------------------------------------------------------------------------

%%%%% Acknowledgments %%%%%
\section*{Acknowledgments} 
P.E., P.K., D.S.\ and M.W.\ acknowledge financial support by the German Science 
Foundation DFG, grant STO 876/8-1.
We would like to thank our collaborators 
Herm\`es B\'elusca-Ma\"\i{}to, Amon Ilakovac, and Marija Ma\dj{}or-Bo\v{z}inovi\'c,
as well as Andreas von Manteuffel for insightful ideas 
and valuable discussions.

%%%%% Appendix %%%%%
\begin{appendices}
\addappheadtotoc

% +++++++++++++++++++++++++++++++++++++++++++++++++++++++++++++++++++++++++++++
\clearpage
\section{Explicit Results for the One-Loop Coefficients}\label{App:Results-1LoopCoeffs}

Here, we provide the counterterm coefficients for the general results presented
in Sec.\ \ref{Sec:1-Loop-Ren-Results}.
In these results, for simplicity, we set $\hypRL=\hypLR$ and 
choose them to be hermitian matrices that
do not only respect electric and colour charge conservation but also commute with
both $\hypL$ and $\hypR$, as discussed in Sec.\ \ref{Sec:1-Loop-Ren-Results}.

In order to derive concise expressions for the coefficients presented 
below, we have used the relations in Eq.\ (\ref{Eq:YukawaHyperchargeBRSTCondition})
imposed by BRST invariance,
the commutation properties of the hypercharge matrices mentioned above and
cyclicity of traces.
This leads to identities such as
\begin{equation}
    \begin{aligned}
        \mathrm{Tr}\big(G^a{K^b}^{\dagger}\big) &= 0,
        &
        \quad
        \mathrm{Tr}\big(MG^a{K^b}^{\dagger}\big) &= 0, 
        &
        \quad 
        \mathrm{Tr}\big(M{G^a}^{\dagger}K^b\big) &= 0,\\
        \mathrm{Tr}\big(M_1{G^a}^{\dagger}M_2K^b\big) &= 0,
        &
        \quad
        \mathrm{Tr}\big({G^a}^{\dagger}K^b{G^c}^{\dagger}K^d\big) &= 0,
        &
        \quad 
        \mathrm{Tr}\big({G^a}^{\dagger}G^b{G^c}^{\dagger}K^d\big) &= 0,
    \end{aligned}
\end{equation}
with $M\in\{\hypL,\hypR,\hypL^2,\hypL\hypR,\hypR^2\}$ and $M_1,\,M_2\in\{\hypL,\hypR\}$, 
among others, to illustrate some examples, which where used to simplify the results.

\subsection{Coefficients of Divergent Contributions}
We start with the coefficients of the divergent contributions,
and define the relations
\begin{equation}\label{App-Eq:DivCoeffRelations}
    \begin{aligned}
        \overline{\mathcal{A}}_{\psi\overline{\psi},\mathrm{R},ji}^{1,\text{inv}} &= g^2 \prescript{}{F}{\overline{\mathcal{A}}}_{\psi\overline{\psi},\mathrm{R},ji}^{1,\text{inv}}
        + \prescript{}{Y}{\overline{\mathcal{A}}}_{\psi\overline{\psi},\mathrm{R},ji}^{1,\text{inv}},\\
        \overline{\mathcal{A}}_{\psi\overline{\psi},\mathrm{L},ji}^{1,\text{inv}} &= g^2 \prescript{}{F}{\overline{\mathcal{A}}}_{\psi\overline{\psi},\mathrm{L},ji}^{1,\text{inv}}
        + \prescript{}{Y}{\overline{\mathcal{A}}}_{\psi\overline{\psi},\mathrm{L},ji}^{1,\text{inv}},\\
        \widehat{\mathcal{A}}_{\psi\overline{\psi},ji}^{1,\text{break}} &= 
        g^2 \prescript{}{F}{\widehat{\mathcal{A}}}_{\psi\overline{\psi},\mathrm{LR},ji}^{1,\text{break}}
        + \prescript{}{Y}{\widehat{\mathcal{A}}}_{\psi\overline{\psi},ji}^{1,\text{break}},\\
        \widehat{\mathcal{A}}_{\psi\overline{\psi}B,ji}^{1,\text{break}} &= 
        g^2 \Big(\mathcal{Y}_{LR} \prescript{}{F}{\widehat{\mathcal{A}}}_{\psi\overline{\psi},\mathrm{LR}}^{1,\text{break}} \Big)_{ji}
        - \Big( \prescript{}{YS}{\widehat{\mathcal{A}}}_{\psi\overline{\psi}B,ji}^{1,\text{break}}
        + \prescript{}{YF}{\widehat{\mathcal{A}}}_{\psi\overline{\psi}B,ji}^{1,\text{break}}\Big),\\
        \overline{\mathcal{A}}_{\psi\overline{\psi}\phi,\mathrm{R},ji}^{1,\text{inv},a} &= \prescript{}{Y}{\overline{\mathcal{A}}}_{\psi\overline{\psi}\phi,\mathrm{R},ji}^{1,\text{inv},a} + g^2 \, \Big(\prescript{}{YF}{\overline{\mathcal{A}}}_{\psi\overline{\psi}\phi,\mathrm{R},ji}^{1,\text{inv},a} + \prescript{}{YSF}{\overline{\mathcal{A}}}_{\psi\overline{\psi}\phi,\mathrm{R},ji}^{1,\text{inv},a}\Big),\\
        \overline{\mathcal{A}}_{\psi\overline{\psi}\phi,\mathrm{L},ji}^{1,\text{inv},a} &= \prescript{}{Y}{\overline{\mathcal{A}}}_{\psi\overline{\psi}\phi,\mathrm{L},ji}^{1,\text{inv},a} + g^2 \, \Big(\prescript{}{YF}{\overline{\mathcal{A}}}_{\psi\overline{\psi}\phi,\mathrm{L},ji}^{1,\text{inv},a} + \prescript{}{YSF}{\overline{\mathcal{A}}}_{\psi\overline{\psi}\phi,\mathrm{L},ji}^{1,\text{inv},a}\Big),\\
        \mathcal{A}_{\phi\phi^{\dagger}\phi\phi^{\dagger},abcd}^{1,\text{inv}} &= \Big( g^4 \prescript{}{S}{\mathcal{A}}_{\phi\phi^{\dagger}\phi\phi^{\dagger}}^{1,\text{inv}} + g^2 \prescript{}{S\lambda}{\mathcal{A}}_{\phi\phi^{\dagger}\phi\phi^{\dagger}}^{1,\text{inv}} + \prescript{}{Y}{\mathcal{A}}_{\phi\phi^{\dagger}\phi\phi^{\dagger}}^{1,\text{inv}} + \prescript{}{\lambda}{\mathcal{A}}_{\phi\phi^{\dagger}\phi\phi^{\dagger}}^{1,\text{inv}} \Big)_{abcd},
    \end{aligned}
\end{equation}
in order to consolidate different contributions associated with the same field monomials.
The following presents the coefficients, organised according to the different types of field monomials:

\paragraph{Gauge Boson Coefficients:} 
The coefficients of the gauge bosons are given by those from
purely fermionic contributions
\begin{equation}\label{App-Eq:GaugeBosonCoeffsFermionic}
    \begin{aligned}
        \prescript{}{F}{\overline{\mathcal{A}}}_{BB}^{1,\text{inv}} &= 
        \frac{2}{3} \Big[ \mathrm{Tr}\big(\mathcal{Y}_R^2\big) + \mathrm{Tr}\big(\mathcal{Y}_L^2\big) \Big],\\
        \prescript{}{F}{\widehat{\mathcal{A}}}_{BB,1}^{1,\text{break}} &= 
        \frac{2}{3} \, \mathrm{Tr}\big(\mathcal{Y}_{LR}^2\big),\\
        \prescript{}{F}{\widehat{\mathcal{A}}}_{BB,2}^{1,\text{break}} &= 
        \frac{2}{3} \, \mathrm{Tr}\Big(\big(\mathcal{Y}_R+\mathcal{Y}_L\big)\mathcal{Y}_{LR}\Big),\\
        \prescript{}{F}{\widehat{\mathcal{A}}}_{BB,3}^{1,\text{break}} &= 
        - 2 \, \mathrm{Tr}\big(\mathcal{Y}_{LR}^2\big),\\
        \prescript{}{F}{\widehat{\mathcal{A}}}_{BB,4}^{1,\text{break}} &= 
        - \frac{1}{3} \, \mathrm{Tr}\Big(\big(\mathcal{Y}_R+\mathcal{Y}_L\big)^2\Big),
    \end{aligned}
\end{equation}
and those from scalar contributions
\begin{equation}\label{App-Eq:GaugeBosonCoeffsScalar}
    \begin{aligned}
        \prescript{}{S}{\mathcal{A}}_{BB}^{1,\text{inv}} &= 
        \frac{N_S}{3} \, \mathcal{Y}_{S}^2.
    \end{aligned}
\end{equation}

\paragraph{Fermion Coefficients:} 
The fermion coefficients originate from gauge interactions
\begin{equation}\label{App-Eq:FermionCoeffFermionGauge}
    \begin{aligned}
        \prescript{}{F}{\overline{\mathcal{A}}}_{\psi\overline{\psi},\mathrm{R},ji}^{1,\text{inv}} &=
        \xi \big(\mathcal{Y}_{R}^2\big)_{ji},\\
        \prescript{}{F}{\overline{\mathcal{A}}}_{\psi\overline{\psi},\mathrm{L},ji}^{1,\text{inv}} &=
        \xi \big(\mathcal{Y}_{L}^2\big)_{ji},\\
        \prescript{}{F}{\widehat{\mathcal{A}}}_{\psi\overline{\psi},\mathrm{LR},ji}^{1,\text{break}} &=
        \frac{1}{3} 
        \Big[
        2(2+\xi) \big(\mathcal{Y}_{R}\mathcal{Y}_{L}\big)_{ji}
        -(1-\xi) \big[\big(\mathcal{Y}_{R}+\mathcal{Y}_{L}\big)\mathcal{Y}_{LR}\big]_{ji}
        -(2+\xi) \big(\mathcal{Y}_{LR}^2\big)_{ji}
        \Big]
    \end{aligned}
\end{equation}
and Yukawa interaction
\begin{equation}\label{App-Eq:FermionCoeffYukawa}
    \begin{aligned}
        \prescript{}{Y}{\overline{\mathcal{A}}}_{\psi\overline{\psi},\mathrm{R},ji}^{1,\text{inv}} &=
        \frac{1}{2} \Big[\big({G^a}^{\dagger}G^a\big)_{ji} + \big({K^a}^{\dagger}K^a\big)_{ji}\Big],\\
        \prescript{}{Y}{\overline{\mathcal{A}}}_{\psi\overline{\psi},\mathrm{L},ji}^{1,\text{inv}} &=
        \frac{1}{2} \Big[\big(G^a{G^a}^{\dagger}\big)_{ji} + \big(K^a{K^a}^{\dagger}\big)_{ji}\Big],\\
        \prescript{}{Y}{\widehat{\mathcal{A}}}_{\psi\overline{\psi},ji}^{1,\text{break}} &=
        \frac{1}{2} \Big[\big(G^aK^a\big)_{ji} + \big(K^aG^a\big)_{ji}\Big].
    \end{aligned}
\end{equation}

\paragraph{Fermion-Gauge Boson Coefficients:} 
Alongside the traditional vertex correction diagram
with gauge boson exchange,
the counterterm coefficients of the fermion-gauge boson interactions
arise from both diagrams with Yukawa and scalar-gauge boson as well as
Yukawa and fermion-gauge boson vertices, i.e.\
\begin{equation}\label{App-Eq:FermionGaugeBosonCoeff}
    \begin{aligned}
        \prescript{}{YS}{\widehat{\mathcal{A}}}_{\psi\overline{\psi}B,ji}^{1,\text{break}} &=
        \frac{1}{2} \, \mathcal{Y}_{S} \Big[\big(K^aG^a\big)_{ji} - \big(G^aK^a\big)_{ji}\Big],\\
        \prescript{}{YF}{\widehat{\mathcal{A}}}_{\psi\overline{\psi}B,ji}^{1,\text{break}} &=
        - \frac{1}{2} \Big[\big(G^a\mathcal{Y}_{LR}K^a\big)_{ji} + \big(K^a\mathcal{Y}_{LR}G^a\big)_{ji}\Big],
    \end{aligned}
\end{equation}
respectively.

\paragraph{Yukawa Coefficients:}
The different contributions to the coefficients of the 
divergent Yukawa contributions are provided by
\begin{equation}\label{App-Eq:DivYukawaCoeffs}
    \begin{aligned}
        \prescript{}{Y}{\mathcal{A}}_{\psi\overline{\psi}\phi,\mathrm{R},ji}^{1,\text{inv},a} &=
        \big(G^b{K^a}^{\dagger}K^b\big)_{ji}+\big(K^b{K^a}^{\dagger}G^b\big)_{ji},\\
        \prescript{}{YF}{\mathcal{A}}_{\psi\overline{\psi}\phi,\mathrm{R},ji}^{1,\text{inv},a} &=
        -(3+\xi)\big(\mathcal{Y}_{L}G^a\mathcal{Y}_{R}\big)_{ji},\\
        \prescript{}{YSF}{\mathcal{A}}_{\psi\overline{\psi}\phi,\mathrm{R},ji}^{1,\text{inv},a} &=
        \xi\,\mathcal{Y}_{S}\Big[\big(G^a\mathcal{Y}_{R}\big)_{ji}-\big(\mathcal{Y}_{L}G^a\big)_{ji}\Big],\\
        \prescript{}{Y}{\mathcal{A}}_{\psi\overline{\psi}\phi,\mathrm{L},ji}^{1,\text{inv},a} &=
        \big({G^b}^{\dagger}G^a{K^b}^{\dagger}\big)_{ji}+\big({K^b}^{\dagger}G^a{G^b}^{\dagger}\big)_{ji},\\
        \prescript{}{YF}{\mathcal{A}}_{\psi\overline{\psi}\phi,\mathrm{L},ji}^{1,\text{inv},a} &=
        -(3+\xi)\big(\mathcal{Y}_{R}{K^a}^{\dagger}\mathcal{Y}_{L}\big)_{ji},\\
        \prescript{}{YSF}{\mathcal{A}}_{\psi\overline{\psi}\phi,\mathrm{L},ji}^{1,\text{inv},a} &=
        \xi\,\mathcal{Y}_{S}\Big[\big({K^a}^{\dagger}\mathcal{Y}_{L}\big)_{ji}-\big(\mathcal{Y}_{R}{K^a}^{\dagger}\big)_{ji}\Big].
    \end{aligned}
\end{equation}

\paragraph{Scalar Coefficients:}
The coefficients for the bilinear scalar contributions are given by
\begin{equation}\label{App-Eq:Div2ScalarCoeffs}
    \begin{aligned}
        \prescript{}{S}{\mathcal{A}}_{\phi\phi^{\dagger},ab}^{1,\text{inv}} &=
        -(3-\xi)\mathcal{Y}_{S}^2 \, \delta_{ab},\\
        \prescript{}{Y}{\overline{\mathcal{A}}}_{\phi\phi^{\dagger},ab}^{1,\text{inv}} &=
        \mathrm{Tr}\big({G^a}^{\dagger}G^b\big)+\mathrm{Tr}\big({K^b}^{\dagger}K^a\big),\\
        {\widehat{\mathcal{A}}}_{\phi\phi^{\dagger},ab}^{1,\text{break}} &=
        \frac{2}{3}\Big(\mathrm{Tr}\big({G^a}^{\dagger}G^b\big)+\mathrm{Tr}\big({K^b}^{\dagger}K^a\big)\Big)
        + \frac{1}{3}\Big(\mathrm{Tr}\big({G^a}^{\dagger}{K^b}^{\dagger}\big)+\mathrm{Tr}\big(K^aG^b\big)\Big),\\
        {\widehat{\mathcal{A}}}_{\phi\phi,ab}^{1,\text{break}} &=
        \frac{1}{3}\Big(\mathrm{Tr}\big(G^aG^b\big)+\mathrm{Tr}\big({K^a}^{\dagger}{K^b}^{\dagger}\big)\Big),
    \end{aligned}
\end{equation}
whereas those from quartic scalar contributions read
\begin{equation}\label{App-Eq:DivSSSScoeffs}
    \begin{aligned}
        \prescript{}{S}{\mathcal{A}}_{\phi\phi^{\dagger}\phi\phi^{\dagger},abcd}^{1,\text{inv}} &=
        6 \, \mathcal{Y}_{S}^4 \, \big(\delta_{ad}\,\delta_{bc}+\delta_{ac}\,\delta_{bd}\big),\\
        \prescript{}{S\lambda}{\mathcal{A}}_{\phi\phi^{\dagger}\phi\phi^{\dagger},abcd}^{1,\text{inv}} &=
        \frac{2\xi}{3} \, \mathcal{Y}_{S}^2 \, \big( \lambda_{abcd} - \lambda_{adcb} - \lambda_{cbad} - \lambda_{cdab} \big),\\
        \prescript{}{Y}{\mathcal{A}}_{\phi\phi^{\dagger}\phi\phi^{\dagger},cadb}^{1,\text{inv}} &=
        -2 \Big[ 
        \mathrm{Tr}\big(G^a{G^c}^{\dagger}G^b{G^d}^{\dagger}\big)
        + \mathrm{Tr}\big(G^a{G^d}^{\dagger}G^b{G^c}^{\dagger}\big)
        + \mathrm{Tr}\big(G^a{G^c}^{\dagger}K^d{K^b}^{\dagger}\big)\\
        &+ \mathrm{Tr}\big(G^a{G^d}^{\dagger}K^c{K^b}^{\dagger}\big)
        + \mathrm{Tr}\big(G^b{G^c}^{\dagger}K^d{K^a}^{\dagger}\big)
        + \mathrm{Tr}\big(G^b{G^d}^{\dagger}K^c{K^a}^{\dagger}\big)\\
        &+ \mathrm{Tr}\big(G^a{K^b}^{\dagger}K^c{G^d}^{\dagger}\big)
        + \mathrm{Tr}\big(G^a{K^b}^{\dagger}K^d{G^c}^{\dagger}\big)
        + \mathrm{Tr}\big(G^b{K^a}^{\dagger}K^c{G^d}^{\dagger}\big)\\
        &+ \mathrm{Tr}\big(G^b{K^a}^{\dagger}K^d{G^c}^{\dagger}\big)
        + \mathrm{Tr}\big(K^c{K^a}^{\dagger}K^d{K^b}^{\dagger}\big)
        + \mathrm{Tr}\big(K^d{K^a}^{\dagger}K^c{K^b}^{\dagger}\big)
        \Big],\\
        \prescript{}{\lambda}{\mathcal{A}}_{\phi\phi^{\dagger}\phi\phi^{\dagger},cadb}^{1,\text{inv}} &=
        \frac{2}{9} \big( 2\,\lambda_{calk}\,\lambda_{dbkl} + 2\,\lambda_{dalk}\,\lambda_{cbkl} + \lambda_{kalb}\,\lambda_{ckdl} \big).
    \end{aligned}
\end{equation}

\paragraph{Scalar-Gauge Boson Coefficients:}
Contributions containing a single gauge boson come with coefficients
\begin{equation}\label{App-Eq:SSBCoeff}
    \begin{aligned}
        {\widehat{\mathcal{A}}}&_{\phi\phi^{\dagger}B,ab}^{1,\text{break}}\\
        = \, &\frac{1}{3}
        \Big[
        \mathrm{Tr}\big(\mathcal{Y}_{LR}{K^b}^{\dagger}{G^a}^{\dagger}\big)
        - \mathrm{Tr}\big(\mathcal{Y}_{LR}{G^a}^{\dagger}{K^b}^{\dagger}\big)
        + \mathrm{Tr}\big(\mathcal{Y}_{LR}G^bK^a\big)
        -\mathrm{Tr}\big(\mathcal{Y}_{LR}K^aG^b\big)
        \Big]\\
        + \, &\frac{2}{3}
        \Big[
        \mathrm{Tr}\big(\mathcal{Y}_{LR}G^b{G^a}^{\dagger}\big)
        - \mathrm{Tr}\big(\mathcal{Y}_{LR}{G^a}^{\dagger}G^b\big)
        + \mathrm{Tr}\big(\mathcal{Y}_{LR}{K^b}^{\dagger}K^a\big)
        -\mathrm{Tr}\big(\mathcal{Y}_{LR}K^a{K^b}^{\dagger}\big)
        \Big],\\
        {\widehat{\mathcal{A}}}&_{\phi\phi B,ab}^{1,\text{break}}\\
        = \, &\frac{1}{3}
        \Big[
        \mathrm{Tr}\big(\mathcal{Y}_{LR}{K^b}^{\dagger}{K^a}^{\dagger}\big)
        - \mathrm{Tr}\big(\mathcal{Y}_{LR}{K^a}^{\dagger}{K^b}^{\dagger}\big)
        + \mathrm{Tr}\big(\mathcal{Y}_{LR}G^bG^a\big)
        -\mathrm{Tr}\big(\mathcal{Y}_{LR}G^aG^b\big)
        \Big],
    \end{aligned}
\end{equation}
while contributions involving two gauge bosons feature coefficients
of the form
\begin{equation}\label{App-Eq:SSBBCoeffs}
    \begin{aligned}
        {\widehat{\mathcal{A}}}&_{\phi\phi^{\dagger}BB,ab}^{1,\text{break}}\\ 
        &= 
        \frac{2}{3} \Big[
        \mathrm{Tr}\Big(\mathcal{Y}_{LR}^2\big\{{G^{a}}^{\dagger},{K^{b}}^{\dagger}\big\}\Big)
        + \mathrm{Tr}\Big(\mathcal{Y}_{LR}^2\big\{G^{b},K^{a}\big\}\Big)
        + 2 \, \mathrm{Tr}\Big(\mathcal{Y}_{LR}^2\big\{{G^{a}}^{\dagger},G^{b}\big\}\Big)\\
        &+ 2 \, \mathrm{Tr}\Big(\mathcal{Y}_{LR}^2\big\{{K^{b}}^{\dagger},K^{a}\big\}\Big)
        - 2 \, \mathrm{Tr}\Big(\mathcal{Y}_{LR}{G^{a}}^{\dagger}\mathcal{Y}_{LR}{K^{b}}^{\dagger}\Big)
        - 2 \, \mathrm{Tr}\Big(\mathcal{Y}_{LR}G^{b}\mathcal{Y}_{LR}K^{a}\Big)\\
        &- 4 \, \mathrm{Tr}\Big(\mathcal{Y}_{LR}{G^{a}}^{\dagger}\mathcal{Y}_{LR}G^{b}\Big)
        - 4 \, \mathrm{Tr}\Big(\mathcal{Y}_{LR}{K^{b}}^{\dagger}\mathcal{Y}_{LR}K^{a}\Big)
        \Big],\\
       {\widehat{\mathcal{A}}}&_{\phi\phi BB,ab}^{1,\text{break}} = 
        \frac{2}{3} \Big[
        \mathrm{Tr}\Big(\mathcal{Y}_{LR}^2\big\{G^{a},G^{b}\big\}\Big)
        + \mathrm{Tr}\Big(\mathcal{Y}_{LR}^2\big\{{K^{a}}^{\dagger},{K^{b}}^{\dagger}\big\}\Big)\\
        &\qquad\quad\, - 2\,\mathrm{Tr}\Big(\mathcal{Y}_{LR}G^{a}\mathcal{Y}_{LR}G^{b}\Big)
        - 2\,\mathrm{Tr}\Big(\mathcal{Y}_{LR}{K^{a}}^{\dagger}\mathcal{Y}_{LR}{K^{b}}^{\dagger}\Big)
        \Big],
    \end{aligned}
\end{equation}
where we have used the anticommutator to bring the result in a concise format.

\subsection{Coefficients of Finite Contributions}
In this subsection, we continue with the coefficients of the finite contributions.
We define the following relations to organise the results for finite fermion-gauge boson contributions:
\begin{equation}\label{App-Eq:FiniteFermionGaugeBosonCoeffsFull}
    \begin{aligned}
        \mathcal{F}_{\psi\overline{\psi}B,\mathrm{R},ji}^{1,\text{break}} &= 
        g^2 \Big(\mathcal{Y}_{R} \prescript{}{F}{\mathcal{F}}_{\psi\overline{\psi},\mathrm{R},ji}^{1,\text{break}} 
        + \prescript{}{F}{\mathcal{F}}_{\psi\overline{\psi}B,ji}^{1,\text{break}}\Big)
        + \prescript{}{YF}{\mathcal{F}}_{\psi\overline{\psi}B,\mathrm{R},ji}^{1,\text{break}},\\
        \mathcal{F}_{\psi\overline{\psi}B,\mathrm{L},ji}^{1,\text{break}} &= 
        g^2 \Big(\mathcal{Y}_{L} \prescript{}{F}{\mathcal{F}}_{\psi\overline{\psi},\mathrm{L},ji}^{1,\text{break}} 
        - \prescript{}{F}{\mathcal{F}}_{\psi\overline{\psi}B,ji}^{1,\text{break}}\Big)
        + \prescript{}{YF}{\mathcal{F}}_{\psi\overline{\psi}B,\mathrm{L},ji}^{1,\text{break}}.
    \end{aligned}
\end{equation}

\paragraph{Gauge Boson Coefficients:}
Both the bilinear and quartic finite gauge boson contributions are governed by
\begin{equation}\label{App-Eq:FiniteGaugeBosonCoeffs}
    \begin{aligned}
        \mathcal{F}_{BB}^{1,\text{break}} &= 
        - \frac{1}{3} \, \mathrm{Tr}\Big(\big(\mathcal{Y}_R-\mathcal{Y}_L\big)^2\Big),\\
        \mathcal{F}_{BBBB}^{1,\text{break}} &= 
        \frac{2}{3} \, \mathrm{Tr}\Big(\big(\mathcal{Y}_R-\mathcal{Y}_L\big)^4\Big).
    \end{aligned}
\end{equation}

\paragraph{Fermion Coefficients:}
The coefficients of the finite fermion contributions are given by
\begin{equation}
    \begin{aligned}
        \prescript{}{F}{\mathcal{F}}_{\psi\overline{\psi},\mathrm{R},ji}^{1,\text{break}} &=
        \big(\mathcal{Y}_R-\mathcal{Y}_L\big)_{jk}\Big(\frac{5+\xi}{6}\mathcal{Y}_R+\frac{1-\xi}{3}\mathcal{Y}_{LR}\Big)_{ki},\\
        \prescript{}{F}{\mathcal{F}}_{\psi\overline{\psi},\mathrm{L},ji}^{1,\text{break}} &=
        -\big(\mathcal{Y}_R-\mathcal{Y}_L\big)_{jk}\Big(\frac{5+\xi}{6}\mathcal{Y}_L+\frac{1-\xi}{3}\mathcal{Y}_{LR}\Big)_{ki}.
    \end{aligned}
\end{equation}

\paragraph{Fermion-Gauge Boson Coefficients:}
In addition to the standard vertex correction diagram
involving a gauge boson exchange, which couples via the physical hypercharges
$\hypL$ and $\hypR$,
the counterterms of the finite fermion-gauge boson interactions
are governed by the coefficients
\begin{equation}
    \begin{aligned}
        \prescript{}{F}{\mathcal{F}}_{\psi\overline{\psi}B,ji}^{1,\text{break}} &=
        \frac{5+\xi}{6}\big(\mathcal{Y}_R-\mathcal{Y}_L\big)_{jk}\big(\mathcal{Y}_{LR}^2\big)_{ki},\\
        \prescript{}{YF}{\mathcal{F}}_{\psi\overline{\psi}B,\mathrm{R},ji}^{1,\text{break}} &=
        -\frac{1}{2}\Big[\big({G^a}^{\dagger}\big(\mathcal{Y}_{R}-\mathcal{Y}_{L}\big)G^a\big)_{ji}
        +\big({K^a}^{\dagger}\big(\mathcal{Y}_{R}-\mathcal{Y}_{L}\big)K^a\big)_{ji}\Big],\\
        \prescript{}{YF}{\mathcal{F}}_{\psi\overline{\psi}B,\mathrm{L},ji}^{1,\text{break}} &=
        \frac{1}{2}\Big[\big(G^a\big(\mathcal{Y}_{R}-\mathcal{Y}_{L}\big){G^a}^{\dagger}\big)_{ji}
        +\big(K^a\big(\mathcal{Y}_{R}-\mathcal{Y}_{L}\big){K^a}^{\dagger}\big)_{ji}\Big].
    \end{aligned}
\end{equation}

\paragraph{Yukawa Coefficients:}
The finite Yukawa counterterms are determined by the following two
coefficients:
\begin{equation}\label{App-Eq:FiniteYukawaCoeffs}
\begin{alignedat}{2}
        \mathcal{F}_{\psi\overline{\psi}\phi,ji}^{1,\text{break},a} 
        &=
        &&- \frac{1}{2}\big(G^bG^aK^b+K^bG^aG^b\big)_{ji} 
        + \frac{1-\xi}{3} g^2 \big(\mathcal{Y}_{L}{K^{a}}^{\dagger}\mathcal{Y}_{LR}+\mathcal{Y}_{LR}{K^{a}}^{\dagger}\mathcal{Y}_{R}\big)_{ji}\\
        &
        &&+ \frac{5+\xi}{3} g^2 \big(\mathcal{Y}_{L}{K^{a}}^{\dagger}\mathcal{Y}_{R}+\mathcal{Y}_{LR}{K^{a}}^{\dagger}\mathcal{Y}_{LR}\big)_{ji},\\
        \mathcal{F}_{\psi\overline{\psi}\phi^{\dagger},ji}^{1,\text{break},a}
        &= 
        &&- \frac{1}{2}\big(K^bK^aG^b+G^bK^aK^b\big)_{ji} 
        + \frac{1-\xi}{3} g^2 \big(\mathcal{Y}_{L}{G^{a}}^{\dagger}\mathcal{Y}_{LR}+\mathcal{Y}_{LR}{G^{a}}^{\dagger}\mathcal{Y}_{R}\big)_{ji}\\
        &
        &&+ \frac{5+\xi}{3} g^2 \big(\mathcal{Y}_{L}{G^{a}}^{\dagger}\mathcal{Y}_{R}+\mathcal{Y}_{LR}{G^{a}}^{\dagger}\mathcal{Y}_{LR}\big)_{ji}.
\end{alignedat}
\end{equation}

\paragraph{Scalar Coefficients:}
The finite scalar counterterms that can contribute to global hypercharge violation 
feature the general coefficients
\begin{equation}\label{App-Eq:Finite2ScalarCoeff}
    \begin{aligned}
        \mathcal{F}_{\phi\phi,ab}^{1,\text{break}} &= 
        -\frac{1}{3}\Big(\mathrm{Tr}\big(G^aG^b\big)+\mathrm{Tr}\big({K^a}^{\dagger}{K^b}^{\dagger}\big)\Big),
    \end{aligned}
\end{equation}
and
\begin{align}
    \begin{split}\label{App-Eq:Finite4ScalarCoeff-SSSS}
        \mathcal{F}_{\phi\phi\phi\phi,abcd}^{1,\text{break}} &=
        \bigg\{\frac{1}{3}\mathrm{Tr}\big(G^{a}G^{b}G^{c}G^{d}\big)
        +\frac{2}{3}\Big[\mathrm{Tr}\big(G^{a}G^{b}G^{c}{K^{d}}^{\dagger}\big)+\mathrm{Tr}\big(G^{a}G^{b}{K^{c}}^{\dagger}G^{d}\big)\\
        &+\mathrm{Tr}\big(G^{a}{K^{b}}^{\dagger}G^{c}G^{d}\big)+\mathrm{Tr}\big({K^{a}}^{\dagger}G^{b}G^{c}G^{d}\big)\Big]\\
        &-\frac{2}{3}\Big[
        \mathrm{Tr}\big(G^{a}G^{b}{K^{c}}^{\dagger}{K^{d}}^{\dagger}\big)+\mathrm{Tr}\big(G^{a}{K^{b}}^{\dagger}G^{c}{K^{d}}^{\dagger}\big)
        +\mathrm{Tr}\big(G^{a}{K^{b}}^{\dagger}{K^{c}}^{\dagger}G^{d}\big)\\
        &+\mathrm{Tr}\big({K^{a}}^{\dagger}G^{b}G^{c}{K^{d}}^{\dagger}\big)+\mathrm{Tr}\big({K^{a}}^{\dagger}G^{b}{K^{c}}^{\dagger}G^{d}\big)
        +\mathrm{Tr}\big({K^{a}}^{\dagger}{K^{b}}^{\dagger}G^{c}G^{d}\big)\Big]\\
        &+\frac{2}{3}\Big[\mathrm{Tr}\big(G^{a}{K^{b}}^{\dagger}{K^{c}}^{\dagger}{K^{d}}^{\dagger}\big)
        +\mathrm{Tr}\big({K^{a}}^{\dagger}G^{b}{K^{c}}^{\dagger}{K^{d}}^{\dagger}\big)
        +\mathrm{Tr}\big({K^{a}}^{\dagger}{K^{b}}^{\dagger}G^{c}{K^{d}}^{\dagger}\big)\\
        &+\mathrm{Tr}\big({K^{a}}^{\dagger}{K^{b}}^{\dagger}{K^{c}}^{\dagger}G^{d}\big)\Big]
        + \frac{1}{3}\mathrm{Tr}\big({K^{a}}^{\dagger}{K^{b}}^{\dagger}{K^{c}}^{\dagger}{K^{d}}^{\dagger}\big)\bigg\}\\
        &+ \big(\text{$23$ permutations of the indices $(a,b,c,d)$}\big),
    \end{split}\\[1.5ex]
    \begin{split}\label{App-Eq:Finite4ScalarCoeff-SdaggerSSS}
        \mathcal{F}_{\phi^{\dagger}\phi\phi\phi,abcd}^{1,\text{break}} &=
        \bigg\{\frac{4}{3}\Big[
        \mathrm{Tr}\big({G^{a}}^{\dagger}G^{b}G^{c}G^{d}\big)
        +\mathrm{Tr}\big({G^{a}}^{\dagger}G^{b}{K^{c}}^{\dagger}{K^{d}}^{\dagger}\big)
        +\mathrm{Tr}\big({G^{a}}^{\dagger}{K^{b}}^{\dagger}G^{c}{K^{d}}^{\dagger}\big)\\
        &+\mathrm{Tr}\big({G^{a}}^{\dagger}{K^{b}}^{\dagger}{K^{c}}^{\dagger}G^{d}\big)
        +\mathrm{Tr}\big(K^{a}G^{b}G^{c}{K^{d}}^{\dagger}\big)
        +\mathrm{Tr}\big(K^{a}G^{b}{K^{c}}^{\dagger}G^{d}\big)\\
        &+\mathrm{Tr}\big(K^{a}{K^{b}}^{\dagger}G^{c}G^{d}\big)
        +\mathrm{Tr}\big(K^{a}{K^{b}}^{\dagger}{K^{c}}^{\dagger}{K^{d}}^{\dagger}\big)
        -\mathrm{Tr}\big({G^{a}}^{\dagger}G^{b}G^{c}{K^{d}}^{\dagger}\big)\\
        &-\mathrm{Tr}\big({G^{a}}^{\dagger}G^{b}{K^{c}}^{\dagger}G^{d}\big)
        -\mathrm{Tr}\big({G^{a}}^{\dagger}{K^{b}}^{\dagger}G^{c}G^{d}\big)
        -\mathrm{Tr}\big(K^{a}G^{b}{K^{c}}^{\dagger}{K^{d}}^{\dagger}\big)\\
        &-\mathrm{Tr}\big(K^{a}{K^{b}}^{\dagger}G^{c}{K^{d}}^{\dagger}\big)
        -\mathrm{Tr}\big(K^{a}{K^{b}}^{\dagger}{K^{c}}^{\dagger}G^{d}\big)
        \Big]\\
        &+\frac{2}{3}\Big[
        \mathrm{Tr}\big({G^{a}}^{\dagger}{K^{b}}^{\dagger}{K^{c}}^{\dagger}{K^{d}}^{\dagger}\big)
        +\mathrm{Tr}\big(K^{a}G^{b}G^{c}G^{d}\big)\Big]\bigg\}\\
        &+ \big(\text{$5$ permutations of the indices $(b,c,d)$}\big).
    \end{split}
\end{align}

\paragraph{Scalar-Gauge Boson Coefficients:}
Finally, the coefficients for the finite contributions to
scalar-gauge boson counterterms are given by
\begin{equation}\label{App-Eq:FiniteScalarGaugeBosonCoeffs}
    \begin{alignedat}{2}
        \mathcal{F}_{\phi\phi^{\dagger}B,ab}^{1,\text{break}} &= 
        &&-\frac{1}{3}\Big[ 
        \mathrm{Tr}\Big(\mathcal{Y}_{L}\big[{G^a}^{\dagger}G^b-{K^b}^{\dagger}K^a\big]\Big)
        + \mathrm{Tr}\Big(\mathcal{Y}_{R}\big[K^a{K^b}^{\dagger}-G^b{G^a}^{\dagger}\big]\Big)\\
        &
        &&+\mathcal{Y}_{S} \Big( \mathrm{Tr}\big({G^a}^{\dagger}G^{b}\big) + \mathrm{Tr}\big({K^{b}}^{\dagger}K^{a}\big) \Big)
        \Big],\\
        \mathcal{F}_{\phi\phi^{\dagger}BB,ab}^{1,\text{break}} &= 
        &&-\frac{2}{3}\Big[
        \mathrm{Tr}\Big(\mathcal{Y}_{L}^2\big[{G^a}^{\dagger}G^b+{K^b}^{\dagger}K^a\big]\Big)
        + 3 \, \mathrm{Tr}\Big(\mathcal{Y}_{L}^2\big[G^b{G^a}^{\dagger}+K^a{K^b}^{\dagger}\big]\Big)\\
        &
        &&- 4 \, \mathrm{Tr}\Big(\mathcal{Y}_{R}\mathcal{Y}_{L}\big[{G^a}^{\dagger}G^b+G^b{G^a}^{\dagger}+{K^b}^{\dagger}K^a+K^a{K^b}^{\dagger}\big]\Big)\\
        &
        &&+ 3 \, \mathrm{Tr}\Big(\mathcal{Y}_{R}^2\big[{G^a}^{\dagger}G^b+{K^b}^{\dagger}K^a\big]\Big)
        + \mathrm{Tr}\Big(\mathcal{Y}_{R}^2\big[G^b{G^a}^{\dagger}+K^a{K^b}^{\dagger}\big]\Big)\\
        &
        &&+ \mathrm{Tr}\Big(\big[\mathcal{Y}_{R}+\mathcal{Y}_{L}\big]{G^a}^{\dagger}\big[\mathcal{Y}_{R}+\mathcal{Y}_{L}\big]G^{b}\Big)
        - 4 \, \mathrm{Tr}\Big(\mathcal{Y}_{R}{G^a}^{\dagger}\mathcal{Y}_{L}G^b\Big)\\
        &
        &&+ \mathrm{Tr}\Big(\big[\mathcal{Y}_{R}+\mathcal{Y}_{L}\big]{K^b}^{\dagger}\big[\mathcal{Y}_{R}+\mathcal{Y}_{L}\big]K^a\Big)
        - 4 \, \mathrm{Tr}\Big(\mathcal{Y}_{R}{K^b}^{\dagger}\mathcal{Y}_{L}K^a\Big)
        \Big].
    \end{alignedat}
\end{equation}

% -----------------------------------------------------------------------------

\end{appendices}

%%%%% Bibliography %%%%%
\printbibliography

\end{document}